\documentclass[12pt]{article}
\usepackage{epsfig}
\usepackage{titlesec}
\usepackage{amssymb,amsmath,amsfonts,latexsym,graphicx,epsfig}
\usepackage{young}
\newcommand{\MeV}{\mathop{\rm MeV}\nolimits}
\newcommand{\GeV}{\mathop{\rm GeV}\nolimits}

\DeclareMathAlphabet\mathbfcal{OMS}{cmsy}{b}{n}
\newcommand{\tr}{\mathop{\rm tr}\nolimits}
\newcommand{\real}{\mathop{\rm Re}\nolimits}

\begin{document}

\begin{titlepage}

\pagestyle{empty}
\begin{flushright}
MITP/21-070\\
%\today
\end{flushright}
\vskip 0.7in
\begin{center}
{\large{\bf
Topological tensor invariants and the current algebra approach:\\[5pt]
Analysis of 196 nonleptonic two-body decays of single\\[8pt]
and double charm baryons -- a review}}
\end{center}

\vspace{12pt}

\begin{center}
{\large \bf Stefan~Groote$^1$, J\"urgen~G.~K\"orner$^2$}\\[.4truecm]
$^1$F\"u\"usika Instituut, Tartu \"Ulikool,\\
  W.~Ostwaldi 1, EE-50411 Tartu, Estonia\\[.1truecm]
$^2$PRISMA Cluster of Excellence, Institut f\"ur Physik,\\
  Johannes-Gutenberg-Universit\"at,\\
  Staudinger Weg 7, D-55099 Mainz, Germany\\[.3truecm]
\end{center}

\begin{abstract}
We shed new light on the standard current algebra approach to the nonleptonic
two-body decays of single and double heavy charm baryons. By making use of the
completeness relation for the flavor wave functions of the ground state
baryon {\bf 20'} representation we are able to rewrite the results of the
current algebra approach in terms of the seven topological tensor invariants
describing the decays. The representation of the current algebra results in
terms of topological tensor invariants depends only on the initial and final
state of the process. The summation over intermediate states inherent to the
current algebra result is automatically taken care of in the tensor
representation. In this way one arrives at a new, quick and very compact
assessment of the results of the current algebra/pole model calculation. For
example, one can quickly identify the decays in which the p.v.\ $S$-wave
amplitude is predicted to vanish implying zero polarization asymmetries in
these decays. We provide tables of the values of the seven topological tensor
invariants for all Cabibbo favored and singly and doubly Cabibbo suppressed
nonleptonic single and double charm baryon decays. In total we treat 196 charm
baryon decays. We also discuss the charm preserving $\Delta C=0$ decays of
single charm baryons and the usual hyperon decays.
\end{abstract}

\end{titlepage}

%%%%%%%%%%%%%%%%%%%%%%%%%%%%%%%%%%%%%%%%%%%%%%%%%%%%%%%%%%%%%%%%%%%%%%%%%%%%%%%
\section{\label{intro}Introduction}
%%%%%%%%%%%%%%%%%%%%%%%%%%%%%%%%%%%%%%%%%%%%%%%%%%%%%%%%%%%%%%%%%%%%%%%%%%%%%%%
The experimental landscape of charm baryon decays has considerably changed
in 2015 through the detection of $\Lambda_{c}^+$ pair production by the BESIII
collaboration at the Beijing $e^+\,e^-$ collider BEPCII~\cite{Ablikim:2015flg}.
Not only was it now possible to determine the absolute branching ratios of
various nonleptonic $\Lambda_{c}$ decays through tagging techniques (see also
Ref.~\cite{Zupanc:2013iki}), but the BESIII collaboration could also determine
the asymmetry parameters of the nonleptonic two-body decays
$\Lambda^+_{c} \to pK_S^0,\,\,\Lambda^0 \pi^+,\,\Sigma^+\pi^0$ and
$\Sigma^0\pi^+$~\cite{Ablikim:2019zwe} (see also the review~\cite{Li:2021iwf}).
The decay asymmetry parameter of the decay $\Xi_{c}^0\to \Xi^-\pi^+$ was
recently measured by the Belle collaboration with the result 
$\alpha_{\Xi_{c}^0\to \Xi^-\pi^+}=-0.60\pm0.04$~\cite{Li:2021uhk}.

Both the absolute branching ratios and the asymmetry parameters provide
essential input for attempts to theoretically understand the nonleptonic
two-body decays of charm baryons. The first Cabibbo suppressed decay
$\Lambda_{c} \to p \,\phi$ was measured by the CLEO collaboration in
1995~\cite{Alexander:1995hd}, followed by more precise measurements of this
mode later on~\cite{Abe:2001mb,Ablikim:2016tze}. At a later stage the BESIII
collaboration measured the singly Cabibbo suppressed decay
$\Lambda_{c}^+ \to p\,\eta $~\cite{Ablikim:2017ors}, later on made more
precise in Ref.~\cite{Li:2021uzo} including an improved upper limit on the
Cabibbo suppressed decay $\Lambda_{c}^+ \to p\,\pi^0 $. By using double tag
techniques, the BESIII collaboration also identified the decay
$\Lambda_{c}^+ \to n K_s^0\pi^+$ involving a neutron~\cite{Ablikim:2016mcr}
which raises the hope that final states including a neutron can be measured in
the future. A considerably larger data sample of $\Lambda_{c}^+$ pairs is
expected in the near future at BEPCII. The rate of $\Lambda_{c}^+$ pairs is to
be increased by a factor of 16 from a total of about $10^5$ to $1.6\cdot 10^6$
$\Lambda_{c}^+$ pairs~\cite{Ablikim:2019hff} which will considerably enlarge the
data sample of $\Lambda^+_{c}$ decays. At the same time an approved energy
upgrade will increase the energy of BEPCII to $4.9\GeV$, opening the
possibility of $\Sigma_{c}$ pair production. If the energy of the BEPCII could
be further increased to a little above $4.95\GeV$ one could also detect the
production of $\Xi_{c}^+$ and $\Xi_{c}^0$ pairs.

Absolute values of the branching fractions of the $\Xi_{c}^+$ and $\Xi_{c}^0$
have recently become available again from the BELLE collaboration through
tagging techniques in the decays $B^- \to \bar\Lambda_{c}^-\,\Xi_{c}^0$ and
$B^0 \to \bar\Lambda_{c}^-\,\Xi_{c}^+$~\cite{Li:2018qak,Li:2019atu}. Using
similar techniques one could also determine the absolute values of the
branching fractions of the decays of the $\Omega_{c}^0$ e.g.\ in the
kinematically accessible decay
$B^-\to \Omega_{c}^0\,\overline{\Xi_{c}^+}$~\cite{Korner:1988mx}.
This would be quite welcome since the absolute branching ratios of the decays
of the $\Omega_{c}^0$ are experimentally unknown at present.

The LHCb collaboration has shown that it is now even possible to detect doubly
Cabibbo suppressed (DCS) nonleptonic charm baryon decays by their measurement
of the decay $\Xi_{c}^+ \to p \phi$~\cite{Aaij:2019kss}. Furthermore, the LHCb
collaboration recently provided results on a branching ratio measurement of
the charm conserving $\Delta C=0$ singly Cabibbo suppressed (SCS) decay
$\Xi_{c}^0 \to \Lambda_{c}^+\,\pi^-$~\cite{Aaij:2020wtg}. Finally, the LHCb
collaboration has extended the scope of possible charm baryon decay
measurements by presenting results on the nonleptonic double charm baryon
two-body decay $\Xi^{++}_{cc} \to \Xi^+_{c} \,\pi^+$~\cite{Aaij:2018gfl}.

It is apparent that the experimentalists have provided us with a rich sample
of results on exclusive charm baryons decays with the hope of much more to
come. This has led to an upsurge of interest in charm baryon decays in the
theoretical community documented by the many recent papers on this subject.

\begin{figure}\begin{center}
\epsfig{figure=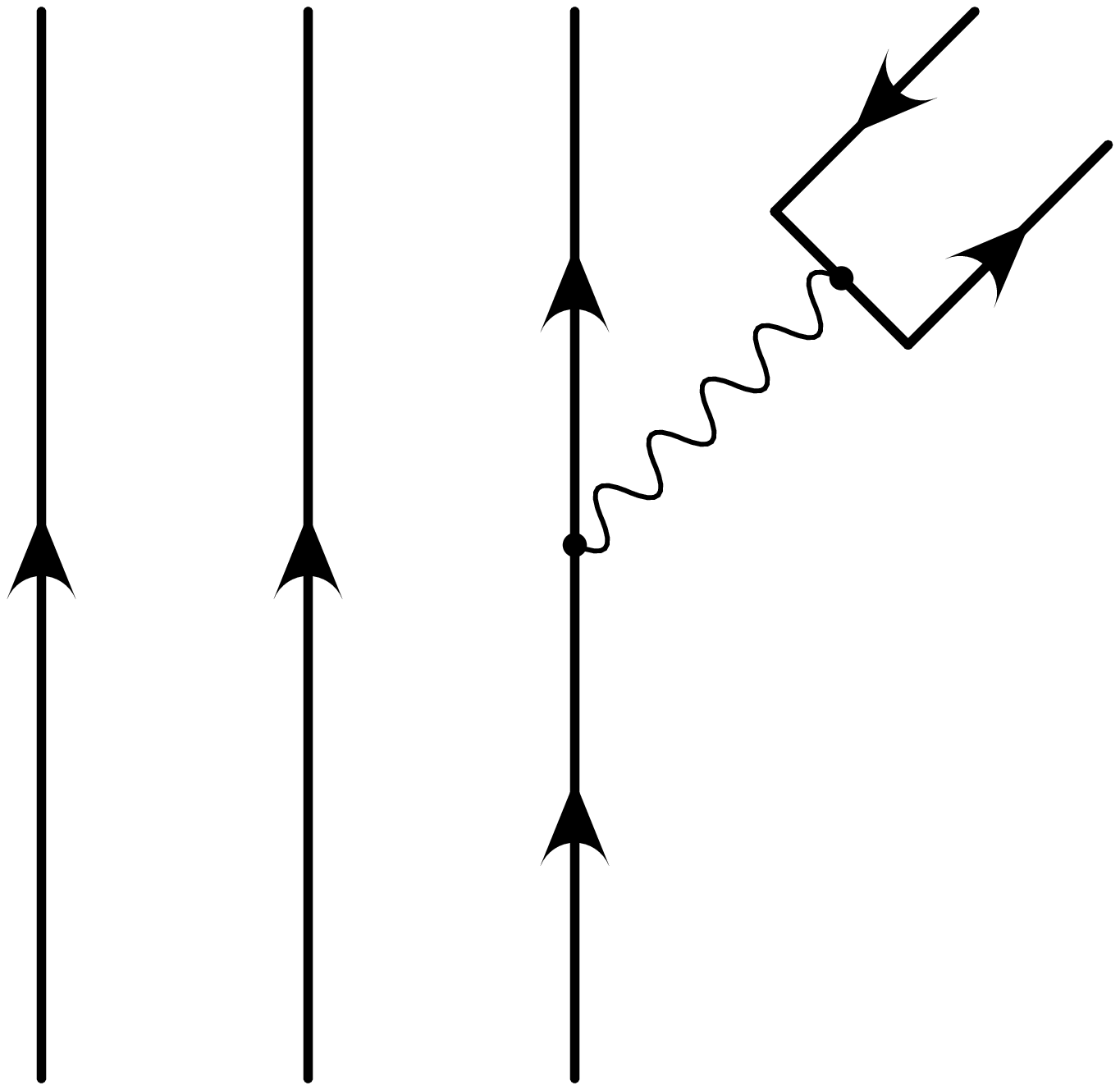,scale=0.18}\qquad
\epsfig{figure=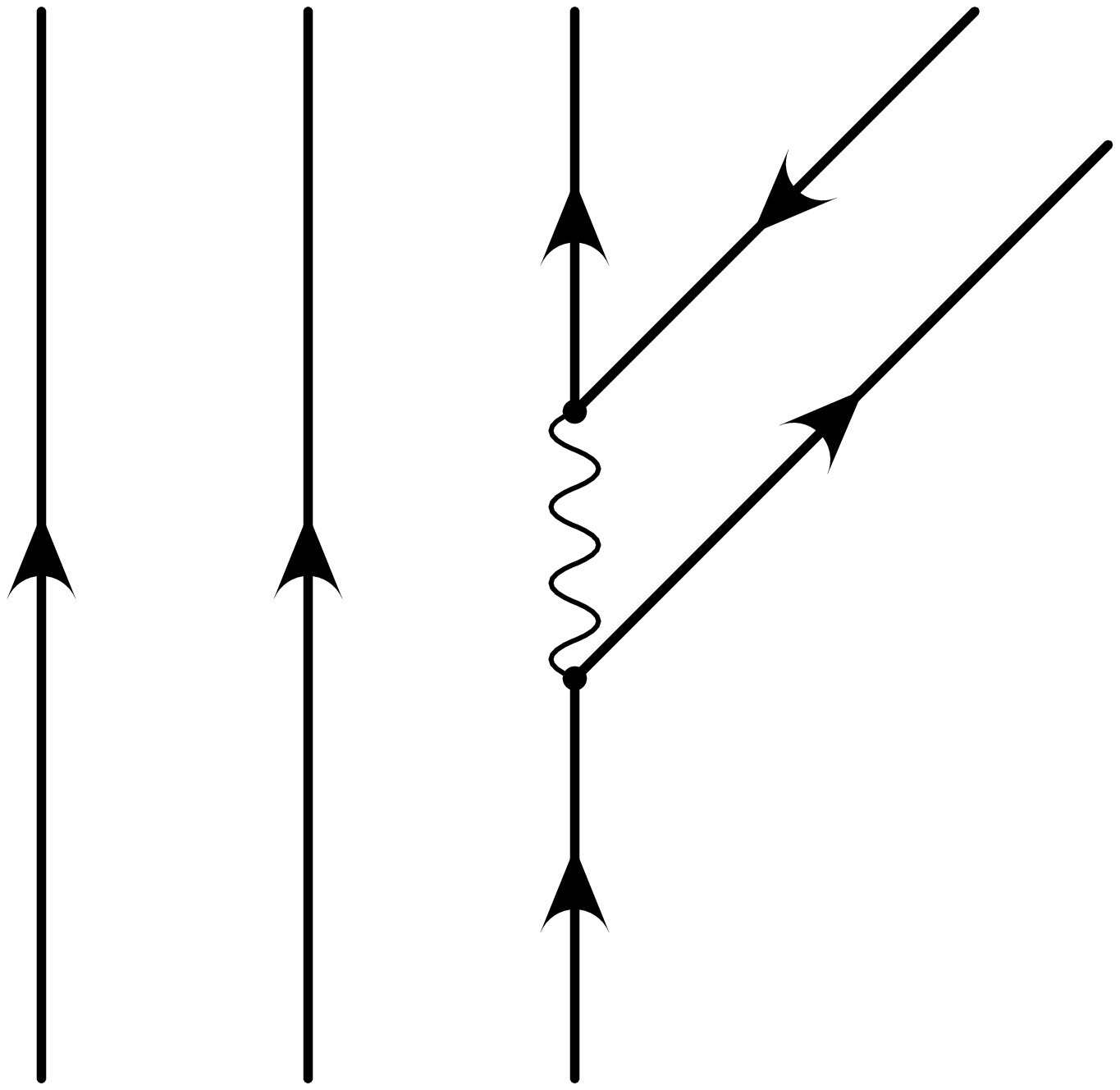,scale=0.18}\qquad
\epsfig{figure=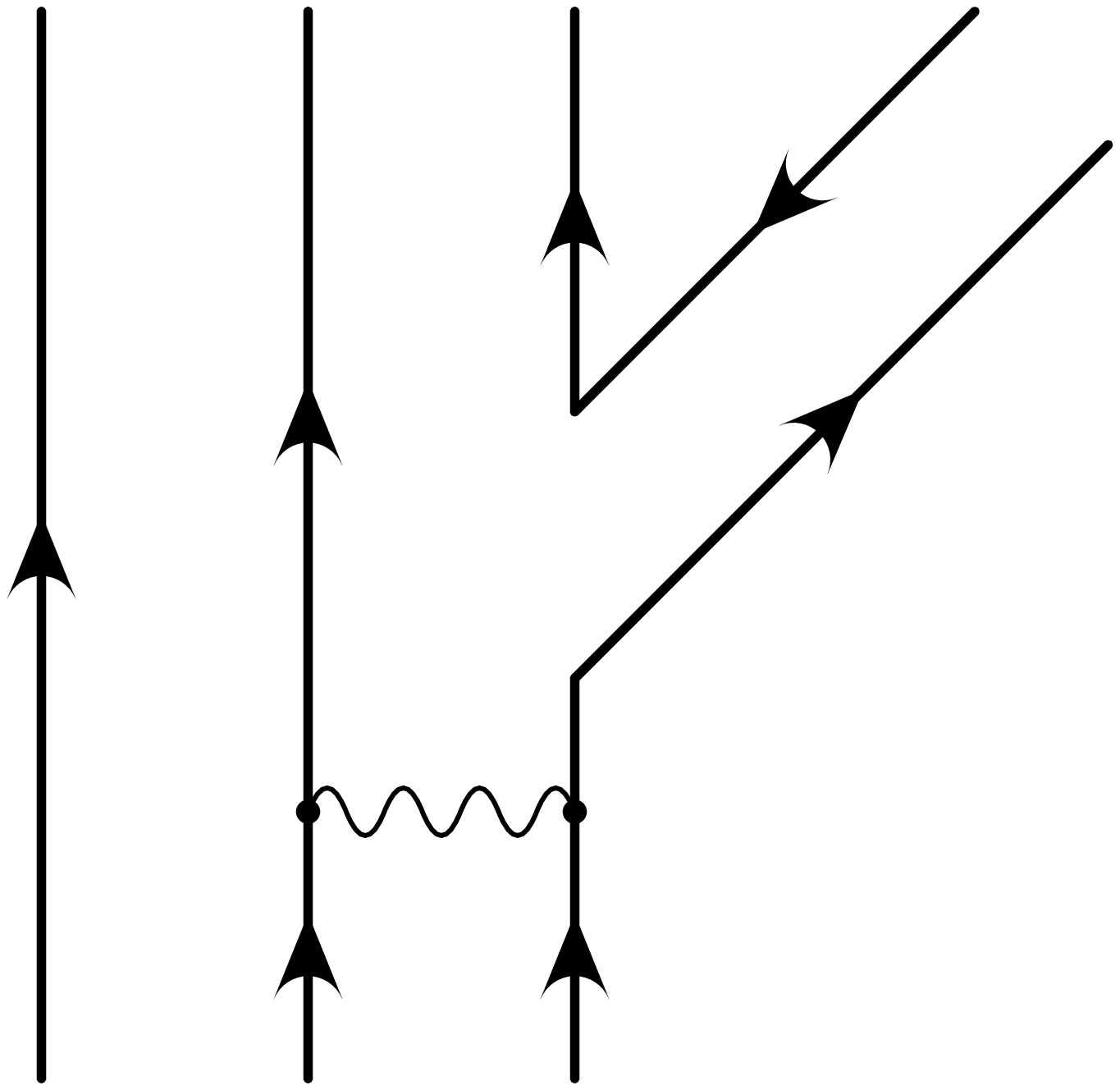,scale=0.18}\qquad
\epsfig{figure=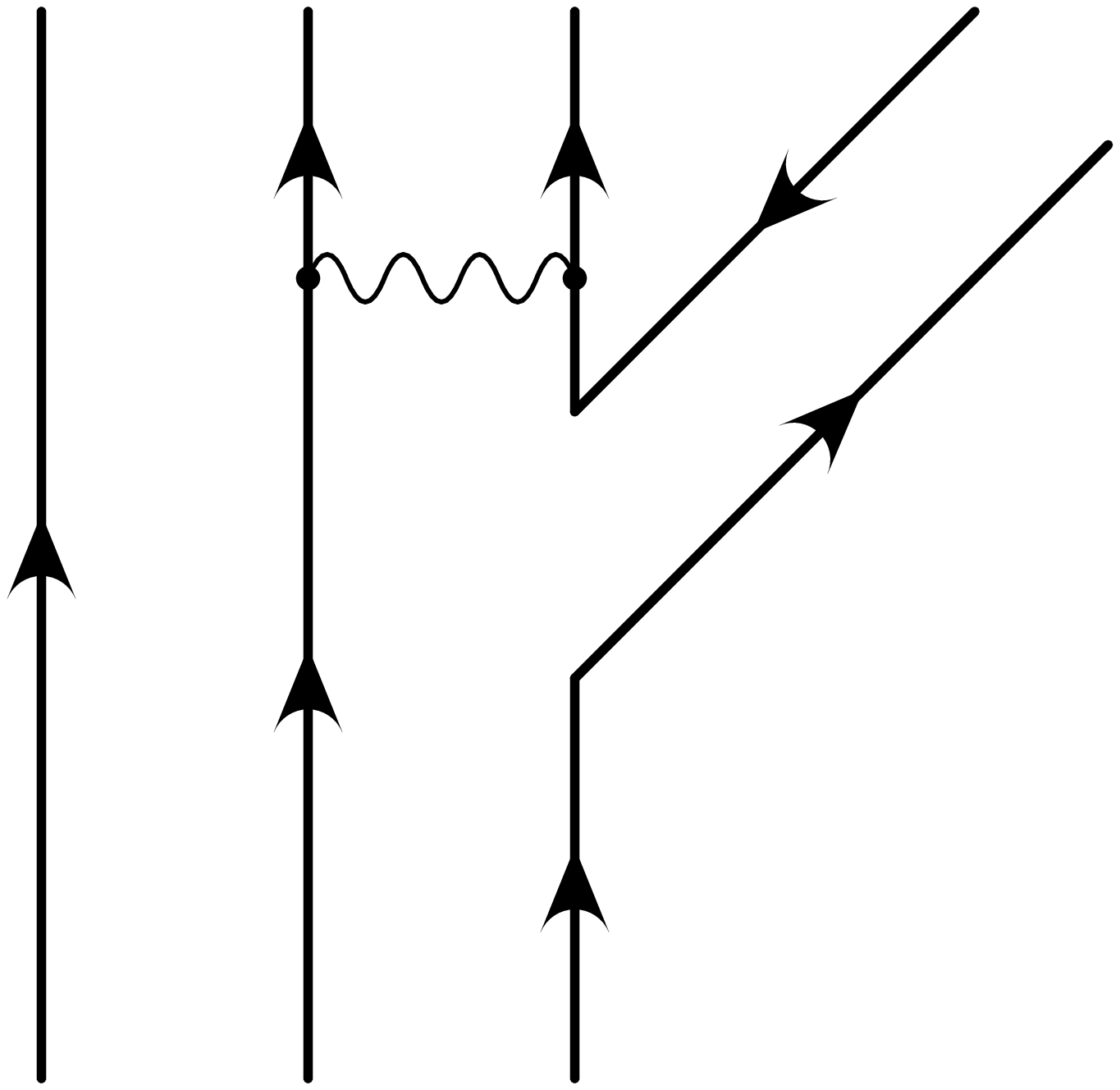,scale=0.18}\qquad
\epsfig{figure=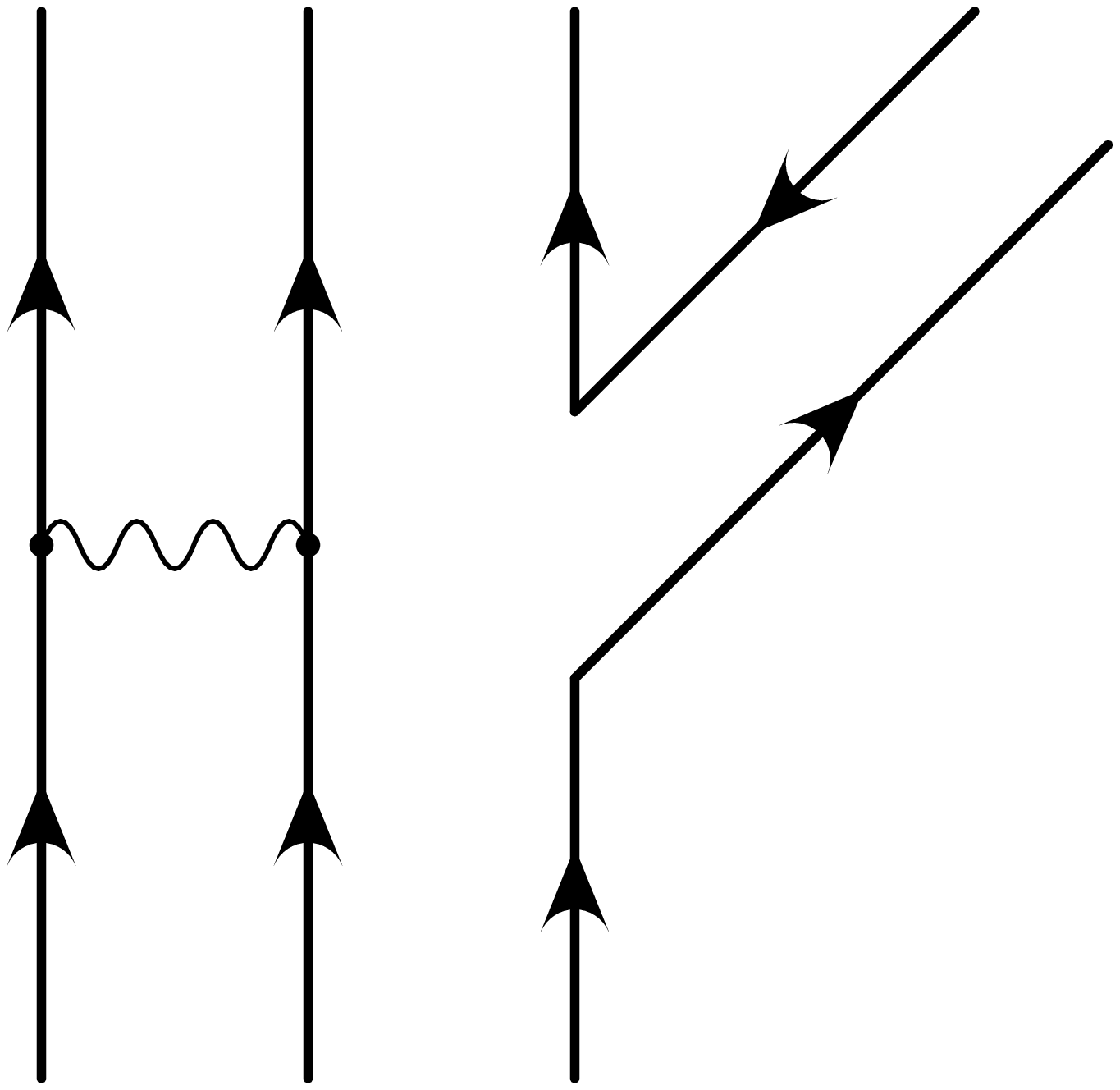,scale=0.18}\\[7pt]
\mbox{Ia}\kern78pt\mbox{Ib}\kern78pt\mbox{IIa}\kern78pt
\mbox{IIb}\kern78pt\mbox{III}\kern36pt\strut\\[12pt]
\caption{\label{topo}The topological diagrams contributing to the nonleptonic
charm baryon decays}
\end{center}\end{figure}

Basic to the current--current (more precisely coined current$\times$current)
quark model description of nonleptonic charm baryon two-body decays are the
five generic quark diagrams depicted in Fig.~\ref{topo} which are also termed
topological diagrams. Diagrams Ia and Ib are usually referred to as
factorizable diagrams or tree diagrams while diagrams IIa, IIb and III are
referred to as nonfactorizable or $W$-exchange diagrams. In this paper we
concentrate on the structure of the $W$-exchange diagrams IIa, IIb and III
which we relate to the $s$- and $u$-channel contributions of the current
algebra aproach.

The theoretical papers on exclusive two-body charm baryons decays roughly fall
into three classes which differ by their dynamical input. In the simplest
case one exploits SU(3) symmetry to relate the amplitudes of different charm
baryon decays~\cite{Savage:1989qr,Savage:1991wu,Pakvasa:1990if,%
Sheikholeslami:1991ab,Verma:1995dk,Sharma:1996sc,Lu:2016ogy,Geng:2018plk,%
Geng:2018bow,Geng:2019xbo,Jia:2019zxi}. An open issue is whether to apply
SU(3) symmetry to the invariant or to the helicity amplitudes of these decays.
This will make a big difference for decays involving the $\eta$ and $\eta'$
mesons. The SU(3) approach has been extended to include SU(3) symmetry
breaking effects~\cite{Savage:1991wu,Geng:2018bow}. Even more insight can be
gained by incorporating  the topological diagram approach~\cite{Korner:1978tc,%
Kohara:1991ug,Chau:1995gk,He:2018joe} into the SU(3) analysis where the
topological diagrams are characterized by the five diagrams Fig.~\ref{topo}
described above. It is, however, not sufficient to simply associate a given
decay with its accompanying topological diagrams, but one has to calculate the
appropiate weight factors for each decay. The SU(3) fit in
Ref.~\cite{Zhao:2018mov} using topological diagrams must therefore be
considered to be invalid.

In the constituent quark model approach one attempts to calculate the
nonleptonic transitions represented by Fig.~\ref{topo} in terms of constituent
quark model transitions. While diagrams Ia, Ib and IIb are directly accessible
to a constituent quark model calculation~\cite{Maiani:1978tu,Niu:2020gjw},
diagrams IIa and III involve the creation of a quark pair from the vacuum in
addition to the constituent quark model transitions. In the covariantized
constituent quark model (CCQM) calculations~\cite{Korner:1978tc,Korner:1992wi}
the quark pair creation from the vacuum is effectively described by the
$^3P_0$ model where the strength of the $^3P_0$ interaction is fixed by a fit
to the data. Both of the above papers are based on an approach developed much
earlier by F.~Hussain and P.~Rotelli~\cite{Hussain:1966} and by J.~G.\
K\"orner and T.~Gudehus~\cite{Korner:1972} for studying nonleptonic hyperon
decays. In the covariant confined quark model calculations of
Refs.~\cite{Ivanov:1997hi,Ivanov:1997ra,Gutsche:2019iac} the five diagrams in
Fig.~\ref{topo} are interpreted as multiloop Feynman diagrams with nonlocal
interactions describing the three- and four-point hadron-quark vertices.

Even more dynamics is added in the standard current algebra plus pole model
approach~\cite{Pakvasa:1990if,Cheng:1991sn,Cheng:1993gf,Xu:1992vc,%
Uppal:1994pt,Zenczykowski:1993hw,Zenczykowski:1993jm,Datta:1995mn,%
Fayyazuddin:1996iy,Sharma:1998rd,Sharma:2017txj,Dhir:2018twm,Cheng:2018hwl}.
This approach has its roots in the description of nonleptonic hyperon decays
developed in the 1960's~\cite{Brown:1966zz,Sugawara:1965zza,Suzuki:1965zz,%
Hara:1997wc} which is based on current algebra and the soft pion theorem. Even
if the criterion of a soft pseudoscalar meson is not always satisfied in the
nonleptonic decays of single and double charm baryons (in exception are the
$\Delta C=0$ single charm baryon decays), the current algebra approach to
charm baryon decays provides a definite calculational framework which can be
tested against experiment. In a recent series of papers Cheng and
collaborators presented a detailed analysis of the nonleptonic decays of
single and double charm baryons using a variant of the current algebra
approach in which the purported current algebra contributions of the
topological diagram III in Fig.~\ref{topo} are dropped~\cite{Cheng:2020wmk,%
Zou:2019kzq,Meng:2020euv,Hu:2020nkg}. The authors of Ref.~\cite{Niu:2020gjw}
make use of the current algebra approach in addition to calculating the
diagram I and IIb transitions in the constituent quark model. This possibly
amounts to a double counting of the contribution from the $W$-exchange diagram
IIb.

Our paper is structured as follows. In Sec.~\ref{effective} we write down the
effective Hamiltonian for the nonleptonic current--current transition. In
Sec.~\ref{topotensor} we discuss issues of SU(4) and SU(3) relevant to the
nonleptonic single and double charm baryon decays. In particular, we introduce
the 20 ground state baryons comprising the ${\bf 20'}$ representation of SU(4)
and write down the third rank flavor tensors that are associated with them. We
present the orthonormality and completeness relation of the third rank flavor
tensors of the ${\bf 20'}$ representation. We introduce a minimal set of
seven topological tensor invariants in flavor space, built from the tensor
representations of the ground state baryons, the ground state pseudoscalar
mesons, and the effective Hamiltonian. The seven tensor invariants are each
associated with one of the corresponding topological quark diagrams.

In Sec.~\ref{tables} we present tables of the values of the topological tensor
invariants for the antitriplet and sextet single charm baryon decays, the
$\Delta C=0$ single charm baryon decays, the triplet double charm baryon
decays to the antitriplet and sextet single charm baryons, and the double
charm baryon decays to the light baryon octet and heavy $D,D_s$ mesons, each
for the Cabibbo favored (CF), singly Cabibbo suppressed (SCS) and doubly
Cabibbo suppressed (DCS) decays. Altogether we list the values of the
topological tensor invariants for 196 single and double nonleptonic charm
baryon decays. For each class of decays we explicate the linear relations
among the associated tensor invariants. For completeness we also list the
values of the tensor invariants for the ordinary nonleptonic hyperon decays.

In Sec.~\ref{current} we briefly recapitulate the current algebra plus pole
model approach to nonleptonic charm baryon decays where we closely follow the
presentation and notation of Refs.~\cite{Cheng:2020wmk,Zou:2019kzq,%
Meng:2020euv,Hu:2020nkg}. Using the completeness relation for the ground state
baryons we explicitly derive the relation of the so-called $s$-channel and
$u$-channel contributions of the current algebra approach to the topological
tensor invariants. The representation of the current algebra results in terms
of topological tensor invariants depends only on the initial and final state
of the process. The summation over intermediate states inherent to the current
algebra result is automatically taken care of in the tensor representation. In
Sec.~\ref{sample} we discuss several explicit examples, leading to the
exposure of general features of the topological tensor and current algebra
approaches in Sec.~\ref{features}. Sec.~\ref{summary} contains our summary and
outlook, while technical issues are treated in five Appendices.

In Appendix~\ref{tensors} we list explicit values of the tensor components of
the ground state baryons, the ground state pseudoscalar mesons, and the
effective Hamiltonian. Appendix~\ref{complete} contains the derivation of the
completeness relation for the members of the ${\bf 20'}$ representation.
In Appendix~\ref{weak-strong} we define a set of flavor tensor invariants
describing the strong and weak transitions $\langle B_fM|B_i\rangle$ and
$\langle B_f|{\cal H}_{\rm eff}|B_i\rangle$ involving the charm baryon states,
the values of which are needed for the comparison of the current algebra and
the topological diagram approaches. We provide tables of the values of the
strong and weak tensor invariants needed for the example cases discussed in
the main text.  In Appendix~\ref{scrutiny} we clarify the implications and
shortcomings of the variant of the current algebra approach introduced in
Refs.~\cite{Cheng:2020wmk,Zou:2019kzq,Meng:2020euv,Hu:2020nkg}. In
Appendix~\ref{spinkinematics} we are dealing with the spin kinematics.

%%%%%%%%%%%%%%%%%%%%%%%%%%%%%%%%%%%%%%%%%%%%%%%%%%%%%%%%%%%%%%%%%%%%%%%%%%%%%%
\section{\label{effective}The effective weak current--current Hamiltonian}
%%%%%%%%%%%%%%%%%%%%%%%%%%%%%%%%%%%%%%%%%%%%%%%%%%%%%%%%%%%%%%%%%%%%%%%%%%%%%%
The $\Delta C=1$ effective Hamiltonian for the Cabibbo favored (CF) decays
is given by
\begin{equation}\label{CF}
  {\cal H}_{\rm eff}=\frac{G_F}{2\sqrt{2}} V_{cs}V^*_{ud}\left(c_+{\cal O}_+
  +c_-{\cal O}_-\right)+H.c.
\end{equation}
where the four-quark operators read
\begin{equation}
{\cal O}_\pm=(\bar s c)(\bar u d) \pm (\bar u c)(\bar s d)
\end{equation}
with $(\bar q_1 q_2)=\bar q_1\gamma_\mu(1-\gamma_5)q_2$.

The $\Delta C=1$ singly Cabibbo suppressed (SCS) decays are induced by two
distinct effective Hamiltonians which are labeled by $a$ and $b$. They read
\begin{eqnarray}
{\cal H}_{\rm eff}(a)&=&\frac{G_F}{2\sqrt{2}} V_{cs}V^*_{us}
  \left(c_+{\cal O}_+(a)+c_-{\cal O}_-(a)\right)+H.c.\label{SCSa}\\
{\cal H}_{\rm eff}(b)&=&\frac{G_F}{2\sqrt{2}} V_{cd}V^*_{ud}
  \left(c_+{\cal O}_+(b)+c_-{\cal O}_-(b)\right)+H.c.\label{SCSb}
\end{eqnarray}
with ${\cal O}_\pm(a)=(\bar s c)(\bar u s) \pm (\bar u c)(\bar s s)$ and
${\cal O}_\pm(b)=(\bar d c)(\bar u d) \pm (\bar u c)(\bar d d)$, respectively.
Using e.g.\ the Wolfenstein parametrization one finds that
$V_{cs}V^*_{us}=-V_{cd}V^*_{ud} + {\cal O}(\lambda^4)$ from unitarity, i.e.\
to a very good approximation one has $V_{cs}V^*_{us}=-V_{cd}V^*_{ud}$ which we
shall always use.  

The $\Delta C=1$ doubly Cabibbo suppressed decays (DCS) are induced by the
effective Hamiltonian
\begin{equation}\label{DCS}
{\cal H}_{\rm eff}(c)\ =\ \frac{G_F}{2\sqrt{2}}V_{cd}V^*_{us}
  \left(c_+{\cal O}_+(c)+c_-{\cal O}_-(c)\right)+H.c.
  \end{equation}
where ${\cal O}_\pm(c)=(\bar d c)(\bar u s) \pm (\bar u c)(\bar d s)$.

Finally, the $\Delta C=0$ charm baryon decays are induced by the two SCS
effective Hamiltonians
\begin{eqnarray}
{\cal H}_{\rm eff}(a')&=&\frac{G_F}{2\sqrt{2}} V_{us}V^*_{ud}
  \left(c_+{\cal O}_+(a')+c_-{\cal O}_-(a')\right)+H.c.\\
{\cal H}_{\rm eff}(b')&=&\frac{G_F}{2\sqrt{2}} V_{cd}V^*_{cs}
  \left(c_+{\cal O}_+(b')+c_-{\cal O}_-(b')\right)+H.c.
  \end{eqnarray}
with ${\cal O}_\pm(a')=(\bar u s)(\bar d u) \pm (\bar d s)(\bar u u)$ and
${\cal O}_\pm(b')=(\bar d c)(\bar c s) \pm (\bar c c)(\bar d s)$, respectively.
From unitarity one has $V_{cd}V^*_{cs}=-V_{us}V^*_{ud}$ which again is true
up to ${\cal O}(\lambda^4)$ in the Wolfenstein parametrization.

In this paper we will mostly be concerned with the so-called $W$-exchange
contributions which, according to the K\"orner--Pati--Woo (KPW)
theorem~\cite{Korner:1970xq,Pati:1970fg}, are related to the operators
${\cal O}_-=(\bar q_1q_2)(\bar q_3q_4)-(\bar q_3q_2)(\bar q_1q_4)$ only, i.e.\
induced by the effective Hamiltonian
\begin{equation}\label{Heffm}
{\cal H}_{\rm eff}({\cal O}_-)=\frac{G_F}{2\sqrt{2}}V^{\phantom*}_{q_2q_1}
  V^*_{q_3q_4}\left(c_-{\cal O}_-\right)+H.c.
\end{equation}

\begin{table}[ht]
\caption{\label{Tablemass}
  Ground state charm baryon states with $J^P=1/2^+$. The square and curly
  brackets $[ab]$ and $\{ab\}$
  denote antisymmetric and symmetric flavor label combinations.}
\begin{center}
\begin{tabular}{llclrrl}
\hline
\hline\\[-11pt]
Notation & Quark & SU(3) & $(I,I_3)$ & $S$ & $C$ &\quad Mass [MeV] \\
&content&&&&&\\[3pt] \hline \\[-11pt]
$\Lambda_{c}^+$ & $c[ud]$ & $\overline{\bf 3}$ & $(0,0)$ &
  $0$ & $1$ & \quad $2286.46\pm 0.14$  \\
$\Xi_{c}^+$& $c[su]$ & $\overline{\bf 3}$ &  $(1/2,1/2)$ &
  $-1$ & $1$ & \quad $2467.95\pm 0.19$ \\
$\Xi_{c}^0$ & $c[sd]$ & $\overline{\bf 3}$ & $(1/2,-1/2)$ &
  $-1$ & $1$ & \quad $2470.99\pm 0.40$ \\[3pt] \hline \\[-11pt] 
$\Sigma_{c}^{++}$ & $cuu$ & ${\bf 6}$ & $(1,1)$ & $0$ & $1$
&\quad $2453.97\pm 0.14$ \\
$\Sigma_{c}^{+}$ & $\{cud\}$ & ${\bf 6}$ & $(1,0)$ & $0$ & $1$
& \quad $2452.9\pm 0.4$ \\
$\Sigma_{c}^{0}$ & $cdd$ & ${\bf 6}$ & $(1,-1)$ & $0$ & $1$
&\quad $2453.75\pm 0.14$ \\
$\Xi_{c}^{\prime+}$ & $c\{su\}$ & ${\bf 6}$ & $(1/2,1/2)$ &
  $-1$ & $1$ & \quad $2578.4\pm 0.5$ \\
$\Xi_{c}^{\prime\,0}$ & $c\{sd\}$ & ${\bf 6}$ & $(1/2,-1/2)$ &
  $-1$ & $1$ & \quad $2579.2\pm 0.5$ \\
$\Omega_{c}^0$ & $css$ & ${\bf 6}$ & $(0,0)$ &
  $-2$ & $1$ &\quad $2695.2\pm 1.7$ \\[3pt]\hline \\[-11pt]
$\Xi_{cc}^{++}$ & $ccu$ & ${\bf 3}$ & $(1/2,1/2)$ &
  $0$ & $2$ & \quad $3621.2\pm 0.7$ \\
$\Xi_{cc}^+$ & $ccd$ & ${\bf 3}$ & $(1/2,-1/2)$ &
  $0$ & $2$ & \quad $3621$ \\
$\Omega_{cc}^+$ & $ccs$ & ${\bf 3}$ & $(0,0)$ &
  $-1$ & $2$ &\quad $3710$ \\[3pt]
\hline
\hline
\end{tabular}
\end{center}
\end{table}

%%%%%%%%%%%%%%%%%%%%%%%%%%%%%%%%%%%%%%%%%%%%%%%%%%%%%%%%%%%%%%%%%%%%%%%%%%%%%%
\section{\label{topotensor}SU(4), SU(3) and topological tensor invariants}
%%%%%%%%%%%%%%%%%%%%%%%%%%%%%%%%%%%%%%%%%%%%%%%%%%%%%%%%%%%%%%%%%%%%%%%%%%%%%%
The 20 ground state baryons with spin parity quantum numbers $J^P=1/2^+$ make
up the {\bf 20'} representation of SU(4) associated with the Young tableaux
\[\begin{Young}
  &\cr
  \cr
  \end{Young}.\]
In the $C=0$ sector one has the usual light baryon octet comprised of the
eight light baryons ($p$, $n$, $\Lambda^0$, $\Sigma^+$, $\Sigma^0$, $\Sigma^-$,
$\Xi^+$, $\Xi^0$). The $C=1$ single charm sector contains the antitriplet
charm baryons ($\Lambda_{c}^+$, $\Xi_{c}^+$, $\Xi_{c}^0$) and the sextet charm
baryons ($\Sigma_{c}^+$, $\Sigma_{c}^0$, $\Sigma_{c}^-$, $\Xi_{c}^{\prime+}$,
$\Xi_{c}^{\prime\,0}$, $\Omega_{c}^0$).

In HQET (or in the spectator quark model) the nine $C=1$ antitriplet and
sextet single charm baryons can be viewed as the 21 members of the ${\bf 21}$
representation of the approximate light spin--flavor symmetry group SU(6) with
the subgroup SU(2)$\times$SU(3). The decomposition reads 
\begin{equation}
  {\bf 21} \subset {\bf 1}\otimes\overline{\bf 3} \oplus {\bf 3}\otimes{\bf 6}
\end{equation}
where ${\bf 1}$ and ${\bf 3}$ denote the $J=0$ and $J=1$ representations of
spin SU(2) with dimensions $2J+1$. The SU(6) spin--flavor wave functions of
the antitriplet and sextet single charm baryons can be found e.g.\ in
Ref.~\cite{Zou:2019kzq}.

Finally, the $C=2$ double charm baryons are made up by the triplet of the
double heavy charm baryons ($\Xi_{cc}^{++}$, $\Xi_{cc}^{+}$, $\Omega_{cc}^+$).
In Table~\ref{Tablemass} we list the quark content, the quantum numbers and,
when available, the experimental mass values~\cite{Zyla:2020zbs} of the 12
charm baryon states. The masses of the double charm states $\Xi^+_{cc}$ and
$\Omega^+_{cc}$ have not been measured yet. For $\Xi^+_{cc}$ we assume
equality of the mass value with $\Xi^{++}_{cc}$, dropping, however, the fifth
digit and the error specification. For $\Omega^+_{cc}$ we list the mass
prediction of Ref.~\cite{Korner:1992wi,Korner:1994nh} based on the one-gluon
exchange model of de Rujula, Georgi and Glashow which features a Breit--Fermi
spin--spin interaction term~\cite{DeRujula:1975qlm}. In as much as the model
prediction (dated from 1992) for the mass of the $\Xi_{cc}$ iso-doublet of
$3621\MeV$ is very close to the measured value of
$3621.2\pm 0.7\MeV$~\cite{Zyla:2020zbs}, we feel quite confident about the
predicted mass of the $\Omega_{cc}^+$ listed in Table~\ref{Tablemass} based on
the same model calculation~\cite{Korner:1992wi,Korner:1994nh}. 

In the charm sector the single charm baryons $\Sigma_{c}^{++,+,0}$ decay
dominantly via one-pion emission while the dominant decay modes of the 
$\Xi_{c}^{\prime+}$ and $\Xi_{c}^{\prime\,0}$ are the one-photon emission
modes. The remaining seven single and double charm baryons
($\Lambda_{c}^+$, $\Xi_{c}^+$, $\Xi_{c}^0$, $\Omega_{c}^0$, $\Xi_{cc}^{++}$,
$\Xi_{cc}^{+}$, $\Omega_{cc}^+$) decay via weak interactions. An important
class of these weak charm baryon decays are their nonleptonic two-body decays
into a ground state baryon with $J^P=1/2^+$ and a pseudoscalar meson with
$J^P=0^-$. The two-body nonleptonic decays of the seven single and double
charm baryons ($\Lambda_{c}^+$, $\Xi_{c}^+$, $\Xi_{c}^0$, $\Omega_{c}^0$,
$\Xi_{cc}^{++}$, $\Xi_{cc}^{+}$, $\Omega_{cc}^+$) are the subject of this
paper.

In this paper we will not be concerned with the 20 $C=0,1,2,3$ $J^P=3/2^+$
ground state baryons which, in SU(4), belong to the ${\bf 20}$ representation
with the associated Young tableaux
\[\begin{Young}
  &&\cr 
\end{Young}\,\,.\]
The properties of the ${\bf 20}$ representation are discussed in
Ref.~\cite{Korner:1994nh}. In principle, the states of the ${\bf 20}$
representation could contribute as intermediate states in the current algebra
plus pole model approach discussed in Sec.~\ref{current} but are prevented
from contributing as intermediate states by the KPW
theorem~\cite{Korner:1970xq,Pati:1970fg} which states that the contraction of
the flavor antisymmetric current--current operator with a flavor symmetric
final state configuration is zero in the SU(3) limit.

It is quite illuminating to represent the baryon flavor wave functions as
third rank tensors $B_{abc}$ in flavor space instead of the second rank
tensors $B_a^b$ frequently used in the literature. In this way one can keep
track of the quark content of the baryons. Moreover, in the transitions
involving baryons one can identify the flavor flow in the transitions which
directly leads to the concept of topological diagrams and the topological
tensor invariants associated with them. Explicit representations of the third
rank flavor tensors of the twenty $J^P=1/2^+$ ground state baryons are given
in Appendix~\ref{tensors}. For the ground state baryons and mesons we have
used the phase conventions of Lichtenberg~\cite{Lichtenberg:1978pc}.

The flavor wave functions of the $J^P=1/2^+$ ground state baryons satisfy a
Jacobi-type identity which reads
\begin{equation}\label{jacobi}
B^\ell_{a[bc]}+B^\ell_{b[ca]}+B^\ell_{c[ab]}=0 \, .
\end{equation}
for each of the flavor wave functions of the ground state multiplet, where
$\ell=1,\ldots,8$ in SU(3) and $\ell=1,\ldots,20$ in SU(4). They are
normalized and orthogonal according to
\begin{equation}\label{orthogonality}
\sum_{k,m,n}B_\ell^{k[mn]}B^{\ell'}_{k[mn]} = \delta^{\ell'}_\ell.
\end{equation}
Furthermore, they satisfy the completeness relation
\begin{equation}
\label{completeness}
\sum_\ell B^\ell_{k[mn]}B_\ell^{b[cd]}
=\frac26(\delta_k^b \delta_m^c \delta_n^d- \delta_k^b \delta_m^d \delta_n^c)
-\frac16(\delta_m^b \delta_n^c \delta_k^d -\delta_m^b \delta_n^d \delta_k^c) 
-\frac16(\delta_n^b \delta_k^c \delta_m^d -\delta_n^b \delta_k^d \delta_m^c).
\end{equation}
Contracting the completeness relation~(\ref{completeness}) with
$\delta^k_b\,\delta^m_{c}\,\delta^n_d$ one obtains the dimension $d=N(N^2-1)/3$
of the ground state representation in SU($N$) which agrees with the dimension
of the representation calculated according to the hook rule with the result
\begin{equation}
  {\rm dim}\left(\kern-9pt\raise-12pt
  \hbox{\begin{Young}
  &\cr
  \cr
  \end{Young}}\right)= \frac{N(N+1)(N-1)}{3\cdot1\cdot1}=\frac13 N(N^2-1).
\end{equation}
One thus obtains $d=8$ and $d=20$ for SU(3) and SU(4), respectively, as
expected. 

The completeness relation~(\ref{completeness}) is central to the chain of
reasoning in Sec.~\ref{current} which, in the limit of SU(3) and the absence
of hyperfine interactions, links the results of the current algebra approach
to a linear superposition of topological reduced matrix elements with
coefficients given by the topological tensor invariants. For the interested
reader we provide an explicit proof of the completeness relation in
Appendix~\ref{complete}. Using the completeness relation we never invoke SU(4)
symmetry but rather make use of the SU(3) symmetry of the separate $C=0$,
$C=1$ and $C=2$ sectors of the completeness relation~(\ref{completeness}). For
a given flavor configuration defined by the in and out states, the sum over
intermediate states $\ell$ in Eq.~(\ref{completeness}) extends either over one
of the ground state baryons or, in the case of the flavor degenerate pairs of
states $(\Lambda^0,\Sigma^0)$, $(\Lambda_{c}^+,\Sigma_{c}^+)$,
$(\Xi_{c}^0,\Xi^{\prime\,0}_{c})$ and $(\Xi_{c}^+,\Xi^{\prime+}_{c})$, over
maximally two ground state baryons. In most of the cases analyzed in this
paper, due to symmetry and constituent quark model considerations only one
state of the flavor degenerate pair of states contribute as intermediate state.
When both flavor degenerate states contribute to a given intermediate state we
assume mass degeneracy of the two states to sum the two contributions. More
details can be found in Sec.~\ref{current}.

In flavor space the ground state to ground state transitions
$1/2^+\rightarrow1/2^++0^-(1^-)$ induced by the antisymmetric flavor-changing
tensor $H^{[q_2q_4]}_{[q_1q_3]}$ representing the effective Hamiltonian
${\cal H}_{\rm eff}({\cal O}_-)$ in Eq.~(\ref{Heffm}) are represented by the
seven topological tensor invariants
\begin{eqnarray}\label{tinv}
I^-_1(\ell,\ell')=B_\ell^{a[bc]}B^{\ell'}_{a[bc']}M^d_{d'}H^{[c'd']}_{[cd]}&&
I^-_2(\ell,\ell')=B_\ell^{a[bc]}B^{\ell'}_{b[c'a]}M^d_{d'}H^{[c'd']}_{[cd]}
  \nonumber\\
I_3(\ell,\ell')=B_\ell^{a[bc]}B^{\ell'}_{a[b'c']}M^d_{c}H^{[c'b']}_{[db]}&&
I_4(\ell,\ell')=B_\ell^{b[ca]}B^{\ell'}_{a[b'c']}M^d_{c}H^{[c'b']}_{[db]}
  \nonumber\\
\hat I_3(\ell,\ell')=B_\ell^{a[bc]}B^{\ell'}_{a[b'c']}M^{c'}_dH^{[db']}_{[cb]}&&
\hat I_4(\ell,\ell')=B_\ell^{a[bc]}B^{\ell'}_{b'[c'a]}M^{c'}_dH^{[db']}_{[cb]}
  \nonumber\\
I_5(\ell,\ell')=B_\ell^{a[bc]}B^{\ell'}_{a'[b'c']}M^{c'}_{c}H^{[a'b']}_{[ab]}&&
\end{eqnarray}
for the baryonic transitions $\ell'\to\ell$, where the fourth rank flavor
changing tensor $H^{[q_2q_4]}_{[q_1q_3]}$ takes the values $\pm 1$ as
specified in Appendix~\ref{tensors}. A summation over doubly occurring indices
in Eq.~(\ref{tinv}) is implied. In the present application in which we
consider various SU(3) subsector transitions, the summation runs over the
three light quarks ($u=1,d=2,s=3$) and the charm quark which we label as
$c=4$. After having used the Jacobi identity~(\ref{jacobi}) and/or by
rearranging tensor labels including the exchange of dummy summation indices,
the set of seven tensor invariants comprise the minimal set of possible tensor
contractions. For example, the tensor invariant
$I_5'(\ell,\ell')=B_\ell^{a[bc]}B^{\ell'}_{b'[c'a']}M^{c'}_{c}H^{[a'b']}_{[ab]}$
is redundant since it can be seen to be equal to $I_5(\ell',\ell)$ by simply
rearranging the tensor labels. The list~(\ref{tinv}) contains only connected
tensor contractions. We have thus not included the tadpole-type unconnected
tensor contraction $B_\ell^{a[bc]}B^{\ell'}_{a[bc]}M^d_{d'}H^{[cd']}_{[cd]}$
which vanishes for the $\Delta C=1$ transitions since
$H^{[cd']}_{[cd]}=0$ (see Appendix~\ref{tensors}). 

\begin{table}[ht]
\caption{\label{Heff}SU(3) and SU(2) properties of the effective $\Delta C=1$
  and $\Delta C=0$ Hamiltonian ${\cal H}_{\rm eff}({\cal O}_-)$ in
  Eq.~(\ref{Heffm}), represented in flavor space by the fourth rank tensor
  $H^{[q_2q_4]}_{[q_1q_3]}$}
\begin{center}
\begin{tabular}{ccccccccc}
\hline \hline \\[-11pt]
   && $Y$ & $I_3$ & $SU(3)$ & $\Delta I$
 & $\Delta U$ & $\Delta U_3$ & $\Delta V$ \\[3pt] \hline \\[-11pt]
CF & $H^{[su]}_{[cd]}$ &
$ 2/3 $&$ -1 $&$ {\bf 6} $&$ 1 $&$ 1 $&$ 1 $& $0$ \\ 
SCS & $H^{[su]}_{[cs]}$&
$ -1/3 $&$ -1/2 $&$  {\bf 6}\oplus\overline{\bf 3} $
&$ 1/2 $&$ 1,0 $&$ 0,\,0 $&$1/2$\\
&$H^{[du]}_{[cd]}$
&$ -1/3 $&$ -1/2 $&$ {\bf 6}\oplus\overline{\bf 3} $
&$ 1/2 $&$ 1,0 $&$ 0, 0 $&$1/2$\\ 
&$H^{[su]}_{[cs]}-H^{[du]}_{[cd]}$&
$  -1/3 $&$  -1/2 $&$ {\bf 6} $&$ 1/2 $&$  1 $&$  0 $&$ 1/2 $ \\ 
%% decay 5
DCS &  $H^{[du]}_{[cs]}$ &
$ -4/3 $&$ 0 $&$ {\bf 6}  $&$  0 $&$ 1 $&$  -1 $ & $1$\\ 
%% decay 6
$\Delta  C=0 $& $H^{[ud]}_{[su]}$&
$ -1 $&$ 1/2 $&$ {\bf 8} $ & $  1/2 $&$ 1$ & $ -1 $ & $1/2$\\ 
%% decay 7
& $H^{[dc]}_{[cs]}$ &
$ -1 $&$ 1/2 $&$ {\bf 8} $&$  1/2 $&$ 1 $&$ -1 $&$1/2$\\[3pt]
\hline
\hline
\end{tabular}
\end{center}
\end{table}

\begin{figure}\begin{center}
  \epsfig{figure=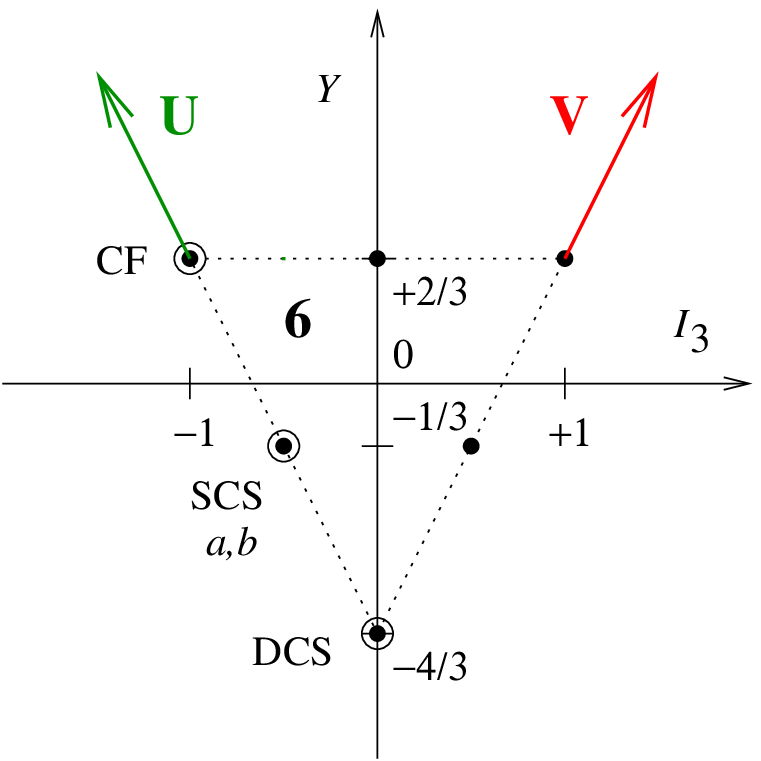, scale=0.90}\qquad
  \raise48pt\hbox{\epsfig{figure=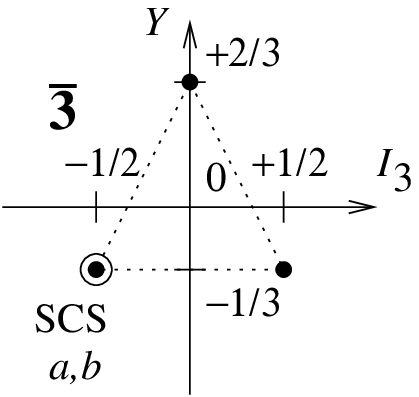, scale=0.90}}\end{center}
\caption{\label{tripsex}Weight diagrams of the sextet (left) and antitriplet
  (right) representation of the effective weak Hamiltonian. The locations of
  the CF, SCS and DCS transitions are marked by a circle dot symbol $\odot$.} 
\end{figure}

In Table~\ref{Heff} the SU(3) and SU(2) properties of the fourth rank flavor
changing tensor are listed. Note that each of the tensors $H^{[su]}_{[cs]}$
and $H^{[du]}_{[cd]}$ inducing the SCS transition transforms separately as
${\bf 6}\oplus\overline{\bf 3}$ in SU(3). Only their difference
$H^{[su]}_{[cs]}-H^{[du]}_{[cd]}$ transforms as the ${\bf 6}$ representation
in SU(3) as can be explicitly verified by using the Clebsch--Gordan tables of
Kaeding~\cite{Kaeding:1995vq}. In Fig.~\ref{tripsex} we display the weight
diagrams of the ${\bf 6}$ and $\overline{\bf 3}$ representations where we have
marked the locations of the CF, SCSa, SCSb and DCS effective quark transitions
in the two weight diagrams. When taking the difference of the two SCS
contributions one remains with a $U$-spin triplet at the left boundary of the
${\bf 6}$ weight diagram. We have somewhat dwelt on this point because we have
not found an adequate discussion concerning this cancellation in the literature.

We mention that the reduction to a minimal set of tensor invariants has not
been attained in Refs.~\cite{Kohara:1991ug,He:2018joe}. As an example we take
the Cabibbo enhanced decays of the antitriplet charm baryons. The authors of
Ref.~\cite{He:2018joe} introduce six and three topological tensor invariants
each for the topology classes IIa and III where only two and one are needed
($I_3$, $I_4$ and $I_5$, resp.) while Kohara~\cite{Kohara:1991ug} defines four
and one topological tensor invariants for the same topology classes IIa and
III where, as before, only the two invariants $I_3$, $I_4$ and the invariant
$I_5$ are needed. This makes it quite cumbersome to compare mutual results.
Note that the distinction of whether the strange quark from the charm quark
transition ends up in the meson or the baryon for the topology class IIa, as
done in Ref.~\cite{He:2018joe}, is not warranted in SU(3).

Let us briefly pause to discuss the charge conjugation properties of the seven
tensor invariants~(\ref{tinv}). In flavor space the charge conjugation
operation raises or lowers the tensor indices, i.e.\ one has
$\bar B_\ell^{a[bc]}=B^\ell_{a[bc]}$, $\bar B^\ell_{a[bc]}=B_\ell^{a[bc]}$,
$\bar M^b_a=M^a_b$, and $\bar H^{[cd]}_{[ab]}=H^{[ab]}_{[cd]}$. As an example
we consider the charge conjugation operation on the tensor invariant $I_3$.
One has
\begin{equation}
\bar I_3(\ell,\ell')=B^\ell_{a[bc]}B_{\ell'}^{a[b'c']}M^c_dH^{[db]}_{[c'b']}
  =\hat I_3(\ell',\ell).
\end{equation}
Similarly one has $\bar I_4(\ell,\ell')=\hat I_4(\ell',\ell)$,
$\bar I^-_{1,2}(\ell,\ell')=I^{-}_{1,2}(\ell',\ell)$, and
$\bar I_5(\ell,\ell')=I_5(\ell',\ell)$. Therefore, there are two antisymmetric
and five symmetric tensor combinations under the flavor charge conjugation
operation. The two antisymmetric combinations are
\begin{equation}
(\bar I_3-\bar{\hat I}_3)(\ell,\ell')=-(I_3-\hat I_3)(\ell',\ell),\qquad
(\bar I_4-\bar{\hat I}_4)(\ell,\ell')=-(I_4-\hat I_4)(\ell',\ell).
\end{equation}
They would contribute to the parity violating (p.v.)\ amplitude $A$ of
Sec.~\ref{current}. The five symmetric combinations read
\begin{eqnarray}
&&(\bar I_3+\bar{\hat I}_3)(\ell,\ell')=+(I_3+\hat I_3)(\ell',\ell)\qquad\quad
(\bar I_4+\bar{\hat I}_4)(\ell,\ell')=+(I_4+\hat I_4)(\ell',\ell)\nonumber\\
&&\quad
\bar I^-_1(\ell,\ell')=I^-_1(\ell',\ell)\qquad
\bar I^-_2(\ell,\ell')=I^-_2(\ell',\ell)\qquad
\bar I_5(\ell,\ell')=I_5(\ell',\ell).
\end{eqnarray}
They would contribute to the parity conserving (p.c.)\ amplitude $B$ of
Sec.~\ref{current}. Note, however, that the dynamics of the tree diagram
contributions leads to an extra factor of $(m_1-m_2)$ in the case of the p.v.\
amplitude $A$ which implies that the tree diagram invariants also contribute
to the amplitude $A$.

Returning to the topological analysis one can associate each of the tensor
invariants~(\ref{tinv}) with one of the topological diagrams in Fig.~\ref{topo} 
by following the flavor flow in the diagrams. It is important to realize that
it is not sufficient to just associate a given decay with a given set of
topological diagrams. Instead, one needs to calculate the projection of that
given decay onto the topological diagrams in terms of the topological tensor
invariants. The association reads
\begin{eqnarray}
&&(I^-_1,I^-_2)\,\,\longleftrightarrow\,\,\mbox{diagrams Ia, Ib} \qquad
\qquad(I_3,I_4)\,\,\longleftrightarrow\,\,\mbox{diagram IIa} \\
&&(\hat I_3,\hat I_4)\,\,\longleftrightarrow\,\,\mbox{diagram IIb} \qquad
\qquad\qquad I_5\,\,\longleftrightarrow\,\,\mbox{diagram III}
\end{eqnarray}
Note that this relation is not one-to-one. For example, from the flavor flow
the decay $\Lambda_{c}^+ \to \Xi^0 K^+$ can be seen to be contributed to by the
topology IIa. However, from an explicit calculation one finds that only the
topological tensor invariant $I_4$ becomes populated. This has important
ramifications for the structure of the decay as will be discussed in
Sec.~\ref{tables}.

To conclude, in flavor space a general amplitude ${\cal A}_{fki}$ describing
the decay $B_i\to B_f+M_k$ can be expanded along seven topological reduced
amplitudes $\mathbfcal{T}_j$ with coefficients given by the seven topological
tensor invariants $I^-_1,\ldots,I_5$. One has
\begin{equation}\label{expansion}
{\cal A}(B_i \stackrel{H}{\longrightarrow} B_fM_k)={\cal A}_{fki}
  =\sum_j I^j_{fki} \,\mathbfcal{T}_j \qquad
  j=1^-,2^-,\,3,\,4,\, \hat3,\,\hat4,\,5
\end{equation}
where the $B_i$ and $B_f$ belong to some given SU(3) representation. The
$\mathbfcal{T}_j$ are the topological invariant amplitudes of the process
while the tensor invariants $I^j_{fki}$ project onto these invariant
amplitudes. Put in a different language, the topological tensor invariants
$I^j_{fki}$ would correspond to SU(3) Clebsch--Gordan coefficients while the
reduced matrix elements $\mathbfcal{T}_j$ would correspond to the SU(3)
invariant amplitudes. All SU(3) relations between given transition amplitudes
including the subclasses of $I$, $U$ and $V$ sum rules are implicit in the
expansion~(\ref{expansion}).

In the general case one is overcounting the number of tensor invariants in the
expansion~(\ref{expansion}), i.e.\ the number of significant SU(3) invariants
can generally be less than the number of seven SU(3) invariants in
Eq.~(\ref{expansion}). In general, the rank of the coefficient matrix
$I^j_{fki}$ linking ${\cal A}_{fki}$ with $\mathbfcal{T}_j$ is less than seven.
This implies that one cannot, in general, determine the reduced topological
amplitudes $\mathbfcal{T}_j$ from a given set of experimentally measured
amplitudes ${\cal A}_{fki}$. In such a case there will be a number of linear
relations among the seven tensor invariants $I^j_{fki}$ which will be commented
on and written down for the different SU(3) transitions treated in this paper.
Some of the linear relations involve tensor invariants of the same topology
class. In such a case the linear relations can be obtained by combining the
information on the symmetry or antisymmetry of the ($u,d,s$) light quark
components of a given class of charm baryons with the use of the Jacobi
identity. The remaining linear relations between tensor invariants involving
different topology classes cannot be obtained by mere tensor manipulations. We
conjecture that one needs Schouten type identities to derive these additional
tensor identities. In this paper we use linear algebra methods to construct
these additional linear relations explicitly. With the help of the linear
relations between tensor invariants one can recombine the topological
invariant amplitudes to a minimal set of significant (nontrivial) topological
invariant amplitudes. However, it should be clear that this set of minimal
reduced amplitudes is not unique.

The flavor space tensor contractions represent a convenient way of calculating
Clebsch--Gordon coefficients entering in nonleptonic charm baryon decays. They
were first introduced in Ref.~\cite{Korner:1978tc} which also contains some
supplementary background material. The values of the topological tensor
invariants for all Cabibbo favored (CF), singly Cabibbo suppressed (SCS) and
doubly Cabibbo suppressed (DCS) nonleptonic charm baryon decays needed in this
paper are for instance given in Tables~\ref{TableCF}, \ref{TableSCSa},
\ref{TableSCSb}, \ref{TableSCS} and~\ref{TableDCS}.

The transition to the isoscalar states are written in terms of the ideally
mixed states $\eta_\omega=\frac{1}{\sqrt{2}}(u\bar u+d\bar d)$ and
$\eta_\phi=s\bar s$. For each decay we have factored out the products of the
denominator factors appearing in the flavor space wave functions, the inverse
of which appear as factors multiplying the amplitude of the respective decays.

%%%%%%%%%%%%%%%%%%%%%%%%%%%%%%%%%%%%%%%%%%%%%%%%%%%%%%%%%%%%%%%%%%%%%%%%%%%%%%%
\section{\label{tables}Tables of the topological tensor invariants}
%%%%%%%%%%%%%%%%%%%%%%%%%%%%%%%%%%%%%%%%%%%%%%%%%%%%%%%%%%%%%%%%%%%%%%%%%%%%%%%
In this section we proceed to discuss the various classes of charm baryon
decays. They are classified according to the SU(3) transitions that specify
the decays. Even though we do not discuss the factorizable contributions, we
also list the values of the tensor invariants $I^-_1$ and $I^-_2$ since they
enter in the linear relations between the seven ${\cal O}^-$ induced tensor
invariants. The corresponding tensor invariants $I^+_1$ and $I^+_2$ induced by
the operator ${\cal O}^+$ can be easily obtained from $I^-_1$ and $I^-_2$ as
described in Appendix~\ref{tensors}.

For the transitions involving the isospin zero neutral meson states we
separately list results for the octet state
$\eta_8=(u \bar u+d \bar d-2 s \bar s)/\sqrt{6}$ and the singlet state
$\eta_1=(u \bar u+d \bar d+ s \bar s)/\sqrt{3}$. The physical $\eta$ and
$\eta'$ states are linear superpositions of these two states according to
\begin{equation}
  \eta=\cos\theta \eta_8 -\sin\theta \eta_1 \qquad
  \eta'=\sin\theta \eta_8 +\cos\theta \eta_1
\end{equation}
with $\theta=-15.4^\circ$~\cite{Feldmann:1998vh,Feldmann:1998sh}. We also list
the transitions into the ideally mixed states
$\eta_\omega=(u\bar u+d\bar d)/\sqrt2$ and $\eta_\phi=s\bar s$ with obvious
applications to the corresponding $1/2^+ \to 1/2^+\,+1^-$ transitions
involving the ideally mixed vector states $\omega$ and $\phi$.

\begin{table}
\caption{\label{TableCF}Values of the seven topological tensor invariants
  for the antitriplet charm baryon decays
  $B_{c}(\overline{\bf 3}) \to B({\bf 8})+M({\bf 8},{\bf 1})$ induced by the CF
  flavor transitions $(c\to s;\,d\to u)$. We have always factored out the
  products of the normalizing denominator factors appearing in the flavor
  space quark model wave functions. The product of the denominator factors
  appear as overall factors of the process specification in column 2.} 
\begin{center}
\begin{tabular}{crclrrrrrrr}
  \hline \hline \\ [-11pt]
  & & & & $I_1^{-}$ & $I^-_2$ & $I_3$ & $I_4$ &
  $\hat{I}_3$ & $\hat{I}_4$ & $I_5$ \\[3pt] \hline \\ [-11pt]
%%% decay 1
CF&$12 \,\Lambda_{c}^+ $ & $\to$ & $ \Lambda^0 \pi^+$
&$ -2 $&$ -2 $&$ -2 $&$ +4 $&$ -2 $&$ +4 $&$ +1 $\\
%%% decay 2
&$4\sqrt{3} \,\Lambda_{c}^+ $ & $\to$ & $ \Sigma^0 \pi^+$
&$  0 $&$  0 $&$ +2 $&$  0 $&$ -2 $&$ +4 $&$ +1 $\\
%%% decay 3
&$4\sqrt{3} \, \Lambda_{c}^+  $ & $\to$ & $ \Sigma^+\pi^0$
&$  0 $&$  0 $&$ -2 $&$  0 $&$ +2 $&$ -4 $&$ -1 $\\
%%% decay 4
&$4\sqrt{3} \,\Lambda_{c}^+ $ & $\to$ & $ \Sigma^+\eta_\omega$
&$  0 $&$  0 $&$ -2 $&$  0 $&$ -2 $&$ +4 $&$ -1 $\\
%%% decay 5
&$2\sqrt{6}\Lambda_{c}^+ $ & $\to$ & $ \Sigma^+ \eta_\phi$
&$  0 $&$  0 $&$ -2 $&$ +2 $&$  0 $&$  0 $&$  0 $\\
%%% decay 6
&$ 12 \Lambda_{c}^+ $ & $\to$ & $ \Sigma^+ \eta_8$
&$  0 $&$  0 $&$ +2 $&$ -4 $&$ -2 $&$ +4 $&$ -1 $\\
%%% decay 7
&$ 6\sqrt{2} \Lambda_{c}^+ $ & $\to$ & $ \Sigma^+ \eta_1$
&$  0 $&$  0 $&$ -4 $&$ +2 $&$ -2 $&$ +4 $&$ -1 $\\
%%% decay 8
&$2\sqrt{6} \,\Lambda_{c}^+$ & $\to$ & $ p \bar{K}^0$
&$ +1 $&$ +1 $&$ +2 $&$ -2 $&$  0 $&$  0 $&$  0 $\\
%%% decay 9
&$2\sqrt{6} \,\Lambda_{c}^+$ & $\to$ & $ \Xi^0 K^+$
&$  0 $&$  0 $&$  0 $&$ -2 $&$  0 $&$  0 $&$ -1 $\\[3pt] \hline\\[-11pt]
%%% decay 10
&$2\sqrt{6} \,\Xi_{c}^+$ & $\to$ & $ \Sigma^+\bar{K}^0$
&$ +1 $&$ +1 $&$  0 $&$  0 $&$ +2 $&$ -4 $&$  0 $\\
%%% decay 11
&$2\sqrt{6} \,\Xi_{c}^+$ & $\to$ & $ \Xi^0\pi^+$
&$ -1 $&$ -1 $&$  0 $&$  0 $&$ -2 $&$ +4 $&$  0 $\\[3pt] \hline\\[-11pt]
%%% decay 12
&$12 \,\Xi_{c}^0$ & $\to$ & $ \Lambda^0 \bar{K}^0$
&$ -1 $&$ -1 $&$ -4 $&$ +2 $&$ +2 $&$ -4 $&$ -1 $\\
%%% decay 13
&$4\sqrt{3} \,\Xi_{c}^0$ & $\to$ & $ \Sigma^0\bar{K}^0$
&$ +1 $&$ +1 $&$  0 $&$ -2 $&$ +2 $&$ -4 $&$ -1 $\\
%%% decay 14
&$2\sqrt{6} \,\Xi_{c}^0$ & $\to$ & $ \Sigma^+K^-$
&$  0 $&$  0 $&$  0 $&$ +2 $&$  0 $&$  0 $&$ +1 $\\
%%% decay 15
&$4\sqrt{3} \,\Xi_{c}^0$ & $\to$ & $ \Xi^0\pi^0$
&$  0 $&$  0 $&$ +2 $&$ -2 $&$ -2 $&$ +4 $&$  0 $\\
%%% decay 16
&$4\sqrt{3} \,\Xi_{c}^0$ & $\to$ & $ \Xi^0\eta_\omega$
&$  0 $&$  0 $&$ +2 $&$ -2 $&$ +2 $&$ -4 $&$  0 $\\
%%% decay 17
&$2\sqrt{6} \,\Xi_{c}^0$ & $\to$ & $ \Xi^0\eta_\phi$
&$  0 $&$  0 $&$ +2 $&$  0 $&$  0 $&$  0 $&$ +1 $\\
%%% decay 18
&$      12  \,\Xi_{c}^0$ & $\to$ & $ \Xi^0\eta_8$
&$  0 $&$  0 $&$ -2 $&$ -2 $&$ +2 $&$ -4 $&$ -2 $\\
%%% decay 19
&$ 6\sqrt{2}  \,\Xi_{c}^0$ & $\to$ & $ \Xi^0\eta_1$
&$  0 $&$  0 $&$ +4 $&$ -2 $&$ +2 $&$ -4 $&$ +1 $\\
%%% decay 20
&$2\sqrt{6} \,\Xi_{c}^0$ & $\to$ & $ \Xi^-\pi^+$
&$ -1 $&$ -1 $&$ -2 $&$ +2 $&$  0 $&$  0 $&$  0 $\\[3pt]
\hline \hline \\[-11pt]
\end{tabular}
\end{center}
\end{table}

\begin{table}
\caption{\label{TableSCSa} Values of the seven topological tensor invariants
  for the SCS antitriplet charm baryon decays
  $B_{c}(\overline{\bf 3}) \to B({\bf 8})+M({\bf 8},{\bf 1})$
  induced by the flavor transition $(c \to s;\,s \to u)$.}
\begin{center}
\begin{tabular}{crclrrrrrrr}
\hline \hline \\[-11pt]\phantom{$\hat{\hat{I}_3}$}& & & &
$I^-_1$ & $I^-_2$ & $I_3$ & $I_4$ & $\hat{I_3}$ & $\hat{I}_4$ & $I_5 $\\[3pt]
\hline \\[-11pt]
%%% decay 1
SCSa &$ 12\, \Lambda^+_{c} $ & $\to$ & $ \Lambda^0  K^+$
&$ +2 $&$ +2 $&$  0 $&$  0 $&$ +2 $&$ -4 $&$  0 $\\
%%% decay 2
&$ 4 {\sqrt{3}}\, \Lambda^+_{c}  $ & $\to$ & $\Sigma^0  K^+$
&$  0 $&$  0 $&$  0 $&$  0 $&$ +2 $&$ -4 $&$  0 $\\
%%% decay 3
&$ 2 \sqrt{6}\, \Lambda^+_{c}  $ & $\to$ & $\Sigma^+ K^0 $
&$  0 $&$  0 $&$  0 $&$  0 $&$ -2 $&$ +4 $&$  0 $\\
%%% decay 4
&$2\sqrt{6}\, \Lambda^+_{c}  $ & $\to$ & $n  \pi^+ $
&$  0 $&$  0 $&$  0 $&$  0 $&$  0 $&$  0 $&$  0 $\\
%%% decay 5
&$ 4 {\sqrt{3}}\, \Lambda^+_{c}  $ & $\to$ & $p  \pi^0 $
&$  0 $&$  0 $&$  0 $&$  0 $&$  0 $&$  0 $&$  0 $\\
%%% decay 6
&$ 4 {\sqrt{3}}\,\Lambda^+_{c}  $ & $  \to $ & $p  \eta_\omega$
&$  0 $&$  0 $&$  0 $&$  0 $&$  0 $&$  0 $&$  0 $\\
%%% decay 7
&$  2\sqrt{6}\, \Lambda^+_{c}  $ & $\to$ & $p  \eta_\phi$
&$ -1 $&$ -1 $&$  0 $&$  0 $&$  0 $&$  0 $&$  0 $\\
%%% decay 8
&$  12\, \Lambda^+_{c}  $ & $\to$ & $p  \eta_8$
&$ +2 $&$ +2 $&$  0 $&$  0 $&$  0 $&$  0 $&$  0 $\\
%%% decay 9
&$  6\sqrt{2}\, \Lambda^+_{c}  $ & $\to$ & $p  \eta_1$
&$ -1 $&$ -1 $&$  0 $&$  0 $&$  0 $&$  0 $&$  0 $\\[3pt] \hline \\[-11pt]
%%% decay 10
&$  2\sqrt{6}\, \Xi^+_{c}  $ & $\to$ & $\Xi^0   K^+$
&$ +1 $&$ +1 $&$  0 $&$ -2 $&$ +2 $&$ -4 $&$ -1 $\\
%%% decay 11
&$ 4 {\sqrt{3}}\, \Xi^+_{c}  $ & $\to$ & $\Sigma^+  \pi^0 $
&$  0 $&$  0 $&$ -2 $&$  0 $&$  0 $&$  0 $&$ -1 $\\
%%% decay 12
&$4{\sqrt{3}}\, \Xi^+_{c} $ &$ \to$ & $\Sigma^+  \eta_\omega $
&$  0 $&$  0 $&$ -2 $&$  0 $&$  0 $&$  0 $&$ -1 $\\
%%% decay 13
&$ 2{\sqrt{6}} \, \Xi^+_{c}  $ & $\to$ & $\Sigma^+  \eta_\phi $
&$ -1 $&$ -1 $&$ -2 $&$ +2 $&$ -2 $&$ +4 $&$  0 $\\
%%% decay 14
&$ 12\, \Xi^+_{c}  $ & $\to$ & $\Sigma^+  \eta_8 $
&$ +2 $&$ +2 $&$ +2 $&$ -4 $&$ +4 $&$ -8 $&$ -1 $\\
%%% decay 15
&$ 6{\sqrt{2}}\, \Xi^+_{c}  $ & $\to$ & $\Sigma^+  \eta_1 $
&$ -1 $&$ -1 $&$ -4 $&$ +2 $&$ -2 $&$ +4 $&$ -1 $\\
%%% decay 16
&$2\sqrt{6}\,\Xi^+_{c}$ & $\to$ & $p  \bar{K}^0 $
&$  0 $&$  0 $&$ +2 $&$ -2 $&$  0 $&$  0 $&$  0 $\\
%%% decay 17
&$12\,\Xi^+_{c}  $ & $\to$ & $\Lambda^0 \pi^+ $
&$  0 $&$  0 $&$ -2 $&$ +4 $&$  0 $&$  0 $&$ +1 $\\
%%% decay 18
&$ 4 {\sqrt{3}}\, \Xi^+_{c}  $ & $\to$ & $\Sigma^0 \pi^+ $
&$  0 $&$  0 $&$ +2 $&$  0 $&$  0 $&$  0 $&$ +1 $\\[3pt] \hline \\[-11pt]
%%% decay 19
&$2\sqrt{6}\,\Xi^0_{c} $ & $\to$ & $\Xi^0  {K}^0 $
&$  0 $&$  0 $&$  0 $&$  0 $&$ +2 $&$ -4 $&$ -1 $\\
%%% decay 20
&$2\sqrt{6}\, \Xi^0_{c} $ & $\to$ & $ \Xi^- K^+ $
&$ +1 $&$ +1 $&$  0 $&$ -2 $&$  0 $&$  0 $&$  0 $\\
%%% decay 21
&$2\sqrt{6}\, \Xi^0_{c} $ & $\to$ & $\Sigma^+ \pi^-$
&$  0 $&$  0 $&$  0 $&$  0 $&$  0 $&$  0 $&$ -1 $\\
%%% decay 22
&$2\sqrt{6}\,\Xi^0_{c} $ &$ \to$ &$\Sigma^- \pi^+ $
&$  0 $&$  0 $&$ +2 $&$  0 $&$  0 $&$  0 $&$  0 $\\
%%% decay 23
&$4{\sqrt{6}}\, \Xi^0_{c} $ & $\to$  & $ \Sigma^0 \pi^0 $
&$  0 $&$  0 $&$ -2 $&$  0 $&$  0 $&$  0 $&$ +1 $\\
%%% decay 24
&$ 4 {\sqrt{6}}\, \Xi^0_{c} $ & $\to$ & $ \Sigma^0   \eta_\omega $
&$  0 $&$  0 $&$ -2 $&$  0 $&$  0 $&$  0 $&$ -1 $\\
%%% decay 25
&$ 4 {\sqrt{3}}\, \Xi^0_{c} $ & $\to$ & $ \Sigma^0   \eta_\phi $
&$ -1 $&$ -1 $&$ -2 $&$ +2 $&$ -2 $&$ +4 $&$  0 $\\
%%% decay 26
&$ 12 {\sqrt{2}}\, \Xi^0_{c} $ & $\to$ & $ \Sigma^0   \eta_8 $
&$ +2 $&$ +2 $&$ +2 $&$ -4 $&$ +4 $&$ -8 $&$ -1 $\\
%%% decay 27
&$ 12\, \Xi^0_{c} $ & $\to$ & $ \Sigma^0   \eta_1 $
&$ -1 $&$ -1 $&$ -4 $&$ +2 $&$ -2 $&$ +4 $&$ -1 $\\
%%% decay 28
&$ 12 {\sqrt{2}}\, \Xi^0_{c} $ & $\to$ & $ \Lambda^0  \pi^0 $
&$  0 $&$  0 $&$ -2 $&$ +4 $&$  0 $&$  0 $&$ +1 $\\
%%% decay 29
&$12{\sqrt{2}} \,\Xi^0_{c} $ & $\to$ & $ \Lambda^0  \eta_\omega $
&$  0 $&$  0 $&$ -2 $&$ +4 $&$  0 $&$  0 $&$ -1 $\\
%%% decay 30
&$12\, \Xi^0_{c} $ & $\to$ & $ \Lambda^0  \eta_\phi $
&$ +1 $&$ +1 $&$ -2 $&$ -2 $&$ -2 $&$ +4 $&$  0 $\\
%%% decay 31
&$12\sqrt{6}\, \Xi^0_{c} $ & $\to$ & $ \Lambda^0  \eta_8 $
&$ -2 $&$ -2 $&$ +2 $&$ +8 $&$ +4 $&$ -8 $&$ -1 $\\
%%% decay 32
&$12\sqrt{3}\,\Xi^0_{c} $ & $\to$ & $ \Lambda^0 \eta_1 $
&$ +1 $&$ +1 $&$ -4 $&$ +2 $&$ -2 $&$ +4 $&$ -1 $\\
%%% decay 33
&$ 2\sqrt{6}\, \Xi^0_{c} $ & $\to$ & $ p   {K}^- $
&$  0 $&$  0 $&$  0 $&$ -2 $&$  0 $&$  0 $&$  0 $\\
%%% decay 34
&$2\sqrt{6}\,\Xi^0_{c} $ & $\to$ & $ n \bar{K}^0 $
&$  0 $&$  0 $&$ +2 $&$  0 $&$  0 $&$  0 $&$  0 $\\[3pt] \hline \hline
\end{tabular} 
\end{center}
\end{table}

\begin{table}
\caption{\label{TableSCSb} Values of the seven topological tensor invariants
  for the SCS antitriplet charm baryon decays
  $B_{c}(\overline{\bf 3}) \to B({\bf 8})+M({\bf 8},{\bf 1})$
  induced by the flavor transition $(c \to d;\,d \to u)$.}
\begin{center}
\begin{tabular}{crclrrrrrrr}
\hline \hline \\[-11pt]
\phantom{$\hat{\hat{I}_3}$}& & & &
$I^-_1$ & $I^-_2$ & $I_3$ & $I_4$ & $\hat{I_3}$ & $\hat{I}_4$ & $I_5 $\\[3pt]
\hline \\[-11pt]
%%% decay 1
SCSb &$ 12\, \Lambda^+_{c} $ & $\to$ & $ \Lambda^0  K^+$
&$  0 $&$  0 $&$ -2 $&$ -2 $&$  0 $&$  0 $&$ -2 $\\
%%% decay 2
&$ 4 {\sqrt{3}}\, \Lambda^+_{c}  $ & $\to$ & $\Sigma^0  K^+$
&$  0 $&$  0 $&$ +2 $&$ -2 $&$  0 $&$  0 $&$  0 $\\
%%% decay 3
&$ 2 \sqrt{6}\, \Lambda^+_{c}  $ & $\to$ & $\Sigma^+ K^0 $
&$  0 $&$  0 $&$ -2 $&$ +2 $&$  0 $&$  0 $&$  0 $\\
%%% decay 4
&$2\sqrt{6}\, \Lambda^+_{c}  $ & $\to$ & $n  \pi^+ $
&$ -1 $&$ -1 $&$  0 $&$ +2 $&$ -2 $&$ +4 $&$ +1 $\\
%%% decay 5
&$ 4 {\sqrt{3}}\, \Lambda^+_{c}  $ & $\to$ & $p  \pi^0 $
&$ +1 $&$ +1 $&$  0 $&$ -2 $&$ +2 $&$ -4 $&$ -1 $\\
%%% decay 6
&$ 4 {\sqrt{3}}\,\Lambda^+_{c}  $ & $  \to $ & $p  \eta_\omega$
&$ -1 $&$ -1 $&$ -4 $&$ +2 $&$ -2 $&$ +4 $&$ -1 $\\
%%% decay 7
&$  2\sqrt{6}\, \Lambda^+_{c}  $ & $\to$ & $p  \eta_\phi$
&$  0 $&$  0 $&$  0 $&$  0 $&$  0 $&$  0 $&$  0 $\\
%%% decay 8
&$  12\, \Lambda^+_{c}  $ & $\to$ & $p  \eta_8$
&$ -1 $&$ -1 $&$ -4 $&$ +2 $&$ -2 $&$ +4 $&$ -1 $\\
%%% decay 9
&$  6\sqrt{2}\, \Lambda^+_{c}  $ & $\to$ & $p  \eta_1$
&$ -1 $&$ -1 $&$ -4 $&$ +2 $&$ -2 $&$ +4 $&$ -1 $\\[3pt] \hline \\[-11pt]
%%% decay 10
&$  2\sqrt{6}\, \Xi^+_{c}  $ & $\to$ & $\Xi^0   K^+$
&$  0 $&$  0 $&$  0 $&$  0 $&$  0 $&$  0 $&$  0 $\\
%%% decay 11
&$ 4 {\sqrt{3}}\, \Xi^+_{c}  $ & $\to$ & $\Sigma^+  \pi^0 $
&$ +1 $&$ +1 $&$  0 $&$  0 $&$  0 $&$  0 $&$  0 $\\
%%% decay 12
&$4{\sqrt{3}}\, \Xi^+_{c} $ & $\to$ & $\Sigma^+  \eta_\omega $
&$ -1 $&$ -1 $&$  0 $&$  0 $&$  0 $&$  0 $&$  0 $\\
%%% decay 13
&$ 2{\sqrt{6}} \, \Xi^+_{c}  $ & $\to$ & $\Sigma^+  \eta_\phi $
&$  0 $&$  0 $&$  0 $&$  0 $&$  0 $&$  0 $&$  0 $\\
%%% decay 14
&$ 12\, \Xi^+_{c}  $ & $\to$ & $\Sigma^+  \eta_8 $
&$ -1 $&$ -1 $&$  0 $&$  0 $&$  0 $&$  0 $&$  0 $\\
%%% decay 15
&$ 6{\sqrt{2}} \, \Xi^+_{c}  $ & $\to$ & $\Sigma^+  \eta_1 $
&$ -1 $&$ -1 $&$  0 $&$  0 $&$  0 $&$  0 $&$  0 $\\
%%% decay 16
&$2\sqrt{6}\,\Xi^+_{c}$ & $\to$ & $p  \bar{K}^0 $
&$  0 $&$  0 $&$  0 $&$  0 $&$ +2 $&$ -4 $&$  0 $\\
%%% decay 17
&$12\,\Xi^+_{c}  $ & $\to$ & $\Lambda^0 \pi^+ $
&$ -1 $&$ -1 $&$  0 $&$  0 $&$ -4 $&$ +8 $&$  0 $\\
%%% decay 18
&$ 4 {\sqrt{3}}\, \Xi^+_{c}  $ & $\to$ & $\Sigma^0 \pi^+ $
&$ -1 $&$ -1 $&$  0 $&$  0 $&$  0 $&$  0 $&$  0 $\\[3pt] \hline \\[-11pt]
%%% decay 19
&$2\sqrt{6}\,\Xi^0_{c} $ & $\to$ & $\Xi^0  {K}^0 $
&$  0 $&$  0 $&$ +2 $&$  0 $&$  0 $&$  0 $&$  0 $\\
%%% decay 20
&$2\sqrt{6}\, \Xi^0_{c} $ & $\to$ & $ \Xi^- K^+ $
&$  0 $&$  0 $&$ -2 $&$  0 $&$  0 $&$  0 $&$  0 $\\
%%% decay 21
&$2\sqrt{6}\, \Xi^0_{c} $ & $\to$ & $\Sigma^+ \pi^-$
&$  0 $&$  0 $&$  0 $&$ +2 $&$  0 $&$  0 $&$  0 $\\
%%% decay 22
&$2\sqrt{6}\,\Xi^0_{c} $ & $\to$ & $\Sigma^- \pi^+ $
&$ -1 $&$ -1 $&$  0 $&$ +2 $&$  0 $&$  0 $&$  0 $\\
%%% decay 23
&$4{\sqrt{6}}\, \Xi^0_{c} $ & $\to$  & $ \Sigma^0 \pi^0 $
&$ +1 $&$ +1 $&$  0 $&$ -4 $&$  0 $&$  0 $&$  0 $\\
%%% decay 24
&$ 4 {\sqrt{6}}\, \Xi^0_{c} $ & $\to$ & $ \Sigma^0   \eta_\omega $
&$ -1 $&$ -1 $&$  0 $&$  0 $&$  0 $&$  0 $&$  0 $\\
%%% decay 25
&$ 4 {\sqrt{3}}\, \Xi^0_{c} $ & $\to$ & $ \Sigma^0   \eta_\phi $
&$  0 $&$  0 $&$  0 $&$  0 $&$  0 $&$  0 $&$  0 $\\
%%% decay 26
&$ 12 {\sqrt{2}}\, \Xi^0_{c} $ & $\to$ & $ \Sigma^0   \eta_8 $
&$ -1 $&$ -1 $&$  0 $&$  0 $&$  0 $&$  0 $&$  0 $\\
%%% decay 27
&$ 12\, \Xi^0_{c} $ & $\to$ & $ \Sigma^0   \eta_1 $
&$ -1 $&$ -1 $&$  0 $&$  0 $&$  0 $&$  0 $&$  0 $\\
%%% decay 28
&$ 12 {\sqrt{2}}\, \Xi^0_{c} $ & $\to$ & $ \Lambda^0  \pi^0 $
&$ -1 $&$ -1 $&$  0 $&$  0 $&$ -4 $&$ +8 $&$  0 $\\
%%% decay 29
&$12{\sqrt{2}} \,\Xi^0_{c} $ & $\to$ & $ \Lambda^0  \eta_\omega $
&$ +1 $&$ +1 $&$ +8 $&$ -4 $&$ +4 $&$ -8 $&$  0 $\\
%%% decay 30
&$12\, \Xi^0_{c} $ & $\to$ & $ \Lambda^0  \eta_\phi $
&$  0 $&$  0 $&$  0 $&$  0 $&$  0 $&$  0 $&$ +2 $\\
%%% decay 31
&$12\sqrt{6}\, \Xi^0_{c} $ & $\to$ & $ \Lambda^0  \eta_8 $
&$ +1 $&$ +1 $&$ +8 $&$ -4 $&$ +4 $&$ -8 $&$ -4 $\\
%%% decay 32
&$12\sqrt{3}\,\Xi^0_{c} $ & $\to$ & $ \Lambda^0 \eta_1 $
&$ +1 $&$ +1 $&$ +8 $&$ -4 $&$ +4 $&$ -8 $&$ +2 $\\
%%% decay 33
&$ 2\sqrt{6}\, \Xi^0_{c} $ & $\to$ & $ p   {K}^- $
&$  0 $&$  0 $&$  0 $&$  0 $&$  0 $&$  0 $&$ +1 $\\
%%% decay 34
&$2\sqrt{6}\,\Xi^0_{c} $ & $\to$ & $ n \bar{K}^0 $
&$  0 $&$  0 $&$  0 $&$  0 $&$ +2 $&$ -4 $&$ -1 $\\[3pt] \hline \hline
\end{tabular} 
\end{center}
\end{table}

\begin{table}
\caption{\label{TableSCS} Values of the seven topological tensor invariants
  for the SCS antitriplet charm baryon decays
  $B_{c}(\overline{\bf 3}) \to B({\bf 8})+M({\bf 8},{\bf 1})$
  induced by both flavor transitions (a) and (b).}
\begin{center}
\begin{tabular}{crclrrrrrrr}
\hline \hline \\[-11pt]\phantom{$\hat{\hat{I}_3}$}& & & &
$I^-_1$ & $I^-_2$ & $I_3$ & $I_4$ & $\hat{I_3}$ & $\hat{I}_4$ & $I_5 $\\[3pt]
\hline \\[-11pt]
%%% decay 1
SCS &$ 12\, \Lambda^+_{c} $ & $\to$ & $ \Lambda^0  K^+$
&$ +2 $&$ +2 $&$ +2 $&$ +2 $&$ +2 $&$ -4 $&$ +2 $\\
%%% decay 2
&$ 4 {\sqrt{3}}\, \Lambda^+_{c}  $ & $\to$ & $\Sigma^0  K^+$
&$  0 $&$  0 $&$ -2 $&$ +2 $&$ +2 $&$ -4 $&$  0 $\\
%%% decay 3
&$ 2 \sqrt{6}\, \Lambda^+_{c}  $ & $\to$ & $\Sigma^+ K^0 $
&$  0 $&$  0 $&$ +2 $&$ -2 $&$ -2 $&$ +4 $&$  0 $\\
%%% decay 4
&$2\sqrt{6}\, \Lambda^+_{c}  $ & $\to$ & $n  \pi^+ $
&$ +1 $&$ +1 $&$  0 $&$ -2 $&$ +2 $&$ -4 $&$ -1 $\\
%%% decay 5
&$ 4 {\sqrt{3}}\, \Lambda^+_{c}  $ & $\to$ & $p  \pi^0 $
&$ -1 $&$ -1 $&$  0 $&$ +2 $&$ -2 $&$ +4 $&$ +1 $\\
%%% decay 6
&$ 4 {\sqrt{3}}\,\Lambda^+_{c}  $ & $  \to $ & $p  \eta_\omega$
&$ +1 $&$ +1 $&$ +4 $&$ -2 $&$ +2 $&$ -4 $&$ +1 $\\
%%% decay 7
&$  2\sqrt{6}\, \Lambda^+_{c}  $ & $\to$ & $p  \eta_\phi$
&$ -1 $&$ -1 $&$  0 $&$  0 $&$  0 $&$  0 $&$  0 $\\
%%% decay 8
&$  12\, \Lambda^+_{c}  $ & $\to$ & $p  \eta_8$
&$ +3 $&$ +3 $&$ +4 $&$ -2 $&$ +2 $&$ -4 $&$ +1 $\\
%%% decay 9
&$  6\sqrt{2}\, \Lambda^+_{c}  $ & $\to$ & $p  \eta_1$
&$  0 $&$  0 $&$ +4 $&$ -2 $&$ +2 $&$ -4 $&$ +1 $\\[3pt] \hline \\[-11pt]
%%% decay 10
&$  2\sqrt{6}\, \Xi^+_{c}  $ & $\to$ & $\Xi^0   K^+$
&$ +1 $&$ +1 $&$  0 $&$ -2 $&$ +2 $&$ -4 $&$ -1 $\\
%%% decay 11
&$ 4 {\sqrt{3}}\, \Xi^+_{c}  $ & $\to$ & $\Sigma^+  \pi^0 $
&$ -1 $&$ -1 $&$ -2 $&$  0 $&$  0 $&$  0 $&$ -1 $\\
%%% decay 12
&$4{\sqrt{3}}\, \Xi^+_{c} $ &$ \to$ & $\Sigma^+  \eta_\omega $
&$ +1 $&$ +1 $&$ -2 $&$  0 $&$  0 $&$  0 $&$ -1 $\\
%%% decay 13
&$ 2{\sqrt{6}} \, \Xi^+_{c}  $ & $\to$ & $\Sigma^+  \eta_\phi $
&$ -1 $&$ -1 $&$ -2 $&$ +2 $&$ -2 $&$ +4 $&$  0 $\\
%%% decay 14
&$ 12\, \Xi^+_{c}  $ & $\to$ & $\Sigma^+  \eta_8 $
&$ +3 $&$ +3 $&$ +2 $&$ -4 $&$ +4 $&$ -8 $&$ -1 $\\
%%% decay 15
&$ 6{\sqrt{2}}\, \Xi^+_{c}  $ & $\to$ & $\Sigma^+  \eta_1 $
&$  0 $&$  0 $&$ -4 $&$ +2 $&$ -2 $&$ +4 $&$ -1 $\\
%%% decay 16
&$2\sqrt{6}\,\Xi^+_{c}$ & $\to$ & $p  \bar{K}^0 $
&$  0 $&$  0 $&$ +2 $&$ -2 $&$ -2 $&$ +4 $&$  0 $\\
%%% decay 17
&$12\,\Xi^+_{c}  $ & $\to$ & $\Lambda^0 \pi^+ $
&$ +1 $&$ +1 $&$ -2 $&$ +4 $&$ +4 $&$ -8 $&$ +1 $\\
%%% decay 18
&$ 4 {\sqrt{3}}\, \Xi^+_{c}  $ & $\to$ & $\Sigma^0 \pi^+ $
&$ +1 $&$ +1 $&$ +2 $&$  0 $&$  0 $&$  0 $&$ +1 $\\[3pt] \hline \\[-11pt]
%%% decay 19
&$2\sqrt{6}\,\Xi^0_{c} $ & $\to$ & $\Xi^0  {K}^0 $
&$  0 $&$  0 $&$ -2 $&$  0 $&$ +2 $&$ -4 $&$ -1 $\\
%%% decay 20
&$2\sqrt{6}\, \Xi^0_{c} $ & $\to$ & $ \Xi^- K^+ $
&$ +1 $&$ +1 $&$ +2 $&$ -2 $&$  0 $&$  0 $&$  0 $\\
%%% decay 21
&$2\sqrt{6}\, \Xi^0_{c} $ & $\to$ & $\Sigma^+ \pi^-$
&$  0 $&$  0 $&$  0 $&$ -2 $&$  0 $&$  0 $&$ -1 $\\
%%% decay 22
&$2\sqrt{6}\,\Xi^0_{c} $ &$ \to$ &$\Sigma^- \pi^+ $
&$ +1 $&$ +1 $&$ +2 $&$ -2 $&$  0 $&$  0 $&$  0 $\\
%%% decay 23
&$4{\sqrt{6}}\, \Xi^0_{c} $ & $\to$  & $ \Sigma^0 \pi^0 $
&$ -1 $&$ -1 $&$ -2 $&$ +4 $&$  0 $&$  0 $&$ +1 $\\
%%% decay 24
&$ 4 {\sqrt{6}}\, \Xi^0_{c} $ & $\to$ & $ \Sigma^0   \eta_\omega $
&$ +1 $&$ +1 $&$ -2 $&$  0 $&$  0 $&$  0 $&$ -1 $\\
%%% decay 25
&$ 4 {\sqrt{3}}\, \Xi^0_{c} $ & $\to$ & $ \Sigma^0   \eta_\phi $
&$ -1 $&$ -1 $&$ -2 $&$ +2 $&$ -2 $&$ +4 $&$  0 $\\
%%% decay 26
&$ 12 {\sqrt{2}}\, \Xi^0_{c} $ & $\to$ & $ \Sigma^0   \eta_8 $
&$ +3 $&$ +3 $&$ +2 $&$ -4 $&$ +4 $&$ -8 $&$ -1 $\\
%%% decay 27
&$ 12\, \Xi^0_{c} $ & $\to$ & $ \Sigma^0   \eta_1 $
&$  0 $&$  0 $&$ -4 $&$ +2 $&$ -2 $&$ +4 $&$ -1 $\\
%%% decay 28
&$ 12 {\sqrt{2}}\, \Xi^0_{c} $ & $\to$ & $ \Lambda^0  \pi^0 $
&$ +1 $&$ +1 $&$ -2 $&$ +4 $&$ +4 $&$ -8 $&$ +1 $\\
%%% decay 29
&$12{\sqrt{2}} \,\Xi^0_{c} $ & $\to$ & $ \Lambda^0  \eta_\omega $
&$ -1 $&$ -1 $&$ -10$&$ +8 $&$ -4 $&$ +8 $&$ -1 $\\
%%% decay 30
&$12\, \Xi^0_{c} $ & $\to$ & $ \Lambda^0  \eta_\phi $
&$ +1 $&$ +1 $&$ -2 $&$ -2 $&$ -2 $&$ +4 $&$ -2 $\\
%%% decay 31
&$12\sqrt{6}\, \Xi^0_{c} $ & $\to$ & $ \Lambda^0  \eta_8 $
&$ -3 $&$ -3 $&$ -6 $&$ +12 $&$  0 $&$  0 $&$ +3 $\\
%%% decay 32
&$12\sqrt{3}\,\Xi^0_{c} $ & $\to$ & $ \Lambda^0 \eta_1 $
&$  0 $&$  0 $&$ -12$&$ +6 $&$ -6 $&$ +12 $&$ -3 $\\
%%% decay 33
&$ 2\sqrt{6}\, \Xi^0_{c} $ & $\to$ & $ p   {K}^- $
&$  0 $&$  0 $&$  0 $&$ -2 $&$  0 $&$  0 $&$  -1 $\\
%%% decay 34
&$2\sqrt{6}\,\Xi^0_{c} $ & $\to$ & $ n \bar{K}^0 $
&$  0 $&$  0 $&$ +2 $&$  0 $&$ -2 $&$ +4 $&$ +1 $\\[3pt] \hline \hline
\end{tabular} 
\end{center}
\end{table}

\begin{table}[ht]
\caption{\label{TableDCS}
  Values of the seven topological tensor invariants for the DCS antitriplet
  charm baryon decays
  $B_{c}(\overline{\bf 3}) \to B({\bf 8})+M({\bf 8},{\bf 1})$
  induced by the flavor transition $(c \to d;\,s \to u)$.}
\begin{center}
\begin{tabular}{crclrrrrrrr}
\hline \hline \\[-11pt]
 & & & & $I^-_1$ & $I^-_2$ & $I_3$ & $I_4$
 & $\hat{I}_3$ & $\hat{I}_4$ & $I_5$ \\ [0.5ex]\hline \\[-11pt]
%%% decay 1
DCS &$2\sqrt{6} \,\Lambda_{c}^+ $ & $\to$ & $ p K^0$
&$ -1 $&$ -1 $&$  0 $&$  0 $&$ -2 $&$ +4 $&$  0 $\\
%%% decay 2
&$2\sqrt{6} \,\Lambda_{c}^+ $ & $\to$ & $ n K^+$
&$ +1 $&$ +1 $&$  0 $&$  0 $&$ +2 $&$ -4 $&$  0 $\\[3pt] \hline \\[-11pt]
%%% decay 3
&$12 \,\Xi_{c}^+ $ & $\to$ & $ \Lambda^0 K^+$
&$ +1 $&$ +1 $&$ -2 $&$ -2 $&$ +4 $&$ -8 $&$ -2 $\\
%%% decay 4
&$4\sqrt{3} \,\Xi_{c}^+ $ & $\to$ & $ \Sigma^0 K^+$
&$ +1 $&$ +1 $&$ +2 $&$ -2 $&$  0 $&$  0 $&$  0 $\\
%%% decay 5
&$2\sqrt{6} \,\Xi_{c}^+ $ & $\to$ & $ \Sigma^+ K^0$
&$ -1 $&$ -1 $&$ -2 $&$ +2 $&$  0 $&$  0 $&$  0 $\\
%%% decay 6
&$4\sqrt{3} \,\Xi_{c}^+ $ & $\to$ & $ p \pi^0$
&$  0 $&$  0 $&$  0 $&$ -2 $&$  0 $&$  0 $&$ -1 $\\
%%% decay 7
&$2\sqrt{6} \,\Xi_{c}^+ $ & $\to$ & $ n \pi^+$
&$  0 $&$  0 $&$  0 $&$ +2 $&$  0 $&$  0 $&$ +1 $\\
%%% decay 8
&$4\sqrt{3} \,\Xi_{c}^+ $ & $\to$ & $ p \eta_\omega$
&$  0 $&$  0 $&$ -4 $&$ +2 $&$  0 $&$  0 $&$ -1 $\\
%%% decay 9
&$2\sqrt{6} \,\Xi_{c}^+ $ & $\to$ & $ p \eta_\phi$
&$  0 $&$  0 $&$  0 $&$  0 $&$ -2 $&$ +4 $&$  0 $\\
%%% decay 10
&$12 \,\Xi_{c}^+ $ & $\to$ & $ p \eta_8$
&$  0 $&$  0 $&$ -4 $&$ +2 $&$ +4 $&$ -8 $&$ -1 $\\
%%% decay 11
&$6\sqrt{2} \,\Xi_{c}^+ $ & $\to$ & $ p \eta_1$
&$  0 $&$  0 $&$ -4 $&$ +2 $&$ -2 $&$ +4 $&$ -1 $\\[3pt] \hline \\[-11pt]
%%% decay 12
&$12 \,\Xi_{c}^0 $ & $\to$ & $ \Lambda^0 K^0$
&$ +1 $&$ +1 $&$ -2 $&$ -2 $&$ +4 $&$ -8 $&$ -2 $\\
%%% decay 13
&$4\sqrt{3} \,\Xi_{c}^0 $ & $\to$ & $ \Sigma^0 K^0$
&$ -1 $&$ -1 $&$ -2 $&$ +2 $&$  0 $&$  0 $&$  0 $\\
%%% decay 14
&$2\sqrt{6} \,\Xi_{c}^0 $ & $\to$ & $ \Sigma^- K^+$
&$ +1 $&$ +1 $&$ +2 $&$ -2 $&$  0 $&$  0 $&$  0 $\\
%%% decay 15
&$2\sqrt{6} \,\Xi_{c}^0 $ & $\to$ & $ p \pi^-$
&$  0 $&$  0 $&$  0 $&$ -2 $&$  0 $&$  0 $&$ -1 $\\
%%% decay 16
&$4\sqrt{3} \,\Xi_{c}^0 $ & $\to$ & $ n \pi^0$
&$  0 $&$  0 $&$  0 $&$ +2 $&$  0 $&$  0 $&$ +1 $\\
%%% decay 17
&$4\sqrt{3} \,\Xi_{c}^0 $ & $\to$ & $ n \eta_\omega$
&$  0 $&$  0 $&$ -4 $&$ +2 $&$  0 $&$  0 $&$ -1 $\\
%%% decay 18
&$2\sqrt{6}\Xi_{c}^0 $ & $\to$ & $ n \eta_\phi$
&$  0 $&$  0 $&$  0 $&$  0 $&$ -2 $&$ +4 $&$  0 $\\
%%% decay 19
&$12 \,\Xi_{c}^0 $ & $\to$ & $ n \eta_8$
&$  0 $&$  0 $&$ -4 $&$ +2 $&$ +4 $&$ -8 $&$ -1 $\\
%%% decay 20
&$6 \sqrt{2} \,\Xi_{c}^0 $ & $\to$ & $ n \eta_1$
&$  0 $&$  0 $&$ -4 $&$ +2 $&$ -2 $&$ +4 $&$ -1 $\\[3pt]
\hline 
\hline
\end{tabular} 
\end{center}
\end{table}

%%%%%%%%%%%%%%%%%%%%%%%%%%%%%%%%%%%%%%%%%%%%%%%%%%%%%%%%%%%%%%%%%%%%%%%%%%%%%%%
\subsection{The charm baryon decays $B_{c}(\overline{\bf 3})
  \stackrel{\!\!H(\bf 6)}{\longrightarrow}B({\bf 8}) + M({\bf 8,1})$}
%%%%%%%%%%%%%%%%%%%%%%%%%%%%%%%%%%%%%%%%%%%%%%%%%%%%%%%%%%%%%%%%%%%%%%%%%%%%%%%
The nonleptonic two-body decays of the three antitriplet charm baryons
($\Lambda_{c}^+,\Xi_{c}^+,\Xi_{c}^0$) belong to this class of decays. The SU(3)
decomposition of the transitions read (we use the notation of Kaeding for
multiple SU(3) representations~\cite{Kaeding:1995vq})
\begin{equation}\label{decomp1}
  \overline{\bf 3} \to {\bf 6}\otimes{\bf 8}\otimes {\bf 8}
  =3\cdot\overline{\bf 3}
  \oplus 4\cdot{\bf  6} \oplus 5\cdot\overline{\bf 15}\oplus
  \overline{\bf 15}\,'\oplus\overline{\bf 21}\oplus\overline{\bf 24}\oplus
  2\cdot\overline{\bf 42}\oplus\overline{\bf 60}
\end{equation}
As explained in Sec.~\ref{topotensor}, the effective Hamiltonian inducing the
$W$-exchange contributions transforms as a sextet in SU(3). For the SCS decays
this comes about by a cancellation of the antitriplet contributions when the
two effective Hamiltonians ${\cal H}_{\rm eff}(c \to s;\,s \to u)$ and
${\cal H}_{\rm eff}(c \to d;\,d \to u)$ are subtracted from one another. The
$\overline{\bf 3}$ representation appears three times in the decomposition.
Therefore, there are three SU(3) invariant amplitudes. This in turn implies
that there are four linear relations among the seven topological tensor
invariants. They read
\begin{equation}\label{linrel1}
  I^-_1=I^-_2,\qquad 2\hat{I}_3+\hat{I}_4=0,\qquad I_3+I_4=2I_5,
  \qquad 2I^-_1=I_3+\hat{I}_3.
\end{equation}
The relation $I^-_1=I^-_2$ can be obtained by rewriting the tensor labels of
the parent baryon in the tensor invariant $I^-_2$ in the way $B^1_{b[c'a]}
\to -B^1_{a[c'b]} \to B^1_{a[bc']}$ by realizing that the indices $a$ and $b$
refer to light quarks. Similarly one can proof $2\hat{I}_3+\hat{I}_4=0$ by
realizing that $a$ and $c'$ are the light quarks in the parent baryon
$B^1_{b'[c'a]}$ and using the Jacobi identity
$B^1_{b'[c'a]}+B^1_{c'[ab']}+B^1_{a[b'c']}=0$.

A more direct access to a minimal set of independent tensor contractions is
to switch to second rank tensor representations of the antitriplet charm
baryons, the octet baryons and the effective
Hamiltonian~\cite{Savage:1989qr,Jia:2019zxi}. As a result one now has only
three independent tensor contractions, the number of which agrees with the
above number of SU(3) invariants. The linear relations~(\ref{linrel1}) can be
seen to be in agreement with the analysis of Jia et al.~\cite{Jia:2019zxi}.

Eqs.~(\ref{linrel1}) are very useful when checking the results for the
topological tensor invariants $I_j(fki)$ that will be listed later on. Note
in particular that the linear relations~(\ref{linrel1}) do not separately hold
for the contributions of the two transitions
($a$) $c\!\to \! s;\,\,s\!\to \! u$ and ($b$) $c\!\to \!  d;\,\,d\!\to \! u$.
They can be seen to hold true only for the sum of the two contributions. Note
that the last equation does not hold for all decay processes listed in
Tables~\ref{TableCF} to~\ref{TableDCS}, in particular not for decays into
$\eta_1$ and, related to that, into $\eta_\omega$ and $\eta_\phi$, as those
are members of (a mixture of $M({\bf 8})$ and) $M({\bf 1})$.

Therefore, we also need to discuss the transition $B_{c}(\overline{\bf 3})
\stackrel{\!\!H(\bf 6)}{\longrightarrow}B({\bf 8}) + M({\bf 1})$ involving the
singlet meson state $M({\bf 1})$. In this case the SU(3) decomposition reads
\begin{equation}\label{decomp2}
\overline{\bf 3} \to {\bf 6}\otimes{\bf 8}\otimes {\bf 1}=\overline{\bf 3}
  \oplus {\bf  6} \oplus \overline{\bf 15}\oplus \overline{\bf 24}.
\end{equation}
Thus there is only one SU(3) invariant amplitude describing the transition
$B_{c}(\overline{\bf 3})
\stackrel{\!\!H(\bf 6)}{\longrightarrow}B({\bf 8}) + M({\bf 1})$ which means
that all transitions involving a singlet meson state $ M({\bf 1})$ are
proportional to one another which is manifest in the decays into $\eta_1$. In
the corresponding rows of Tables~\ref{TableCF} to~\ref{TableDCS} one has
\begin{equation}\label{linrel2}
I^-_1=I^-_2=0,\qquad I_3=-2I_4=2\hat{I}_3=-\hat{I}_4=4I_5.
\end{equation}
The vanishing of the tree diagram invariants can be argued again directly, as
the single quark line transition
${\bf 1} \to {\bf 6}\otimes {\bf 1} \otimes {\bf 3}= {\bf 8} \oplus {\bf 10}$
into singlet meson states vanishes in SU(3).

We list the values of the seven topological tensor invariants separately for
the Cabibbo favored (CF) (Table~\ref{TableCF}), singly Cabibbo suppressed (SCS)
(Tables~\ref{TableSCSa}, \ref{TableSCSb} and~\ref{TableSCS}), and doubly
Cabibbo suppressed (DCS) (Table~\ref{TableDCS}) antitriplet charm baryon
decays. The SCS decays are induced by the two SCS transitions ($a$)
$c\to s;\,s\to u$ and ($b$) $c\to d;\,d\to u$ which are listed separately in
Tables~\ref{TableSCSa} and~\ref{TableSCSb}. As explained in
Sec.~\ref{effective}, in the SU(3) limit dealt with in this paper one has to
subtract the two contributions ($a$) and ($b$). The result of this subtraction
is shown in Table~\ref{TableSCS}.

As discussed in Sec.~\ref{topotensor}, Kohara has introduced four reduced
matrix elements $d_1,d_2,d_3,d_4$ for the topology class IIa where only two
are needed~\cite{Kohara:1991ug}. Using the results of e.g.\
Table~\ref{TableCF} for the decays involving octet mesons one finds that the
four topological reduced matrix elements of Kohara are related to our
topological reduced matrix elements by $d_1=2\mathbfcal{T}_3$,
$d_2=-2(\mathbfcal{T}_3-\mathbfcal{T}_4)$, $d_3=2\mathbfcal{T}_4$, and
$d_4=2\mathbfcal{T}_3-\mathbfcal{T}_4$. This shows again that the set
$d_1,\,d_2,\,d_3,\,d_4$ is redundant since one has $d_1=d_3+d_4$ and
$d_2=-d_4$. We do not agree with Ref.~\cite{Kohara:1991ug} on the
contributions of the $d_i$ to the decays involving the SU(3) singlet meson
$\eta_1$.

\begin{table}
\caption{\label{TableOmegac} Values of the seven topological tensor invariants
  for the nonleptonic CF, SCS and DCS decays of the sextet charm baryon state
  $\Omega^0_{c}$ belonging to the class of decays
  $B_{c}({\bf 6}) \to  B({\bf 8}) + M({\bf 8},{\bf 1})$. The SCS decays are
  induced by the flavor transition ($a$) $(c \to s;\,s \to u)$ and ($b$)
  $(c \to d;\,d \to u)$. The subscripts $a$ and $b$ of the tensor invariants
  identify the origin of the respective flavor transitions where the relative
  sign of the two transitions has been accounted for.}
\begin{center}
\begin{tabular}{crclccccccc}
\hline \hline \\[-11pt]
 & & & & $I^-_1$ & $I^-_2$ & $I_3$ & $I_4$
 & $\hat{I}_3$ & $\hat{I}_4$ & $I_5$ \\[3pt] \hline \\ [-11pt]
%% decay 1
 CF&$2 \Omega_{c}^0 $ & $\to$ & $ \Xi^0 \bar K^0$
&$-1$&$+1$&$0$&$0$&$-2$&$0$&$0$\\[3pt] \hline\\[-11pt]
%% decay 2
SCS&$2\sqrt{2} \Omega_{c}^0 $ & $\to$ & $ \Xi^0 \pi^0$
&$+1_b$&$-1_b$&$+2_a$&$-2_a$&$0$&$0$&$0$\\
%% decay 3
&$2\sqrt{2} \Omega_{c}^0 $ & $\to$ & $ \Xi^0 \eta_\omega$
&$-1_b$&$+1_b$&$+2_a$&$-2_a$&$0$&$0$&$0$\\
%%% decay 4
&$2 \Omega_{c}^0 $ & $\to$ & $ \Xi^0 \eta_\phi$
&$+1_a$&$-1_a$&$+2_a$&$0$&$+2_a$&$0$&$+1_a$\\
%% decay 5
&$2\sqrt{6} \Omega_{c}^0 $ & $\to$ & $ \Xi^0 \eta_8$
&$-2_a-1_b$&$+2_a+1_b$&$-2_a$&$-2_a$&$-4_a$&$0$&$-2_a$\\
%% decay 6
&$2\sqrt{3} \Omega_{c}^0 $ & $\to$ & $ \Xi^0 \eta_1$
&$+1_a-1_b$&$-1_a+1_b$&$+4_a$&$-2_a$&$+2_a$&$0$&$+1_a$\\
%% decay 7
&$2\sqrt{6} \Omega_{c}^0 $ & $\to$ & $ \Lambda^0 \bar K^0$
&$0$&$0$&$-4_a$&$+2_a$&$+4_b$&$0$&$-1_a$\\
%% decay 8
&$2\sqrt{2} \Omega_{c}^0 $ & $\to$ & $ \Sigma^0 \bar K^0$
&$0$&$0$&$0$&$-2_a$&$0$&$0$&$-1_a$\\
%% decay 9
&$2 \Omega_{c}^0 $ & $\to$ & $ \Sigma^+ K^-$
&$0$&$0$&$0$&$+2_a$&$0$&$0$&$+1_a$\\
%% decay 10
&$2 \Omega_{c}^0 $ & $\to$ & $ \Xi^- \pi^+$
&$-1_b$&$+1_b$&$-2_a$&$+2_a$&$0$&$0$&$0$\\[3pt] \hline \\[-11pt]
%% decay 11
DCS&$2 \Omega_{c}^0 $ & $\to$ & $ p K^-$
&$  0 $&$  0 $&$  0 $&$  0 $&$  0 $&$  0 $&$ +1 $\\
%% decay 12
&$2 \Omega_{c}^0 $ & $\to$ & $ n \bar K^0 $
&$  0 $&$  0 $&$  0 $&$  0 $&$  0 $&$  0 $&$ -1 $\\
%% decay 13
&$4\sqrt{3} \Omega_{c}^0 $ & $\to$ & $ \Lambda^0 \pi^0$
&$  0 $&$  0 $&$  0 $&$  0 $&$  0 $&$  0 $&$  0 $\\
%% decay 14
&$4\sqrt{3} \Omega_{c}^0 $ & $\to$ & $ \Lambda^0 \eta_\omega$
&$  0 $&$  0 $&$ +8 $&$ -4 $&$  0 $&$  0 $&$  0 $\\
%% decay 15
&$2\sqrt{6} \Omega_{c}^0 $ & $\to$ & $ \Lambda^0 \eta_\phi $
&$  0 $&$  0 $&$  0 $&$  0 $&$ +4 $&$  0 $&$ +2 $\\
%% decay 16
&$12 \Omega_{c}^0 $ & $\to$ & $ \Lambda^0 \eta_8$
&$  0 $&$  0 $&$ +8 $&$ -4 $&$ -8 $&$  0 $&$ -4 $\\
%% decay 17
&$6\sqrt{2} \Omega_{c}^0 $ & $\to$ & $ \Lambda^0 \eta_1$
&$  0 $&$  0 $&$ +8 $&$ -4 $&$ +4 $&$  0 $&$ +2 $\\
%% decay 18
&$4 \Omega_{c}^0 $ & $\to$ & $ \Sigma^0 \pi^0$
&$  0 $&$  0 $&$  0 $&$ -4 $&$  0 $&$  0 $&$  0 $\\
%% decay 19
&$4 \Omega_{c}^0 $ & $\to$ & $ \Sigma^0 \eta_\omega $
&$  0 $&$  0 $&$  0 $&$  0 $&$  0 $&$  0 $&$  0 $\\
%% decay 20
&$2\sqrt{2} \Omega_{c}^0 $ & $\to$ & $ \Sigma^0 \eta_\phi $
&$  0 $&$  0 $&$  0 $&$  0 $&$  0 $&$  0 $&$  0 $\\
%% decay 21
&$4\sqrt{3} \Omega_{c}^0 $ & $\to$ & $ \Sigma^0 \eta_8$
&$  0 $&$  0 $&$  0 $&$  0 $&$  0 $&$  0 $&$  0 $\\
%% decay 22
&$2\sqrt{6} \Omega_{c}^0 $ & $\to$ & $ \Sigma^0 \eta_1$
&$  0 $&$  0 $&$  0 $&$  0 $&$  0 $&$  0 $&$  0 $\\
%% decay 23
&$2 \Omega_{c}^0 $ & $\to$ & $ \Sigma^+ \pi^-$
&$  0 $&$  0 $&$  0 $&$ +2 $&$  0 $&$  0 $&$  0 $\\
%% decay 24
&$2 \Omega_{c}^0 $ & $\to$ & $ \Sigma^- \pi^+$
&$  0 $&$  0 $&$  0 $&$ +2 $&$  0 $&$  0 $&$  0 $\\
%% decay 25
&$2 \Omega_{c}^0 $ & $\to$ & $ \Xi^- K^+$
&$ -1 $&$ +1 $&$ -2 $&$  0 $&$  0 $&$  0 $&$  0 $\\
%% decay 26
&$2 \Omega_{c}^0 $ & $\to$ & $ \Xi^0 K^0$
&$ +1 $&$ -1 $&$ +2 $&$  0 $&$  0 $&$  0 $&$  0 $\\[3pt]
\hline
\hline
\end{tabular} 
\end{center}
\end{table}

%%%%%%%%%%%%%%%%%%%%%%%%%%%%%%%%%%%%%%%%%%%%%%%%%%%%%%%%%%%%%%%%%%%%%%%%%%%%%%%
\subsection{The charm baryon decays
  $B_{c}({\bf6}) \stackrel{H({\bf 6})}{\longrightarrow}
  B({\bf 8}) + M({\bf 8,1})$}
%%%%%%%%%%%%%%%%%%%%%%%%%%%%%%%%%%%%%%%%%%%%%%%%%%%%%%%%%%%%%%%%%%%%%%%%%%%%%%%
The nonleptonic two-body decays of the charm baryon $\Omega_{c}^0$ belong to
this class of decays. The SU(3) decomposition of the direct product
${\bf 6}\otimes{\bf 8}\otimes{\bf 8}$ has been written down before in
Eq.~(\ref{decomp1}). One notes that the ${\bf 6}$ representation appears four
times in the decomposition. The transition
${\bf 6} \to {\bf 6}\otimes{\bf 8}\otimes{\bf 8}$ is thus described by four
SU(3) invariants. This implies that there are $7-4=3$ linear relations among
the 7 tensor invariants. These are
\begin{equation}
I^-_1+I^-_2=0,\qquad \hat{I_4}=0,\qquad 2I^-_1=I_3+\hat{I}_3. 
\end{equation}
The first relation $I^-_1+I^-_2=0$ follows from the Jacobi identity rewriting
$I^-_2$ as
\begin{equation}\label{linrel}
I^-_2(\ell,\ell')=B_\ell^{a[bc]}B^{\ell'}_{b[c'a]}M^d_{d'}H^{[c'd']}_{[cd]}
  =B_\ell^{a[bc]}(-B^{\ell'}_{c'[ab]}-B^{\ell'}_{a[bc']})M^d_{d'}
  H^{[c'd']}_{[cd]}
\end{equation}
and noting that the first term on the r.h.s.\ of Eq.~(\ref{linrel}) vanishes
due to the fact that $c'$ must be the charm quark, and that the light quarks
$a,b$ are symmetric in the ${\bf 6}$ representation. With the same reasoning
one can show that $\hat{I}_4=0$. The relation $2I^-_1=I_3+\hat{I}_3$ cannot be
obtained by tensor label manipulations but can be derived from the values of
the three respective tensor invariants in Table~\ref{TableOmegac}. In the case
of the SCS transitions the suffixes $a$ and $b$ identify the origin of the
respective flavor transitions ($a$) and ($b$) including the relative sign.
Note that we always keep the ($a$)- and ($b$)-type contributions in the SCS
contribution apart for two reasons: in order to incorporate some SU(3)
breaking effects in this way, and for comparison with parts of the literature
in which the two contributions have been kept apart. Note that the tensor
relation $ 2I^-_1=I_3+\hat{I}_3$ does not hold separately for the transitions
($a$) and ($b$) but only for their sum, and it does not hold if the singlet
meson state is involved.

The transitions involving the singlet meson state $\eta_1$ are again described
by a single SU(3) invariant, as can be read off from the
decomposition~(\ref{decomp2}). For these transitions on finds
$I^-_1=I^-_2=\hat I_4=0$ and $I_3=-2I_4=2\hat I_3=4I_5$.

\begin{table}
\caption{\label{deltaC=0} Values of the seven tensor invariants for the
  $\Delta C=0$ SCS decays of the single charm baryons $\Xi_{c}^+$, $\Xi_{c}^0$,
  and $\Omega_{c}^0$ into antitriplet charm baryons induced by the flavor
  transitions ($a'$) $s\to u;\quad u\to d$ and ($b'$) $c\to d;\quad s\to c$.
  The subscripts $a'$ and $b'$ of the tensor invariants identify the origin of
  the respective flavor transitions where the relative sign of the two
  transitions has been accounted for.}
\begin{center}
\begin{tabular}{rclrrrrrrr}
\hline \hline \\[-11pt]
 & & & $I^-_1$ & $I^-_2$ & $I_3$ & $I_4$
 & $\hat{I}_3$ & $\hat{I}_4$ & $I_5$ \\ [3pt] \hline \\[-11pt]
%% decay 1
$12\sqrt{2} \Xi_{c}^+ $ & $\to$ & $ \Lambda_{c}^+ \pi^0$&
$-5_{a'}$ & $+4_{a'}$ & $-2_{b'}$ & $ 4_{b'}$
  & $-8_{a'}$ & $ 4_{a'}$ & $ 1_{b'}$ \\
%% decay 2
$12 \Xi_{c}^0 $ & $\to$ & $ \Lambda_{c}^+ \pi^-$&
$-5_{a'}$ & $+4_{a'}$ & $-2_{b'}$ & $ 4_{b'}$ & $-8_{a'}$ & $ 4_{a'}$ &
$ 1_{b'}$
  \\[3pt]\hline\\[-11pt]
%%% decay 1
$2\sqrt{6}\, \Omega_{c}^0 $ & $\to$ & $ \Xi_{c}^+ \pi^-$&
$ 1_{a'}$ & $-2_{a'}$ & $+2_{b'}$ & $-4_{b'}$ & $ 0  $ & $ 0  $ & $ 0  $ \\
%% decay 2
$4\sqrt{3}\, \Omega_{c}^0 $ & $\to$ & $ \Xi_{c}^0 \pi^0$&
$-1_{a'}$ & $+2_{a'}$ & $-2_{b'}$ & $+4_{b'}$ & $ 0  $ & $ 0  $ & $ 0  $ \\[3pt]
\hline
\hline
\end{tabular} 
\end{center}
\end{table}

%%%%%%%%%%%%%%%%%%%%%%%%%%%%%%%%%%%%%%%%%%%%%%%%%%%%%%%%%%%%%%%%%%%%%%%%%%%%%%%
\subsection{The $\Delta C=0$ singly Cabibbo suppressed charm baryon decays
  $B_{c}(\overline{\bf 3}) \stackrel{H({\bf 8})}{\longrightarrow}
  B_{c}(\overline{\bf 3}) + M({\bf 8})$
  and  $B_{c}({\bf 6}) \stackrel{H({\bf 8})}{\longrightarrow}
  B_{c}(\overline{\bf 3}) + M({\bf 8})$}
%%%%%%%%%%%%%%%%%%%%%%%%%%%%%%%%%%%%%%%%%%%%%%%%%%%%%%%%%%%%%%%%%%%%%%%%%%%%%%%
The $\Delta C=0$ charm baryon decays are induced by the SCS quark flavor
transitions ($a'$) $s\to u;\,u\to d$ and ($b'$) $c\to d;\,s\to c$. The
kinematical constraints of the $\Delta C=0$ decays only allow for the pionic
modes. Therefore, there is no need to discuss the decays into the $\eta$ and
$\eta'$ states or the $K$-meson states. As concerns the application of the
current algebra approach to charm baryon decays, the $\Delta C=0$ charm baryon
decays are the favorites of all charm baryon decays since the emitted pion has
very little energy, i.e.\ the pion satisfies the requirement of the soft pion
theorem. The $\Delta C=0$ charm baryon decays have been discussed before in
Refs.~\cite{Voloshin:2000et,Li:2014ada,Faller:2015oma,Cheng:2015ckx,%
Voloshin:2019ngb}.

There are two isospin decays in each the two classes $B_{c}(\overline{\bf 3})
\stackrel{H({\bf 8})}{\longrightarrow}B_{c}(\overline{\bf 3}) + M({\bf 8})$ and
$B_{c}({\bf 6}) \stackrel{H({\bf 8})}{\longrightarrow} B_{c}(\overline{\bf 3})
+ M({\bf 8})$. The four kinematically accessible decays are listed in
Table~\ref{deltaC=0} together with the values of the topological tensor
invariants. As mentioned in the Introduction, the decay 
$ \Xi_{c}^0 $  $\to$  $ \Lambda_{c}^+ \pi^-$ has recently been
observed~\cite{Aaij:2020wtg}.

For the $\Xi_{c}^{+,0}$ and $\Omega^0$ decays the SU(3) decomposition reads
\begin{eqnarray}
  \label{decomp3}
  \overline{\bf 3} \to {\bf 8}\otimes\overline{\bf 3}\otimes {\bf 8}
  &=&3\cdot\overline{\bf 3}
  \oplus 3\cdot{\bf  6} \oplus 4\cdot\overline{\bf 15}\oplus
         \overline{\bf 15}\,'
  \oplus2 \cdot \overline{\bf 24}\oplus\overline{\bf 42} \\
{\bf 6} \to {\bf 8}\otimes\overline{\bf 3}\otimes {\bf 8}
  &=&3\cdot\overline{\bf 6}
  \oplus 3\cdot{\bf  6} \oplus 4\cdot\overline{\bf 15}\oplus
         \overline{\bf 15}\,'
  \oplus2 \cdot \overline{\bf 24}\oplus\overline{\bf 42}
\end{eqnarray}
One concludes that one should have four linear relations each among the
seven tensor invariants. With the help of tensor manipulations alone one
finds the relations $2I_3=-I_4$ for both transitions $\overline{\bf 3}
\stackrel{H({\bf 8})}{\longrightarrow}\overline{\bf 3}+{\bf 8}$ and
${\bf 6}\stackrel{H({\bf 8})}{\longrightarrow}\overline{\bf 3}+{\bf 8}$. The
data base in Table~\ref{deltaC=0} is not large enough to identify the
remaining tensor identities. From a group theoretical point of view one could,
of course, enlarge the data base by including also off-shell decays. We have
not pursued this possible avenue.

While there are three SU(3) reduced matrix elements each for the two classes
of decays there is only one isospin SU(2) reduced matrix element for each of
the two classes of decays. This becomes evident when doing the same exercise
as in Eq.~(\ref{decomp3}) but now for isospin SU(2). One finds
\begin{eqnarray}
{\bf 2}\to{\bf 2}\otimes{\bf 1}\otimes{\bf 3}={\bf 2}\oplus{\bf 4}&&
\mbox{(in spin notation
$\tfrac 12\to \tfrac 12 \otimes 0 \otimes 1= \tfrac 12 \oplus \tfrac 32$)},
  \nonumber\\
{\bf 1}\to{\bf 2}\otimes{\bf 2}\otimes{\bf 3}={\bf 1}\oplus 2\cdot{\bf 3}
  \oplus{\bf 5}&&
\mbox{($0 \to\tfrac 12\otimes\tfrac 12\otimes 1=0\oplus 2\cdot 1\oplus 2$)}.
\end{eqnarray}
This leads to the $\Delta I=1/2$ isospin relations
\begin{eqnarray}
\sqrt{2}\,M(\Xi_{c}^+\to\Lambda_{c}^+\pi^0)
  &=&M(\Xi_{c}^0\to\Lambda_{c}^+\pi^-)\\
\sqrt{2}\,M(\Omega_{c}^0\to\Xi_{c}^0\pi^0)
  &=&-M(\Omega_{c}^0\to\Xi_{c}^+\pi^-)
\end{eqnarray}
in agreement with the entries in Table~\ref{deltaC=0}.

\begin{table}
\caption{\label{cc3,bar3,8}
  Values of the seven topological tensor invariants for the CF, SCS and DCS
  decays of the triplet double charm baryon states $\Xi_{cc}^{++}$,
  $\Xi_{cc}^+$ and $\Omega_{cc}^+$ into the antitriplet single charm baryon
  states $\Lambda_{c}^+$, $\Xi_{c}^{+}$ and $\Xi_{c}^0$ belonging to the class
  of decays $B_{cc}({\bf 3}) \to B_{c}(\overline{\bf 3}) + M({\bf 8},{\bf 1})$. 
  The notation of the subscripts $a$ and $b$ is explained in
  Table~\ref{TableOmegac}.}
\begin{center}
\begin{tabular}{crclccccccc}
\hline \hline \\[-11pt]
 & & & & $I^-_1$ & $I^-_2$ & $I_3$ & $I_4$
 & $\hat{I}_3$ & $\hat{I}_4$ & $I_5$ \\ [3pt] \hline \\[-11pt]
%%% decay 1 %%%%%%%%%%%%%%%%%%%%%%%%%%%%%%%%%%%%%%%%%%%%%%%%%%%%%%%%%%%%%%%%%
CF&$2\sqrt{6} \,\Xi_{cc}^{++} $ & $\to$ & $ \Xi_{c}^+ \pi^+$
&$ -2 $&$ +1 $&$  0 $&$  0 $&$ -4 $&$ +4 $&$  0 $\\
%% decay 2
&$2\sqrt{6} \,\Xi_{cc}^{+} $ & $\to$ & $ \Xi_{c}^0 \pi^+$
&$ -2 $&$ +1 $&$ -4 $&$ +2 $&$  0 $&$  0 $&$  0 $\\
%% decay 3
&$4\sqrt{3} \,\Xi_{cc}^+ $ & $\to$ & $\Xi_{c}^+ \pi^0$
&$  0 $&$  0 $&$ +4 $&$ -2 $&$ -4 $&$ +4 $&$  0 $\\
%%% decay 4
&$4\sqrt{3} \,\Xi_{cc}^+  $ & $\to$ & $ \Xi_{c}^+\eta_\omega$
&$  0 $&$  0 $&$ +4 $&$ -2 $&$ +4 $&$ -4 $&$  0 $\\
%%% decay 5
&$2\sqrt{6} \,\Xi_{cc}^+ $ & $\to$ & $ \Xi_{c}^+\eta_\phi$
&$  0 $&$  0 $&$ +4 $&$ -2 $&$  0 $&$  0 $&$  0 $\\
%%% decay 6
&$12 \,\Xi_{cc}^+ $ & $\to$ & $ \Xi_{c}^+\eta_8$
&$  0 $&$  0 $&$ -4 $&$ +2 $&$ +4 $&$ -4 $&$  0 $\\
%%% decay 7
&$6\sqrt{2} \,\Xi_{cc}^+ $ & $\to$ & $ \Xi_{c}^+\eta_1$
&$  0 $&$  0 $&$ +8 $&$ -4 $&$ +4 $&$ -4 $&$  0 $\\
%%% decay 8
&$2\sqrt{6}\,\Xi_{cc}^+ $ & $\to$ & $ \Lambda_{c}^+ \bar K^0$
&$ -2 $&$ +1 $&$ -4 $&$ +2 $&$  0 $&$  0 $&$  0 $\\
%%% decay 9
&$2\sqrt{6} \,\Omega_{cc}^+ $ & $\to$ & $ \Xi_{c}^+ \bar K^0$
&$ -2 $&$ +1 $&$  0 $&$  0 $&$ -4 $&$ +4 $&$  0 $\\[3pt] \hline \\[-11pt]
%%% decay 1  %%%%%%%%%%%%%%%%%%%%%%%%%%%%%%%%%%%%%%%%%%%%%%%%%%%%%%%%%%%%%%%
SCS&$2\sqrt{6} \,\Xi_{cc}^{++} $ & $\to$ & $ \Xi_{c}^+ K^+$
&$ +2_a $&$ -1_a $&$  0 $&$  0 $&$ +4_a $&$ -4_a $&$  0 $\\
%%% decay 2
&$2\sqrt{6} \,\Xi_{cc}^{++} $ & $\to$ & $ \Lambda_{c}^+ \pi^+$
&$ +2_b $&$ -1_b $&$  0 $&$  0 $&$ +4_b $&$ -4_b $&$  0 $\\
%%% decay 3
&$2\sqrt{6} \,\Xi_{cc}^{+} $ & $\to$ & $ \Xi_{c}^0 K^+$
&$ +2_a $&$ -1_a $&$ +4_b $&$ -2_b $&$  0 $&$  0 $&$  0 $\\
%%% decay 4
&$4\sqrt{3} \,\Xi_{cc}^{+} $ & $\to$ & $ \Lambda_{c}^+ \pi^0$
&$ +2_b $&$ -1_b $&$  0 $&$  0 $&$ +4_b $&$ -4_b $&$  0 $\\
%%% decay 5
&$4\sqrt{3} \,\Xi_{cc}^{+} $ & $\to$ & $ \Lambda_{c}^{+} \eta_\omega$
&$ -2_b $&$ +1_b $&$ -8_b $&$ +4_b $&$ -4_b $&$ +4_b $&$  0 $\\
%%% decay 6
&$   2\sqrt{6} \,\Xi_{cc}^{+} $ & $\to$ & $ \Lambda_{c}^{+} \eta_\phi$
&$ +2_a $&$ -1_a $&$  0 $&$  0 $&$  0 $&$  0 $&$  0 $\\
%%% decay 7
&$12 \,\Xi_{cc}^{+} $ & $\to$ & $ \Lambda_{c}^{+} \eta_8$
&$ -4_a-2_b $&$ +2_a+1_b $&$ -8_b $&$ +4_b $&$ -4_b $&$ +4_b $&$  0 $\\
%%% decay 8
&$6\sqrt{2} \,\Xi_{cc}^{+} $ & $\to$ & $ \Lambda_{c}^{+} \eta_1$
&$ +2_a-2_b $&$ -1_a+1_b $&$ -8_b $&$ +4_b $&$ -4_b $&$ +4_b $&$  0 $\\
%%% decay 9
&$4\sqrt{3} \,\Omega_{cc}^{+} $ & $\to$ & $ \Xi_{c}^{+} \pi^0$
&$ +2_b $&$ -1_b $&$ +4_a $&$ -2_a $&$  0 $&$  0 $&$  0 $\\
%%% decay 10
&$4\sqrt{3} \,\Omega_{cc}^{+} $ & $\to$ & $ \Xi_{c}^{+} \eta_\omega$
&$ -2_b $&$ +1_b $&$ +4_a $&$ -2_a $&$  0 $&$  0 $&$  0 $\\
%%% decay 11
&$2\sqrt{6} \,\Omega_{cc}^{+} $ & $\to$ & $ \Xi_{c}^{+} \eta_\phi$
&$ +2_a $&$ -1_a $&$ +4_a $&$ -2_a $&$ +4_a $&$ -4_a $&$  0 $\\
%%%decay12
&$12\,\Omega_{cc}^{+} $ & $\to$ & $ \Xi_{c}^{+} \eta_8$
&$ -4_a-2_b $&$ +2_a+1_b $&$ -4_a $&$ +2_a $&$ -8_a $&$ +8_a $&$  0 $\\
%%%decay13
&$6\sqrt{2} \,\Omega_{cc}^{+} $ & $\to$ & $ \Xi_{c}^{+} \eta_1$
&$ +2_a-2_b $&$ -1_a+1_b $&$ +8_a $&$ -4_a $&$ +4_a $&$ -4_a $&$  0 $\\
%%%decay 14
&$2\sqrt{6} \,\Omega_{cc}^{+} $ & $\to$ & $ \Xi_{c}^{0} \pi^+$
&$ -2_b $&$ +1_b $&$ -4_a $&$ +2_a $&$  0 $&$  0 $&$  0 $\\
%%%decay15
&$2\sqrt{6} \,\Omega_{cc}^{+} $ & $\to$ & $ \Lambda_{c}^{+} \bar K^0$
&$  0 $&$  0 $&$ -4_a $&$ +2_a $&$ +4_b $&$ -4_b $&$  0 $\\
[3pt] \hline \\[-11pt]
%%% decay 1 %%%%%%%%%%%%%%%%%%%%%%%%%%%%%%%%%%%%%%%%%%%%%%%%%%%%%%%%%%%%%%%%%%
DCS&$2\sqrt{6} \,\Xi_{cc}^{++} $ & $\to$ & $ \Lambda_{c}^+ K^+$
&$ +2 $&$ -1 $&$  0 $&$  0 $&$ +4 $&$ -4 $&$  0 $\\
%%% decay 2
&$2\sqrt{6} \,\Xi_{cc}^{+} $ & $\to$ & $ \Lambda_{c}^+ K^0$
&$ +2 $&$ -1 $&$  0 $&$  0 $&$ +4 $&$ -4 $&$  0 $\\
%%% decay 3
&$2\sqrt{6} \,\Omega_{cc}^+ $ & $\to$ & $ \Xi_{c}^0 K^+$
&$ -2 $&$ +1 $&$ -4 $&$ +2 $&$  0 $&$  0 $&$  0 $\\
%%% decay 4
&$2\sqrt{6} \Omega_{cc}^+ $ & $\to$ & $ \Xi_{c}^+ K^0$
&$ +2 $&$ -1 $&$ +4 $&$ -2 $&$  0 $&$  0 $&$  0 $\\
%%% decay 5
&$4\sqrt{3} \,\Omega_{cc}^{+} $ & $\to$ & $ \Lambda_{c}^{+} \pi^0$
&$  0 $&$  0 $&$  0 $&$  0 $&$  0 $&$  0 $&$  0 $\\
%%% decay 6
&$4\sqrt{3} \,\Omega_{cc}^+ $ & $\to$ & $ \Lambda_{c}^+\eta_\omega$
&$  0 $&$  0 $&$ +8 $&$ -4 $&$  0 $&$  0 $&$  0 $\\
%%% decay 7
&$2\sqrt{6} \,\Omega_{cc}^+ $ & $\to$ & $ \Lambda_{c}^+\eta_\phi$
&$  0 $&$  0 $&$  0 $&$  0 $&$ +4 $&$ -4 $&$  0 $\\
%%% decay 8
&$\,12 \,\Omega_{cc}^+ $ & $\to$ & $ \Lambda_{c}^+\eta_8$
&$  0 $&$  0 $&$ +8 $&$ -4 $&$ -8 $&$ +8 $&$  0 $\\
%%% decay 9
&$6\sqrt{2} \,\Omega_{cc}^+ $ & $\to$ & $ \Lambda_{c}^+\eta_1$
&$  0 $&$  0 $&$ +8 $&$ -4 $&$ +4 $&$ -4 $&$  0 $\\[3pt]
\hline
\hline
\end{tabular}
\end{center}
\end{table}

%%%%%%%%%%%%%%%%%%%%%%%%%%%%%%%%%%%%%%%%%%%%%%%%%%%%%%%%%%%%%%%%%%%%%%%%%%%%%%%
\subsection{The double charm baryon decays
  $B_{cc}({\bf 3}) \stackrel{H({\bf 6})}{\longrightarrow}
  B_{c}(\overline{\bf 3}) + M({\bf 8,\,1})$}
%%%%%%%%%%%%%%%%%%%%%%%%%%%%%%%%%%%%%%%%%%%%%%%%%%%%%%%%%%%%%%%%%%%%%%%%%%%%%%%
The nonleptonic two-body decays of the triplet of double charm states
$\Xi_{cc}^{++}$, $\Xi_{cc}^{+}$ and $\Omega_{cc}^+$ into the antitriplet
single charm baryons belong to this class of decays. Of the many possible two
body decays only the decay $\Xi^{++}_{cc} \to \Xi_{c}^+ \,\pi^+$ has been
identified to date~\cite{Aaij:2018gfl}.

The ${\bf 3}$ representation occurs twice in the decomposition
\begin{equation}
  {\bf 3} \to {\bf 6}\otimes \overline{\bf 3}\otimes{\bf 8}
    = 2\cdot {\bf3}\oplus 2\cdot{\bf\overline{6}}\oplus
    3\cdot{\bf 15}
  \oplus{\bf 15'}\oplus{\bf 24}\oplus {\bf 42}
\end{equation}
There are thus $7-2=5$ relations for the tensor invariants which read
\begin{equation}
I^-_1+2I^-_2=0,\qquad I_3+2I_4=0,\qquad\hat{I}_3+\hat{I}_4=0,\qquad I_5=0,
  \qquad 2I^-_1=I_3+\hat I_3 
\end{equation}
As before, the last relation does not hold for the decay involving the octet
singlet state. For the transitions involving the singlet meson state one finds
\begin{equation}
  {\bf 3} \to {\bf 6}\otimes \overline{\bf 3}\otimes{\bf 1}
    = {\bf3}\oplus {\bf 15}
\end{equation}
which implies that all transitions involving the singlet meson are proportional
to each other with $I^-_1=I^-_2=I_5=0$ and $I_3=-2I_4=2\hat{I}_3=-2\hat{I}_4$.
In Table~\ref{cc3,bar3,8} we list the values of the tensor invariants for all
the decays of this class.

\begin{table}\kern-24pt
\caption{\label{Tablecc3,6,8}
  Values of the seven tensor invariants for the CF, SCS and DCS decays of the
  triplet double charm baryon states $\Xi_{cc}^{++}$, $\Xi_{cc}^+$, and
  $\Omega_{cc}^+$ into sextet single baryon states,
  $B_{cc}({\bf 3}) \to B_{c}({\bf 6})+ M({\bf 8})$. The subscripts $a$ and $b$
  for the SCS decays has been explained in Table~\ref{TableOmegac}. Decays
  into $\eta_1$ states are not listed since they vanish identically.}
\begin{center}
\begin{tabular}{crclccccccc}
\hline \hline \\[-11pt]
 & & & & $I^-_1$ & $I^-_2$ & $I_3$ & $I_4$
 & $\hat{I}_3$ & $\hat{I}_4$ & $I_5$ \\[3pt] \hline \\[-11pt]
%%% decay 1
CF&$2\sqrt{2} \,\Xi_{cc}^{++} $ & $\to$ & $ \Xi_{c}^{\prime+} \pi^+$
&$  0 $&$ -1 $&$  0 $&$  0 $&$  0 $&$  0 $&$  0 $\\
%%% decay 2
&$2 \,\Xi_{cc}^{++} $ & $\to$ & $ \Sigma_{c}^{++} \bar K^0$
&$  0 $&$ +1 $&$  0 $&$  0 $&$  0 $&$  0 $&$  0 $\\
%%% decay 3
&$2\sqrt{2} \,\Xi_{cc}^+ $ & $\to$ & $ \Xi_{c}^{\prime\,0} \pi^+ $
&$  0 $&$ -1 $&$  0 $&$ +2 $&$  0 $&$  0 $&$  0 $\\
%%% decay 4
&$4 \,\Xi_{cc}^+ $ & $\to$ & $ \Xi_{c}^{\prime+} \pi^0$
&$  0 $&$  0 $&$  0 $&$ -2 $&$  0 $&$  0 $&$  0 $\\
%%% decay 5
&$4 \,\Xi_{cc}^+ $ & $\to$ & $ \Xi_{c}^{\prime+} \eta_\omega$
&$  0 $&$  0 $&$  0 $&$ -2 $&$  0 $&$  0 $&$  0 $\\
%%% decay 6
&$2\sqrt{2} \,\Xi_{cc}^+ $ & $\to$ & $ \Xi_{c}^{\prime+} \eta_\phi$
&$  0 $&$  0 $&$  0 $&$ +2 $&$  0 $&$  0 $&$  0 $\\
%%% decay 7
&$4\sqrt{3} \,\Xi_{cc}^+ $ & $\to$ & $ \Xi_{c}^{\prime+} \eta_8$
&$  0 $&$  0 $&$  0 $&$ -6 $&$  0 $&$  0 $&$  0 $\\
%%% decay 8
&$2 \,\Xi_{cc}^+ $ & $\to$ & $ \Sigma_{c}^{++} K^-$
&$  0 $&$  0 $&$  0 $&$ +2 $&$  0 $&$  0 $&$  0 $\\
%%% decay 9
&$2\sqrt{2} \,\Xi_{cc}^+ $ & $\to$ & $ \Sigma_{c}^+\bar K^0$
&$  0 $&$ +1 $&$  0 $&$ -2 $&$  0 $&$  0 $&$  0 $\\
%%% decay 10
&$2 \,\Xi_{cc}^+ $ & $\to$ & $ \Omega_{c}^0 K^+$
&$  0 $&$  0 $&$  0 $&$ -2 $&$  0 $&$  0 $&$  0 $\\
%%% decay 11
&$2\sqrt{2} \,\Omega_{cc}^+ $ & $\to$ & $ \Xi_{c}^{\prime+} \bar K^0$
&$  0 $&$ +1 $&$  0 $&$  0 $&$  0 $&$  0 $&$  0 $\\
%%% decay 12
&$2 \,\Omega_{cc}^+ $ & $\to$ & $ \Omega_{c}^0 \pi^+$
&$  0 $&$ -1 $&$  0 $&$  0 $&$  0 $&$  0 $&$  0 $\\ [3pt]
\hline \\[-11pt] %%%%%%%%%%%%%%%%%%%%%%%%%%%%%%%%%%%%%%%%%%%%%%%%%%%%%%%%%%%%%
%%% decay 1
SCS&$2\sqrt{2} \,\Xi_{cc}^{++} $ & $\to$ & $ \Xi_{c}^{\prime+} K^+$
&$  0 $&$ +1_a $&$  0 $&$  0 $&$  0 $&$  0 $&$  0 $\\
%%% decay 2
&$2\sqrt{2} \,\Xi_{cc}^{++} $ & $\to$ & $ \Sigma_{c}^+ \pi^+$
&$  0 $&$ +1_b $&$  0 $&$  0 $&$  0 $&$  0 $&$  0 $\\
%%% decay 3
&$2\sqrt{2} \,\Xi_{cc}^{++} $ & $\to$ & $ \Sigma_{c}^{++} \pi^0$
&$  0 $&$ -1_b $&$  0 $&$  0 $&$  0 $&$  0 $&$  0 $\\
%%% decay 4
&$2\sqrt{2} \,\Xi_{cc}^{++} $ & $\to$ & $ \Sigma_{c}^{++} \eta_\omega$
&$  0 $&$ +1_b $&$  0 $&$  0 $&$  0 $&$  0 $&$  0 $\\
%%% decay 5
&$2 \,\Xi_{cc}^{++} $ & $\to$ & $ \Sigma_{c}^{++} \eta_\phi$
&$  0 $&$ -1_a $&$  0 $&$  0 $&$  0 $&$  0 $&$  0 $\\
%%% decay 6
&$2\sqrt{6} \,\Xi_{cc}^{++} $ & $\to$ & $ \Sigma_{c}^{++} \eta_8$
&$  0 $&$ +2_a+1_b $&$  0 $&$  0 $&$  0 $&$  0 $&$  0 $\\
%%% decay 7  leftout
%&$2\sqrt{3} \,\Xi_{cc}^{++} $ & $\to$ & $ \Sigma_{c}^{++} \eta_1$
%&$  0 $&$ -1_a+1_b $&$  0 $&$  0 $&$  0 $&$  0 $&$  0 $\\
%%% decay 8
&$2\sqrt{2} \,\Xi_{cc}^{+} $ & $\to$ & $ \Xi_{c}^{\prime \,0} K^+$
&$  0 $&$ +1_a $&$  0 $&$ +2_b $&$  0 $&$  0 $&$  0 $\\
%%% decay 9
&$2 \,\Xi_{cc}^{+} $ & $\to$ & $ \Sigma_{c}^0 \pi^+$
&$  0 $&$ +1_b $&$  0 $&$ -2_b $&$  0 $&$  0 $&$  0 $\\
%%% decay 10
&$4 \,\Xi_{cc}^{+} $ & $\to$ & $ \Sigma_{c}^+ \pi^0$
&$  0 $&$ -1_b $&$  0 $&$ +4_b $&$  0 $&$  0 $&$  0 $\\
%%% decay 11
&$4 \,\Xi_{cc}^{+} $ & $\to$ & $ \Sigma_{c}^{+} \eta_\omega$
&$  0 $&$ +1_b $&$  0 $&$  0 $&$  0 $&$  0 $&$  0 $\\
%%% decay 12
&$2\sqrt{2} \,\Xi_{cc}^{+} $ & $\to$ & $ \Sigma_{c}^{+} \eta_\phi$
&$  0 $&$ -1_a $&$  0 $&$  0 $&$  0 $&$  0 $&$  0 $\\
%%% decay 13
&$4\sqrt{3} \,\Xi_{cc}^{+} $ & $\to$ & $ \Sigma_{c}^{+} \eta_8$
&$  0 $&$ +2_a+1_b $&$  0 $&$  0 $&$  0 $&$  0 $&$  0 $\\
%%% decay 14 left out
%&$2\sqrt{6} \,\Xi_{cc}^{+} $ & $\to$ & $ \Sigma_{c}^{+} \eta_1$
%&$  0 $&$ -1_a+1_b $&$  0 $&$  0 $&$  0 $&$  0 $&$  0 $\\
%%% decay 15
&$2\sqrt{2} \,\Omega_{cc}^{+} $ & $\to$ & $ \Sigma_{c}^+ \bar{K}^0$
&$  0 $&$0 $&$ 0 $&$ -2_a $&$  0 $&$  0 $&$  0 $\\
%%% decay 16
&$2\sqrt{2} \,\Omega_{cc}^{+} $ & $\to$ & $ \Xi_{c}^{\prime\,0} \pi^+$
&$  0 $&$ +1_b $&$  0 $&$ +2_a $&$  0 $&$  0 $&$  0 $\\
%%% decay 17
&$2 \,\Omega_{cc}^{+} $ & $\to$ & $ \Omega_{c}^{0} K^+$
&$  0 $&$ +1_a $&$  0 $&$ -2_a $&$  0 $&$  0 $&$  0 $\\
%%% decay 18
%&$4 \,\Omega_{cc}^{+} $ & $\to$ & $ \Sigma_{c}^{+} \pi^0$
%&$  0 $&$  0 $&$  0 $&$  0 $&$  0 $&$  0 $&$  0 $\\
%%% decay 17
%&$4 \,\Omega_{cc}^{+} $ & $\to$ & $ \Sigma_{c}^{+} \eta_\omega $
%&$  0 $&$  0 $&$  0 $&$  0 $&$  0 $&$  0 $&$  0 $\\
%%% decay 18
%&$2\sqrt{2} \,\Omega_{cc}^{+} $ & $\to$ & $ \Sigma_{c}^{+} \eta_\phi$
%&$  0 $&$  0 $&$  0 $&$  0 $&$  0 $&$  0 $&$  0 $\\
%%% decay 19
%&$4\sqrt{3} \,\Omega_{cc}^{+} $ & $\to$ & $ \Sigma_{c}^{+} \eta_8$
%&$  0 $&$  0 $&$  0 $&$  0 $&$  0 $&$  0 $&$  0 $\\
%%% decay 20 left out
%&$2\sqrt{6} \,\Omega_{cc}^{+} $ & $\to$ & $ \Sigma_{c}^{+} \eta_1$
%&$  0 $&$  0 $&$  0 $&$  0 $&$  0 $&$  0 $&$  0 $\\
[3pt] \hline \\[-11pt] %%%%%%%%%%%%%%%%%%%%%%%%%%%%%%%%%%%%%%%%%%%%%%%%%%%%%%%%
%%% decay 1
DCS&$2\sqrt{2} \,\Xi_{cc}^{++} $ & $\to$ & $ \Sigma_{c}^{+} K^+$
&$ 0 $&$ +1 $&$  0 $&$  0 $&$  0 $&$  0 $&$  0 $\\
%%% decay 2
&$2 \,\Xi_{cc}^{++} $ & $\to$ & $ \Sigma_{c}^{++} K^0$
&$ 0 $&$ -1 $&$  0 $&$  0 $&$  0 $&$  0 $&$  0 $\\
%%% decay 3
&$2 \,\Xi_{cc}^+ $ & $\to$ & $ \Sigma_{c}^{0} K^+ $
&$ 0 $&$ +1 $&$  0 $&$  0 $&$  0 $&$  0 $&$  0 $\\
%%% decay 4
&$2\sqrt{2} \,\Xi_{cc}^+ $ & $\to$ & $ \Sigma_{c}^{+} K^0$
&$ 0 $&$ -1 $&$  0 $&$  0 $&$  0 $&$  0 $&$  0 $\\
%%% decay 5
&$2\sqrt{2} \,\Omega_{cc}^+ $ & $\to$ & $ \Xi_{c}^{\prime\,0} K^+$
&$ 0 $&$ +1 $&$  0 $&$ -2 $&$  0 $&$  0 $&$  0 $\\
%%% decay 6
&$2\sqrt{2} \,\Omega_{cc}^+ $ & $\to$ & $ \Xi_{c}^{\prime+} K^0$
&$ 0 $&$ -1 $&$  0 $&$ +2 $&$  0 $&$  0 $&$  0 $\\
%%% decay 7
&$4 \,\Omega_{cc}^+ $ & $\to$ & $ \Sigma_{c}^{+} \pi^0$
&$ 0 $&$  0 $&$  0 $&$ -4 $&$  0 $&$  0 $&$  0 $\\
%%% decay 8
&$4 \,\Omega_{cc}^+ $ & $\to$ & $ \Sigma_{c}^{+} \eta_\omega$
&$ 0 $&$  0 $&$  0 $&$  0 $&$  0 $&$  0 $&$  0 $\\
%%% decay 9
&$2\sqrt{2} \,\Omega_{cc}^+ $ & $\to$ & $ \Sigma_{c}^+\eta_\phi$
&$ 0 $&$  0 $&$  0 $&$  0 $&$  0 $&$  0 $&$  0 $\\
%%% decay 10
&$4\sqrt{3} \,\Omega_{cc}^+ $ & $\to$ & $ \Sigma_{c}^+ \eta_8$
&$ 0 $&$  0 $&$  0 $&$  0 $&$  0 $&$  0 $&$  0 $\\[3pt]
\hline
\hline
\end{tabular}
\end{center}
\end{table}

%%%%%%%%%%%%%%%%%%%%%%%%%%%%%%%%%%%%%%%%%%%%%%%%%%%%%%%%%%%%%%%%%%%%%%%%%%%%%%%
\subsection{The double charm baryon decays
  $B_{cc}({\bf 3}) \stackrel{H({\bf 6})}{\longrightarrow}  B_{c}({\bf 6})
  + M({\bf 8,\,1})$}
%%%%%%%%%%%%%%%%%%%%%%%%%%%%%%%%%%%%%%%%%%%%%%%%%%%%%%%%%%%%%%%%%%%%%%%%%%%%%%%
The nonleptonic two-body decays of the double charm states $\Xi_{cc}^{++}$,
$\Xi_{cc}^{+}$ and $\Omega_{cc}^+$ into the sextet single charm baryons belong
to this class of decays. Of the many possible decay modes none has been
identified to date.

The ${\bf 3}$ representation occurs twice in the decomposition
\begin{equation}
  \label{decomp5}
  {\bf 3} \to {\bf 6}\otimes {\bf 6}\otimes {\bf 8}= 2\cdot {\bf 3}
  \oplus2\cdot\overline{\bf 6}\oplus4\cdot{\bf 15}
  \oplus2\cdot {\bf 15'}\oplus2\cdot {\bf 24}\oplus2\cdot{\bf 42}
  \oplus{\bf 48}
\end{equation}
Therefore, there are $7-2=5$ relations among the seven tensor invariants
which can be derived by either tracking the charm quark flavor flow in the
respective topological diagrams or by taking into account that the single
charm baryons $B_{c}({\bf 6})$ are symmetric in the light quark indices.
The relations read
\begin{equation}
I^-_1=I_3=\hat{I}_3=\hat{I}_4=I_5=0,
\end{equation}
i.e.\ the only nonvanishing tensor invariants are $I^-_2$ and $I_4$. In
particular, the tensor invariants $\hat{I}_3$ and $\hat{I}_4$ are zero because
of the KPW theorem. The KPW theorem holds only in the SU(3) limit. Violations
of the KPW theorem due to the SU(3) constituent mass breaking effect
$m_s \neq m_u$ were calculated in Ref.~\cite{Gutsche:2018msz} for the decays
$\Xi_{cc}^{++} \to \Xi_{c}^{\prime+} \pi^+$
and $\Omega_{cc}^+ \to \Xi_{c}^{\prime+} \bar K^0$ and turned out to be of the
order of $(1-4)\%$. The authors of Ref.~\cite{Cheng:2020wmk} obtained a mass
breaking effect of the KPW theorem of the order of 2$\%$ in their bag model
calculation of the same decays.

The singlet meson state ${\bf 1}$ does not contribute to this class of decays
as can be seen from the decomposition
\begin{equation}
  {\bf 3} \to {\bf 6}\otimes {\bf 6}\otimes {\bf 1}
  =\overline{\bf 6}\oplus{\bf 15}\oplus {\bf 15'}.
\end{equation}
In Table~\ref{Tablecc3,6,8} we list the values of the seven topological flavor
invariants.

\begin{table}
\caption{\label{TableBccBD}Values of the seven topological tensor invariants
  for the nonleptonic CF, SCS and DCS decays $B_{cc}({\bf 3}) \to B({\bf 8})
  +M(\overline{\bf 3})$ of the triplet double charm baryon states
  $\Xi_{cc}^{++}$, $\Xi_{cc}^{+}$ and $\Omega^+_{cc}$ into the light baryon
  octet and the charm mesons $D^+$, $D^0$ and $D_s^+$. The notation for the
  SCS decays is explained in the caption of Table~\ref{TableOmegac}.}
\begin{center}
\begin{tabular}{crclccccccc}
\hline \hline \\[-11pt]
 & & & & $I_1^{-}$ & $I^-_2$ & $I_3$ & $I_4$
 & $\hat{I}_3$ & $\hat{I}_4$ & $I_5$ \\[3pt] \hline \\ [-11pt]
%% decay 1
CF&$2 \Xi_{cc}^{++} $ & $\to$ & $ \Sigma^+ D^+$
&$  0 $&$  0 $&$  0 $&$  0 $&$  0 $&$ -2 $&$  0 $\\
%% decay 2
&$2 \Xi_{cc}^{+} $ & $\to$ & $ \Sigma^+ D^0$
&$  0 $&$  0 $&$  0 $&$  0 $&$  0 $&$  0 $&$ +1 $\\
%% decay 3
&$2 \sqrt{2}\Xi_{cc}^{+} $ & $\to$ & $ \Sigma^0 D^+$
&$  0 $&$  0 $&$  0 $&$  0 $&$  0 $&$ -2 $&$ -1 $\\
%% decay 4
&$2 \sqrt{6}\Xi_{cc}^{+} $ & $\to$ & $ \Lambda^0 D^+$
&$  0 $&$  0 $&$  0 $&$  0 $&$  0 $&$ -2 $&$ +1 $\\
%% decay 5
&$2 \Xi_{cc}^{+} $ & $\to$ & $ \Xi^0 D_s^+$
&$  0 $&$  0 $&$  0 $&$  0 $&$  0 $&$  0 $&$ +1 $\\
%% decay 6
&$2 \Omega_{cc}^{+} $ & $\to$ & $ \Xi^0 D^+$
&$  0 $&$  0 $&$  0 $&$  0 $&$  0 $&$ -2 $&$  0 $\\
[3pt] \hline\\[-11pt] %%%%%%%%%%%%%%%%%%%%%%%%%%%%%%%%%%%%%%%%%%%%%%%%%%%%%%%%
%% decay 7 
SCS&$2 \Xi_{cc}^{++} $ & $\to$ & $ p D^+$
&$  0 $&$  0 $&$  0 $&$  0 $&$  0 $&$ +2_b $&$  0   $\\
%% decay 8
&$2 \Xi_{cc}^+ $ & $\to$ & $ p D^0$
&$  0 $&$  0 $&$  0 $&$  0 $&$  0 $&$  0   $&$ -1_b $\\
%% decay 9
&$2 \Xi_{cc}^+ $ & $\to$ & $ n D^+$
&$  0 $&$  0 $&$  0 $&$  0 $&$  0 $&$ +2_b $&$ +1_b $\\
%% decay 10
&$2 \Xi_{cc}^{++} $ & $\to$ & $ \Sigma^+ D_s^+$
&$  0 $&$  0 $&$  0 $&$  0 $&$  0 $&$ +2_a $&$  0   $\\
%% decay 11
&$\sqrt{2} \Xi_{cc}^+ $ & $\to$ & $ \Sigma^0 D_s^+$
&$  0 $&$  0 $&$  0 $&$  0 $&$  0 $&$ +2_a $&$  0   $\\
%% decay 12
&$2\sqrt{6} \Xi_{cc}^+ $ & $\to$ & $ \Lambda^0 D_s^+$
&$  0 $&$  0 $&$  0 $&$  0 $&$  0 $&$ +2_a $&$ -2_b $\\
%% decay 13
&$2 \Omega_{cc}^+ $ & $\to$ & $ \Sigma^+ D^0$
&$  0 $&$  0 $&$  0 $&$  0 $&$  0 $&$  0   $&$ +2_a $\\
%% decay 14
&$2 \sqrt{2}\Omega_{cc}^+ $ & $\to$ & $ \Sigma^0 D^+$
&$  0 $&$  0 $&$  0 $&$  0 $&$  0 $&$  0   $&$ -1_a $\\
%% decay 15
&$2 \sqrt{6}\Omega_{cc}^+ $ & $\to$ & $ \Lambda^0 D^+$
&$  0 $&$  0 $&$  0 $&$  0 $&$  0 $&$ +4_b $&$ -1_a $\\
%% decay 16
&$2 \Omega_{cc}^+ $ & $\to$ & $ \Xi^0 D_s^+$
&$  0 $&$  0 $&$  0 $&$  0 $&$  0 $&$ +2_a $&$ +1_a $\\
[3pt] \hline \\[-11pt] %%%%%%%%%%%%%%%%%%%%%%%%%%%%%%%%%%%%%%%%%%%%%%%%%%%%%%%
%% decay 17
DCS&$2 \Xi_{cc}^{++} $ & $\to$ & $ p D_s^+$
&$  0 $&$  0 $&$  0 $&$  0 $&$  0 $&$ +2 $&$  0 $\\
%% decay 18
&$2 \Xi_{cc}^{+} $ & $\to$ & $ n D_s^+$
&$  0 $&$  0 $&$  0 $&$  0 $&$  0 $&$ +2 $&$  0 $\\
%% decay 19
&$2 \Omega_{cc}^{+} $ & $\to$ & $ p D^0$
&$  0 $&$  0 $&$  0 $&$  0 $&$  0 $&$  0 $&$ +1 $\\
%% decay 20
&$2 \Omega_{cc}^{+} $ & $\to$ & $ n D^+$
&$  0 $&$  0 $&$  0 $&$  0 $&$  0 $&$  0 $&$ +1 $\\
%% decay 21
&$2\sqrt{2} \Omega_{cc}^{+} $ & $\to$ & $ \Sigma^0 D_s^+$
&$  0 $&$  0 $&$  0 $&$  0 $&$  0 $&$  0 $&$  0 $\\
%% decay 22
&$2\sqrt{6} \Omega_{cc}^{+} $ & $\to$ & $ \Lambda^0 D_s^+$
&$  0 $&$  0 $&$  0 $&$  0 $&$  0 $&$ +4 $&$ +2 $\\[3pt]
\hline
\hline
\end{tabular}
\end{center}
\end{table}

%%%%%%%%%%%%%%%%%%%%%%%%%%%%%%%%%%%%%%%%%%%%%%%%%%%%%%%%%%%%%%%%%%%%%%%%%%%%%%%
\subsection{The double charm baryon decays
  $B_{cc}({\bf 3}) \stackrel{H({\bf 6})}{\longrightarrow}  B({\bf 8})
  + M(\overline{\bf 3})$}
%%%%%%%%%%%%%%%%%%%%%%%%%%%%%%%%%%%%%%%%%%%%%%%%%%%%%%%%%%%%%%%%%%%%%%%%%%%%%%%
None of this class of decays into a light baryon and a charm meson in the
final state has been observed. The final states in these decays have a very
distinct signature which could lead to their discovery in the not too distant
future. To our knowledge the only theoretical paper dealing with these decays
is Ref.~\cite{Li:2020qrh}.
As a seed for the $B_{cc}({\bf 3})\to B({\bf 8})+M(\overline{\bf 3})$ decay,
Li et al.\ first consider the short distance decay of a double charm baryon
state into a single charm baryon and a $C=0$ meson~\cite{Li:2020qrh}. The
charm quantum number is then transferred from the baryon to the meson by long
distance final state interactions. The authors of Ref.~\cite{Li:2020qrh}
estimate that the CF decays of this class of decays could be of the order of
$1\%$ with the SCS and DCS decays suppressed by the respective CKM suppression
factors.

For the SU(3) decomposition one obtains
\begin{equation}
  {\bf 3} \to {\bf 6}\otimes {\bf 8}\otimes\overline{\bf 3}
    = 2\cdot {\bf3}\oplus 2\cdot{\bf\overline{6}}\oplus
    3\cdot{\bf 15}
    \oplus{\bf 15'}\oplus{\bf 24}\oplus {\bf 42}
\end{equation}
The decomposition shows that there are two independent SU(3) invariants.
Therefore, one expects $7-2=5$ relations among the seven topological tensor
invariants. By visual inspection of the five topologies one finds that
$I^-_1=I^-_2=I_3=I_4=0$. The topological invariant
$\hat{I}_3(\ell,\ell')=B_\ell^{a[bc]}B^{\ell'}_{a[b'c']}M^{c'}_{d}
H^{[c'b']}_{[db]}$ vanishes since $b'$ and $c'$ are heavy charm quark labels
which are necessarily symmetric such that the initial state tensor
$B^{\ell'}_{a[b'c']}$ vanishes. Altogether one has
\begin{equation}
I^-_1=I^-_2=I_3=I_4=\hat I_3=0.
\end{equation}
The only nonvanishing topological tensor invariants are $\hat{I}_4$ and $I_5$.
In Table~\ref{TableBccBD} we list the values of the nonzero tensor invariants
$\hat{I}_4$ and $I_5$ for the altogether 22 CF, SCS and DCS double charm
decays $B_{cc} \to  B\,D^{+/0}, B\,D_s^+$. Note that the decay
$\Omega_{cc}^{+} \to \Sigma^0 D_s^+$ is forbidden in the SU(3) limit because
of the KPW theorem.

\begin{table}
\caption{\label{hyperon}Values of the seven SU(3) tensor invariants for the
  nonleptonic hyperon decays induced by the transitions $(s \to u;\,u\to d)$.}
\begin{center}
\begin{tabular}{rclccccccccc}
\hline \hline \\[-11pt]
   && & $I^-_1$ & $I^-_2$ & $I_3$ & $I_4$ & $\hat{I}_3$ & $\hat{I}_4$
   & $I_3-\hat{I}_3$ & $I_4-\hat{I}_4$  & $I_5$ \\[3pt] \hline \\[-11pt]
%% decay 1
$2\sqrt{6} \,\Lambda^0$ & $\to$ & $  p \pi^-$
&$ +1 $&$ +1 $&$  0 $&$ -2 $&$ +2 $&$ -4 $&$ -2 $&$ +2 $&$ -1 $\\ 
%% decay 2
$4\sqrt{3} \,\Lambda^0$ & $\to$ & $n \pi^0$
&$ -1 $&$ -1 $&$  0 $&$ +2 $&$ -2 $&$ +4 $&$ +2 $&$ -2 $&$ +1 $\\
%% decay 3
$2\sqrt{2} \, \Sigma^+$ & $\to$ & $p \pi^0$
&$ -1 $&$ +1 $&$  0 $&$ -2 $&$ -2 $&$  0 $&$ +2 $&$ -2 $&$ -1 $\\ 
%% decay 4
$2\,\Sigma^+$ & $\to$ & $n \pi^+$
&$  0 $&$  0 $&$  0 $&$ +2 $&$  0 $&$  0 $&$  0 $&$ +2 $&$ +1 $\\ 
%% decay 5
$2\,\Sigma^-$ & $\to$ & $n  \pi^-$
&$ -1 $&$ +1 $&$  0 $&$  0 $&$ -2 $&$  0 $&$ +2 $&$  0 $&$  0 $\\ 
%% decay 6
$2\sqrt{6} \,\Xi^-$ & $\to$ & $\Lambda^0 \pi^-$
&$ -2 $&$ +1 $&$  0 $&$  0 $&$ -4 $&$ +4 $&$ +4 $&$ -4 $&$  0 $\\ 
%% decay 7
$4\sqrt{3} \,\Xi^0$ & $\to$ & $\Lambda^0 \pi^0$
&$ -2 $&$ +1 $&$  0 $&$  0 $&$ -4 $&$ +4 $&$ +4 $&$ -4 $&$  0 $\\[3pt]
\hline
\hline
\end{tabular}
\end{center}
\end{table}

%%%%%%%%%%%%%%%%%%%%%%%%%%%%%%%%%%%%%%%%%%%%%%%%%%%%%%%%%%%%%%%%%%%%%%%%%%%%%%%
\subsection{The hyperon decays
  $B(8) \stackrel{H(8)}{\longrightarrow}  B(8) + M(8)$}
%%%%%%%%%%%%%%%%%%%%%%%%%%%%%%%%%%%%%%%%%%%%%%%%%%%%%%%%%%%%%%%%%%%%%%%%%%%%%%%
In complete the picture obtained so far, we also consider the decays of
hyperons into light baryons. The decomposition
\begin{equation}\label{decomphyp}
{\bf 8}\to {\bf 8}\otimes{\bf 8}\otimes{\bf  8}
  =2\cdot {\bf 1} \oplus8\cdot {\bf 8}\oplus 4\cdot{\bf 10} \oplus
  4\cdot{\bf\overline{10}}
  \oplus6\cdot{\bf 27}\oplus2\cdot{\bf 35}\oplus2\cdot\overline{\bf 35}
  \oplus {\bf 64}
\end{equation}
suggests that there are eight SU(3) invariant couplings which outnumber the
seven invariant tensor couplings by one. However, one with the help of a
Fierz-type identity one can show that the eight SU(3) couplings are not
independent.

To proof this, consider the usual second rank tensor baryon representation
\begin{equation}
B=\left(\begin{array}{ccc}
    -\frac{\Lambda^0}{\sqrt{6}}+\frac{\Sigma^0}{\sqrt{2}}& -\Sigma^+ & p \\
    \Sigma^- &-\frac{\Lambda^0}{\sqrt{6}}-\frac{\Sigma^0}{\sqrt{2}}  & n \\
    \Xi^- & -\Xi^0 & \frac{2\Lambda^0}{\sqrt{6}}\\
   \end{array}\right),
\end{equation}
together with the representations for mesons and the weak transition tensor
\begin{equation}
\bar M=\left(\begin{array}{ccc}
  \frac{\eta_8}{\sqrt6}+\frac{\pi^0}{\sqrt2} & \pi^- & K^- \\
  -\pi^+ & \frac{\eta_8}{\sqrt6}-\frac{\pi^0}{\sqrt2} & -\bar K^0 \\
  K^+ & K^0 & -\frac{2\eta_8}{\sqrt6}\\
  \end{array}\right),\qquad
S=\left(\begin{array}{ccc}
    0& 0 & 0 \\
    0& 0 & 1 \\
    0& 0 & 0 \\
   \end{array}\right).
\end{equation}
In SU(3) a Levi-Civita tensor with four indices in three flavor dimensions
vanishes. Therefore, one can write down a Fierz-type identity which reads
\begin{equation}\label{fierzident}
\epsilon_{abcd}\epsilon^{efgh}=\left|
  \begin{array}{cccc}
    \delta_a^e & \delta_a^f & \delta_a^g & \delta_a^h\\
    \delta_b^e & \delta_b^f & \delta_b^g & \delta_b^h\\
    \delta_{c}^e & \delta_{c}^f & \delta_{c}^g & \delta_{c}^h\\
    \delta_d^e & \delta_d^f & \delta_d^g & \delta_d^h\\
   \end{array}\right|=0.
\end{equation}
When one contracts the Fierz-type identity with
$B^a_e\bar B^b_f\bar M^c_g S^d_h$ one obtains the relation
\begin{eqnarray}\label{FI2}
0&=&\tr(B \bar B \bar M S)+\tr(B \bar B S \bar M)+\tr(B \bar M S \bar B)
  +\tr(B \bar M \bar B S)+\tr(B S \bar M \bar B)+\strut\nonumber\\&&\strut
  +\tr(B S \bar B \bar M)-\tr(B \bar B)\tr(\bar M S)
  -\tr(B \bar M)\tr(\bar B S)-\tr(B S)\tr(\bar B \bar M).
\end{eqnarray}
We conjecture that the use of this Fierz-type identity would provide for the
missing linear relations between tensor invariants of differing topologies.
In Table~\ref{hyperon} one can identify the $\Delta I=1/2$ (or octet)
rules~\cite{Gaillard:1974nj,Altarelli:1974exa}
\begin{eqnarray}
{\cal A}(\Lambda^0 \to p \pi^-)&=&-\sqrt{2}{\cal A}(\Lambda^0 \to n \pi^0) \\
\sqrt{2}{\cal A}(\Sigma^+ \to p \pi^0)
  &=& -{\cal A}(\Sigma^+ \to n \pi^+) + {\cal A}(\Sigma^- \to n  \pi^-) \\
{\cal A}(\Xi^- \to \Lambda^0 \pi^-)
  &=&\sqrt{2}{\cal A}(\Xi^0 \to \Lambda^0 \pi^0).
\end{eqnarray}
However, for some reason the Lee--Sugawara
relation~\cite{Lee:1964zzc,Sugawara:1964zz} is not satisfied,
%! klaeren!
\begin{equation}
2A(\Xi^- \to \Lambda^0 \pi^-)+A(\Lambda^0 \to p \pi^-)
  \ne \sqrt{3}A(\Sigma^+ \to p \pi^0)
\end{equation}
Topological tensor invariants for this class of processes have been considered
in Ref.~\cite{Wang:2019alu}.

%%%%%%%%%%%%%%%%%%%%%%%%%%%%%%%%%%%%%%%%%%%%%%%%%%%%%%%%%%%%%%%%%%%%%%%%%%%%%%%
\section{\label{current}\bf The current algebra description of\\
  nonleptonic charm baryon decays}
%%%%%%%%%%%%%%%%%%%%%%%%%%%%%%%%%%%%%%%%%%%%%%%%%%%%%%%%%%%%%%%%%%%%%%%%%%%%%%%
In the second part of this general survey we add elements of the current
algebra approach to the information contained in the tables. The current
algebra approach was originally developed for the description of nonleptonic
hyperon decays (see Refs.~\cite{Weinberg:1966fm,Bjorken:1968ej,Marshak:1969,%
Gronau:1972pj} and references therein) and later applied to nonleptonic decays
of charmed baryons~\cite{Hussain:1983pk,Hussain:1984ma}. As will be described
in the following it turns out that the flavor invariants of the $W$-exchange
contributions to the current algebra amplitudes can be expressed in terms of
the topological flavor invariants $I_i$ ($i=3,4,5$) and $\hat{I}_i$ ($i=3,4$)
introduced in Sec.~\ref{topotensor}. We briefly recapitulate the standard
current algebra plus soft pion approach to nonleptonic two-body charm baryon
decays $B(1/2^+) \to B(1/2^+) + M(0^-)$ where we will stay quite close to the
presentation and notation of Refs.~\cite{Cheng:2018hwl,Cheng:2020wmk,%
Zou:2019kzq,Meng:2020euv,Hu:2020nkg}.

We define $S$- and $P$-wave amplitudes $A_{fki}$ and $B_{fki}$, resp., by
writing
\begin{equation}
\label{AandB}
\langle B_fM_k|{\cal H}_{\rm eff}|B_i\rangle
  =\bar{u}_f(A_{fki} - B_{fki}\gamma_5)u_i.
\end{equation}
We follow the convention of Ref.~\cite{Cheng:2018hwl,Cheng:2020wmk,%
Zou:2019kzq,Meng:2020euv,Hu:2020nkg} in that $A_{fki}$ and $B_{fki}$ are
defined with a relative minus sign in Eq.~(\ref{AandB}). Note that the choice
of this sign will affect the sign of the calculated asymmetry parameter of the
nonleptonic two-body decays.

Again in the notation of Ref.~\cite{Cheng:2018hwl,Cheng:2020wmk,Zou:2019kzq,%
Meng:2020euv,Hu:2020nkg} the p.v.\ and p.c.\ amplitudes $A$ and $B$ are given
by 
\begin{equation}
A_{fki}^{\phantom{a}}=A_{fki}^{\rm fac}+A_{fki}^{\rm pole}+A_{fki}^{\rm com},
  \qquad
B_{fki}^{\phantom{a}}=B_{fki}^{\rm fac}+B_{fki}^{\rm pole}+B_{fki}^{\rm com}.
\end{equation}
The factorizing contributions $A_{fki}^{\rm fac}$ and $B_{fki}^{\rm fac}$
related to the topological invariants $I^-_1$ and $I^-_2$ are complemented by
the nonfactorizing contributions, consisting of a pole and a commutator part.

%%%%%%%%%%%%%%%%%%%%%%%%%%%%%%%%%%%%%%%%%%%%%%%%%%%%%%%%%%%%%%%%%%%%%%%%%%%%%%%
\subsection{The parity violating $S$-wave amplitude $A_{fki}^{\rm com}$}
%%%%%%%%%%%%%%%%%%%%%%%%%%%%%%%%%%%%%%%%%%%%%%%%%%%%%%%%%%%%%%%%%%%%%%%%%%%%%%%
For the commutator contribution one obtains
\begin{equation}
\label{com}
 A_{fki}^{\rm com} = \frac{\sqrt{2}}{f_k}
 \langle B_f|[M_k,{\cal H}^{\rm pc}_{\rm eff}]|B_i\rangle,
\end{equation}
where ${\cal H}^{\rm pc}_{\rm eff}$ is the p.c.\ part of the effective
Hamiltonian, $M_k$ is the SU(3) vector charge associated with the pseudoscalar
meson $k$, and where the $f_k$ are the pseudoscalar coupling constants, with
e.g.\ $f_\pi=0.95\,m_\pi$. One then introduces a sum over intermediate baryon
states $\sum_\ell|B_\ell\rangle\langle B_\ell|$ and
$\sum_{\ell'}|B_{\ell'}\rangle\langle B_{\ell'}|$ to rewrite Eq.~(\ref{com}) as
\begin{eqnarray}\label{Acom}
A^{\rm com}_{fki} &=& \frac{\sqrt{2}}{f_k}
  \Big(\sum_\ell\langle B_f|M_k|B_\ell\rangle
  \langle B_\ell|{\cal H}^{\rm pc}_{\rm eff}|B_i\rangle
  -\sum_{\ell'}\langle B_f|{\cal H}^{\rm pc}_{\rm eff}|B_{\ell'}\rangle
  \langle B_{\ell'}|M_k|B_i\rangle\Big)\ =\nonumber\\
  &=&A_{fki}^{\rm com}(s)- A_{fki}^{\rm com}(u)
\end{eqnarray}
Since $M_k$ is a conserved vector charge operator, the sum over intermediate
states extends only over the twenty $J^p=1/2^{\,+}$ ground state baryons of
the SU(4) ${\bf 20'}$ representation.

\begin{figure}\begin{center}
\epsfig{figure=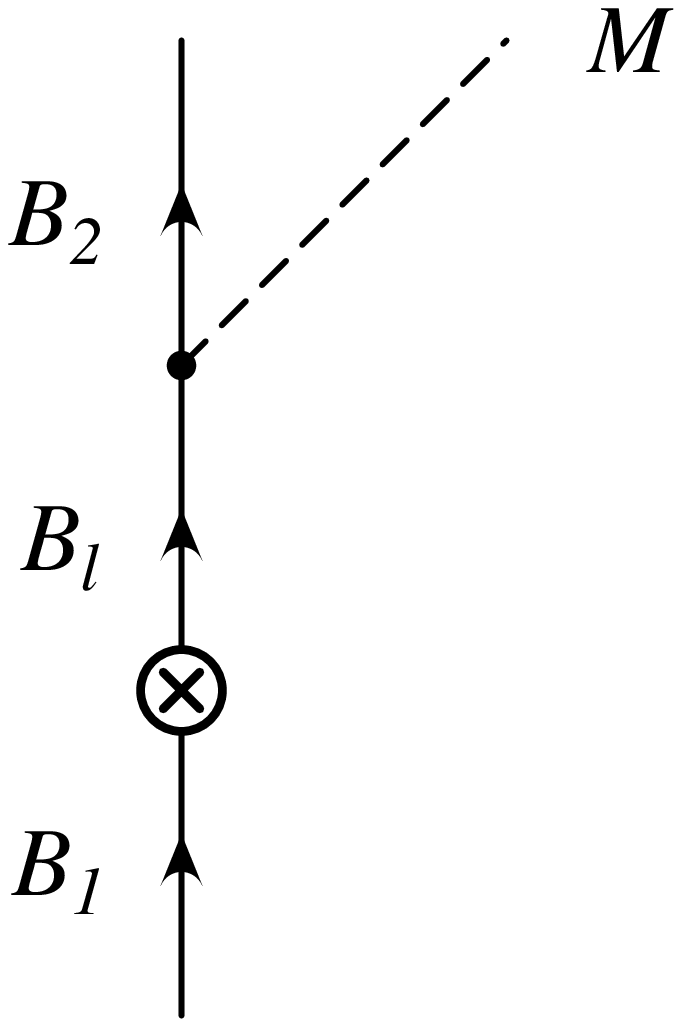, scale=0.40}\qquad
\epsfig{figure=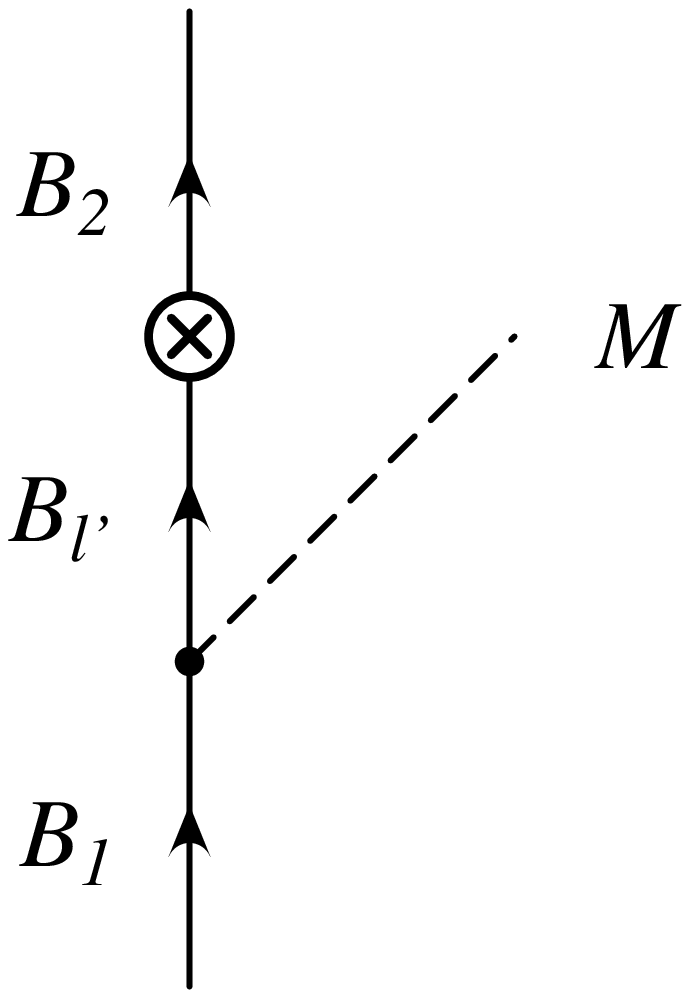, scale=0.40}\end{center}
\caption{\label{contri} $s$-channel contribution (left) and $u$-channel
  contribution (right)}
\end{figure}

The notation ``$s$'' and ``$u$'' stands for the $s$- and $u$-channel
contributions in the current algebra approach where the $s$-channel
contribution refers to the process where one has a weak transition at the
first stage followed by meson emission and vice-versa for the $u$-channel
contribution, as depicted in Fig.~\ref{contri}. One can associate the
topological diagrams with the $s$-channel and $u$-channel contributions by
cutting the diagrams IIa, IIb and III in Fig.~\ref{topo} at the appropriate
places. Diagrams IIa and IIb are clearly associated with the $s$- and
$u$-channel contributions, respectively, whereas diagram III can be seen to be
associated with both $s$- and $u$-channel contributions depending on where
diagram III is cut. As it turns out, diagram III contributes in equal amounts
to the $s$- and $u$-channel parts for both $S$- and $P$-wave transitions.
Altogether the $s$-channel contribution correspond to the contribution of
diagram IIa and one half of diagram III, while the $u$-channel contribution
corresponds to diagram IIb and one half of diagram III.

Our aim is to proof that the commutator contributions $A_{fki}^{\rm com}(s)$
and $A_{fki}^{\rm com}(u)$ can be rewritten in terms of a linear superposition
of the topological tensor invariants $I_3$, $I_4$, $\hat{I}_3$, $\hat{I}_4$
and $I_5$ introduced in Sec.~\ref{topotensor}. This is achieved by making use
of the completeness relation~(\ref{completeness}). The same can be done for the
pole term contributions $A_{kfi}^{\rm pole}(s)$ and  $A_{fki}^{\rm pole}(u)$
in the absence of hyperfine mass splittings for the flavor degenerate members
of the ${\bf 20'}$ representation, as will be written down later.

In Appendix~\ref{weak-strong} we demonstrate that the $f$-type baryon matrix
element $\langle B_f|M_k|B_\ell\rangle$ of the conserved vector charge can be
expressed in terms of two basic SU(3) contractions $\tilde I_1$ and
$\tilde I_2$. One has
\begin{equation}
\langle B_f|M_k|B_\ell\rangle:=(I^f)_{fk\ell}
  =4(\tilde I_1)_{fk\ell}+2(\tilde I_2)_{fk\ell}
\end{equation}
and similarly for $\langle B_{\ell'}|M_k|B_i\rangle$. The two basic tensor
contractions $\tilde I_1$ and $\tilde I_2$ are given by
\begin{equation}
\label{tildeI12}
(\tilde I_1)_{fk\ell}=B_f^{a[bc]} B^\ell_{a[bc']} (M_k)^{c'}_{c} \quad
(\tilde I_2)_{fk\ell}=B_f^{a[bc]} B^\ell_{b[c'a]} (M_k)^{c'}_{c}.
\end{equation}

The matrix element $\langle B_\ell|{\cal H}^{\rm pc}_{\rm eff}|B_i\rangle$ of
the effective Hamiltonian (called $a_{\ell\,i}$ in Ref.~\cite{Zou:2019kzq})
splits into a dynamical piece and a symmetry factor. In the bag model
calculation the factors $a_{\ell\,i}$ can be expressed in terms of a single
tensor contraction. In the normalization of Ref.~\cite{Zou:2019kzq} one writes
\begin{equation}
\langle B_\ell |{\cal H}^{\rm pc}_{\rm eff}|B_{\ell'}\rangle
  := a_{\ell\ell'} = 6\bar X_2(4\pi)(I^{\rm pc})_{\ell\ell'}
\end{equation}
with
\begin{equation}
(I^{\rm pc})_{\ell\ell'}=B_\ell^{a[bc]} B^{\ell'}_{a[b'c']}H^{[b'c']}_{[bc]}.
\end{equation}
In addition, one has
\begin{equation}
A_{fki}^{\rm com}(s)-A_{fki}^{\rm com}(u)=\frac{1}{f_k}6\bar X_2(4\pi)
  \left(\hat A_{fki}^{\rm com}(s)-\hat A_{fki}^{\rm com}(u)\right) 
\end{equation}
with
\begin{equation}
\hat A_{fki}^{\rm com}(s)-\hat A_{fki}^{\rm com}(u)
  =\sum_\ell (I^f)_{fk\ell}(I^{\rm pc})_{\ell\,i}
  -\sum_{\ell'}(I^{\rm pc})_{f\ell'}(I^f)_{\ell'ki}.
\end{equation}
In Ref.~\cite{Zou:2019kzq} the values of the bag integrals as e.g.\ $X_2$ are
given as dimensionless numbers. This is puzzling at first sight since the
amplitude $A^{\rm com}_{fki}$ is dimensionless. This conundrum is resolved
after a literature search. As stated some time ago in Ref.~\cite{Cheng:1991sn},
the values of the bag integrals are given in units of $c_-\,G_F\,\GeV^3$
which should be augmented by the CKM factor $C_{\rm CKM}$.

The relation to the topological tensor invariants is obtained by using the
completeness relation~(\ref{completeness}) in order to perform the sum over
the intermediate baryon states in Eq.~(\ref{Acom}). We begin with the
$s$-channel contribution $A_{fki}^{\rm com}(s)$ for which we go through the
derivation step by step. As discussed in Appendix~\ref{weak-strong}, the
flavor invariant contribution $I^f$ can be expressed as a linear superposition
of the two building blocks $\tilde I_1$ and $\tilde I_2$ in the form
$I^f=4\tilde I_1+2\tilde I_2$. Let us write out the contributions of the two
building blocks $\tilde I_1$ and $\tilde I_2$ to the sum over intermediate
states in $A_{fki}^{\rm com}(s)$. The result of taking the sum over the
intermediate states via the completeness relation~(\ref{completeness}) can be
expressed in terms of the topological tensor invariants $I_3$, $I_4$ and
$I_5$. One has
\begin{eqnarray}\label{istate}
\sum_\ell(\tilde I_1)_{fk\ell}I^{\rm pc}_{\ell\,i}
  &=&B_f^{a[bc]}(M_k)^{c'}_{c} \Big(\sum_\ell B^\ell_{a[bc']}  
  B_\ell^{r[st]}\Big) B^i_{r[a'b']}H^{[a'b']}_{[st]}
  =\tfrac23 I_3-\tfrac13 I_4+\tfrac 23 I_5,\nonumber\\
\sum_\ell(\tilde I_2)_{fk\ell}I^{\rm pc}_{\ell\,i}
  &=&B_f^{a[bc]}(M_k)^{c'}_{c} \Big(\sum_\ell B^\ell_{b[c'a]}
  B_\ell^{r[st]}\Big) B^i_{r[a'b']} H^{[a'b']}_{[st]}
  =-\tfrac 13 I_3+\tfrac 23 I_4+\tfrac 23 I_5.\qquad
\end{eqnarray}
The contribution of $I_4$ cancels in the sum $I^f=4\tilde I_1+2\tilde I_2$ of
the two contributions~(\ref{istate}), and one arrives at
\begin{equation}
\hat A_{fki}^{\rm com}(s)=4\,\sum_\ell(\tilde I_1)_{fk\ell}I^{\rm pc}_{\ell\,i}
  +2\,\sum_\ell(\tilde I_2)_{fk\ell}I^{\rm pc}_{\ell\,i}=2I_3+4I_5.
\end{equation}
Doing the same exercise for the $u$-channel contribution $A_{fki}^{\rm com}(u)$
in Eq.~(\ref{Acom}) one obtains
\begin{eqnarray}
\sum_{\ell'}(I^{\rm pc})_{f\ell'}(\tilde I_1)_{\ell'ki}
  &=& B_f^{a[bc]} H^{b'c'}_{bc} \Big(\sum_{\ell'} B^\ell_{a[b'c']}  
  B_\ell^{r[st]}\Big) B^i_{r[sa']}(M_k)^{a'}_t
  \ =\ \tfrac 23 \hat I_3 - \tfrac 13 \hat I_4+ \tfrac 23 I_5,\nonumber\\
\sum_{\ell'}(I^{\rm pc})_{f\ell' }(\tilde I_2)_{\ell'i k}
  &=& B_f^{a[bc]} H^{[a'b']}_{[st]}\Big(\sum_\ell B^\ell_{b[c'a]}
  B_\ell^{r[st]}\Big) B^i_{s[a'r]}(M_k)^{a'}_t
  \ =\ -\tfrac 13 \hat I_3 + \tfrac 23 \hat I_4+\tfrac 23 I_5,\nonumber\\
\end{eqnarray}
leading to
\begin{equation}
\hat A_{fki}^{\rm com}(u)
  =4\sum_{\ell'}(I^{\rm pc})_{f\ell'}(\tilde I_1)_{\ell'ki}
  +2\sum_{\ell'}(I^{\rm pc})_{f\ell'}(\tilde I_2)_{\ell'ki}=2\hat I_3+4I_5.
\end{equation}
Calculating the difference of the $s$- and $u$-channel contributions
$\hat A_{fki}^{\rm com}(s)$ and $\hat A_{fki}^{\rm com}(u)$ according to
Eq.~(\ref{Acom}), one obtains the remarkable result that the contribution of
the topological tensor invariant $I_5$ cancels out. The final result is
\begin{equation}
\hat A_{fki}^{\rm com}=\hat A_{fki}^{\rm com}(s)-\hat A_{fki}^{\rm com}(u)
  =2(I_3-\hat I_3).
\end{equation}
We mention that the result $A_{fki}^{\rm com}\sim I_3-\hat I_3$ was already
been derived in the early paper~\cite{Korner:1978tc}.

The p.v.\ pole contributions not dealt with in detail in this paper read
\begin{equation}
A_{fki}^{\rm pole}=-\sum_\ell\frac{g_{fk\ell}\,b_{\ell\,i}}{m_i-m_\ell}
  -\sum_{\ell'}\frac{b_{f\ell'}\,g_{\ell'ki}}{m_f-m_{\ell'}}
  \ =\ A_{fki}^{\rm pole}(s)+A_{fki}^{\rm pole}(u),
\end{equation}
where
$b_{\ell\ell'}=\langle B_\ell|{\cal H}^{\rm pv}_{\rm eff}|B_{\ell'}\rangle$
are the p.v.\ matrix elements which are much smaller than the p.c.\ matrix
elements $a_{\ell\ell'}$~\cite{Ebert:1983yh,Cheng:1985dw}. For the same reason,
one also skips
\begin{equation}
B^{\rm com}_{fki}=\frac{\sqrt{2}}{f_k}
  \Big(\sum_\ell\langle B_f|M_k|B_\ell\rangle
  \langle B_\ell|{\cal H}^{\rm pv}_{\rm eff}|B_i\rangle
  -\sum_{\ell'}\langle B_f|{\cal H}^{\rm pv}_{\rm eff}|B_{\ell'}\rangle
  \langle B_{\ell'}|M_k|B_i\rangle\Big)
\end{equation}

%%%%%%%%%%%%%%%%%%%%%%%%%%%%%%%%%%%%%%%%%%%%%%%%%%%%%%%%%%%%%%%%%%%%%%%%%%%%%%%
\subsection{The parity conserving $P$-wave amplitude $B_{fki}^{{\rm pole}}$ }
%%%%%%%%%%%%%%%%%%%%%%%%%%%%%%%%%%%%%%%%%%%%%%%%%%%%%%%%%%%%%%%%%%%%%%%%%%%%%%%
Using again the notation of Refs.~\cite{Cheng:2018hwl,Cheng:2020wmk,%
Zou:2019kzq,Meng:2020euv,Hu:2020nkg}, the pole contribution to the p.c.\
$P$-wave amplitude $B$ is given by
\begin{equation}\label{pole}
B_{fki}^{\rm pole} = \sum_\ell\frac{g_{fk\ell}\,a_{\ell\,i}}{m_i-m_\ell}
   +\sum_{\ell'}\frac{a_{f\ell'}\,g_{\ell'ki}}{m_f-m_{\ell'}}
   = B_{fki}^{\rm pole}(s) + B_{fki}^{\rm pole}(u).  
\end{equation}
Again, the first and second terms in Eq.~(\ref{pole}) represent the $s$- and
$u$-channel pole contributions, respectively. The sum over the intermediate
state labels $\ell$ and $\ell'$ extends over all 20 $J^P=1/2^+$ ground state
baryons that have the correct quantum numbers to contribute to the $s$- and
$u$-channel pole contributions~(\ref{pole}). In the present application one
has zero, one or maximally two contributions in the sum over intermediate
ground state baryons. The latter case occurs when pairs of flavor degenerate
states contribute to the sum as e.g.\ the states ($\Lambda^0$, $\Sigma^0$) in
the $C=0$ sector or ($\Lambda_{c}^+$, $\Sigma_{c}^+$),
($\Xi_{c}^0$, $\Xi\,^{\prime \,0}_{c}$) and ($\Xi_{c}^+$,
$\Xi\,^{\prime \,+}_{c}$) in the $C=1$ sector. The members of the ground state
$J^P=3/2^+$ 20-plet do not contribute to the sum over intermediate states
since the weak transitions $a_{\ell\,i}$ and $a_{f\ell'}$ with $\ell,\ell'$
from the $J^P=3/2^+$ 20-plet vanish in the SU(3) limit because of the KPW
theorem.

The $s$- and $u$-channel pole contributions involve the strong coupling
coefficients denoted by $g_{fk\ell}$ and the weak parity conserving
transition matrix element $a_{\ell\,i}$ which also appeared in
$A^{\rm com}_{fki}$. One then uses the generalized Goldberger--Treiman relation
\begin{equation}
g_{fk\ell}=\frac{\sqrt{2}}{f_k}(m_f+m_\ell)g^A_{fk\ell}
\end{equation}
to express the strong coupling coefficients $g_{fk\ell}$ by the appropriate
axial vector elements $g^A_{fk\ell}$. One has
\begin{equation}
\label{sandu}
B_{fki}^{\rm pole}(s)+B_{fki}^{\rm pole}(u)
  = \frac{\sqrt{2}}{f_k}\left(\sum_\ell \,g^A_{fk\ell}
\,\frac{m_f+m_\ell}{m_i-m_\ell}\, a_{\ell\,i}+  \sum_{\ell'} a_{f\ell'}
 \frac{m_i+m_{\ell'}}{m_f-m_{\ell'}}\, g^A_{\ell'ki}\right). 
\end{equation}
As in the p.v.\ case we factor out the parameters depending on the dynamic
model to remain with a purely flavor dependent contribution $\hat B$. In the
bag model calculation of Ref.~\cite{Cheng:2018hwl,Cheng:2020wmk,Zou:2019kzq,%
Meng:2020euv,Hu:2020nkg} this is achieved by writing
\begin{eqnarray}
B_{fki}^{\rm pole}(s)+B_{fki}^{\rm pole}(u)
  &=& \frac{1}{f_k}6 \bar X_2(4\pi)\frac 23\bar Z(4\pi)
  \left(\hat B_{fki}^{\rm pole}(s)+\hat B_{fki}^{\rm pole}(u)\right)
\end{eqnarray}
where
\begin{equation}
\hat B_{fki}^{\rm pole}(s)+\hat B_{fki}^{\rm pole}(u)
  =\sum_\ell I^{\rm CQM}_{fk\ell}I^{\rm pc}_{\ell\,i}R_{fi}(B_\ell)
  +\sum_\ell I^{\rm pc}_{f\ell'}I^{\rm CQM}_{\ell'ki}R_{if}(B_{\ell'})
\end{equation}
with the compact notation for the two mass ratio expressions
\begin{equation}
\label{massratios}
R_{fi}(B_\ell)=\frac{m_f+m_\ell}{m_i-m_\ell}=R_s(B_\ell),\qquad
R_{if}(B_{\ell'})=\frac{m_i+m_{\ell'}}{m_f-m_{\ell'}}=R_u(B_{\ell'}), 
\end{equation}
the second ones used as pseudonyms. The flavor invariant $I^{\rm CQM}_{fki}$
denotes the flavor structure of the matrix element
$\langle B_{\ell'}M_k|B_i\rangle$ in the constituent quark model. In the bag
model the flavor structure of the strong coupling is given by a $d/f$ ratio of
$d/f=3/2$ (cf.\ Appendix~\ref{weak-strong}). In terms of the two building
blocks $\tilde I_1$ and $\tilde I_2$ introduced in Eq.~(\ref{tildeI12}) the
strong matrix element is given by
\begin{equation}
   I^{\rm CQM}_{fki}=(4\tilde I_1+5\tilde I_2)_{fki}
   =B_f^{a[bc]}\Big(4B^i_{a[bc']}+5B^i_{b[c'a]}\Big)(M_k)^{c'}_{c}
\end{equation}
Different from the p.v.\ $S$-wave case the sum over intermediate states
cannot be taken because of the mass ratio factors which differ for the
contributions of the hyperfine doublet partners. However, if one neglects the
hyperfine mass splitting, the sum over intermediate states can be performed as
in the p.v.\ $S$-wave case. To proceed, we define average values of the masses
of the set of hyperfine doublet partners $\{B_\ell\}$ denoted by $\bar m_\ell$
and correspondingly average values for the mass ratio factors by writing
\begin{equation}
\bar R_{fi}=\frac{m_f+\bar m_\ell}{m_i-\bar m_\ell}=\bar R_s(\{B_\ell\}),\qquad
\bar R_{if}=\frac{m_i+\bar m_{\ell'}}{m_f-\bar m_{\ell'}}
  =\bar R_u(\{B_{\ell'}\}).
\end{equation}
The average mass ratio factors $\bar R_{fi}$ and $\bar R_{if}$ can then be
factored out from the sum over intermediate states and one obtains
\begin{equation}\label{Bsum}
\hat B_{fki}^{\rm pole}(s)+\hat B_{fki}^{\rm pole}(u)
  =\bar R_{fi}\sum_\ell I^{\rm CQM}_{fk\ell}I^{\rm pc}_{\ell\,i}
  +\bar R_{if}\sum_{\ell'}I^{\rm pc}_{f\ell'}I^{\rm CQM}_{\ell'ki}.
\end{equation}
The sum over the intermediate states can now be performed, using again the
completeness relation~(\ref{completeness}). The task is simplified by the fact
that the strong transition $I^{\rm CQM}=4\tilde I_1+5\tilde I_2$ is a linear
superposition of the same two building blocks $\tilde I_1$ and $\tilde I_2$
that were used in the evaluation of the corresponding sums in the amplitude
$A$. The result of calculating the sum~(\ref{Bsum}) is then given by
\begin{eqnarray}\label{Bsums}
\sum_\ell I^{\rm CQM}_{fk\ell}I^{\rm pc}_{\ell\,i}
  &=& I_3+2I_4+6I_5, \nonumber \\
\sum_{\ell'}I^{\rm pc}_{f\ell'}I^{\rm CQM}_{\ell'ki}
  &=& \hat I_3+2\hat I_4+6I_5.
\end{eqnarray}
The flavor content of the pole model contributions is thus given by
\begin{equation}
\hat B_{fki}^{\rm pole}(s)+\hat B_{fki}^{\rm pole}(u)
  =(I_3+2I_4+6I_5)\bar R_{fi}+(\hat I_3+2\hat I_4+6I_5)\bar R_{if}.
\end{equation}
We have found again the quite remarkable fact that, in the absence of
hyperfine mass splittings for the flavor degenerate members of the ${\bf 20'}$
representation, the pole model representation of the p.c.\ $P$-wave amplitude
$B$ depends only on the initial and final states of the nonleptonic decays and
not on the detailed structure of the intermediate states.

Eqs.~(\ref{Bsums}) can be generalized to the case where the strong $d/f$ ratio
takes on general values (see Appendix~\ref{weak-strong}). One obtains 
\begin{eqnarray}
\sum_\ell I^{\rm gen}_{fk\ell}I^{\rm pc}_{\ell\,i}(s)
  &=& \tfrac53 d(I_3 + 2 I_4 + 6 I_5)-(\tfrac53 d - f)(2I_3+4I_5),\nonumber\\
\sum_{\ell'} I^{\rm pc}_{f\ell'}I^{\rm gen}_{\ell'ki}(u)
  &=& \tfrac23 d(\hat I_3 + 2 \hat I_4 + 6 I_5)
  -(\tfrac23 d - f)(2\hat I_3+4I_5).
\end{eqnarray}
One can check that one recovers Eqs.~(\ref{Bsums}) for $d/f=3/2$, i.e.\ for
$d=3/5$ and $f=2/5$.

%%%%%%%%%%%%%%%%%%%%%%%%%%%%%%%%%%%%%%%%%%%%%%%%%%%%%%%%%%%%%%%%%%%%%%%%%%%%%%%
\section{\label{sample}Some more sample decays}
%%%%%%%%%%%%%%%%%%%%%%%%%%%%%%%%%%%%%%%%%%%%%%%%%%%%%%%%%%%%%%%%%%%%%%%%%%%%%%%
Initiated by a thorough analysis of the methods applied in
Refs.~\cite{Cheng:2020wmk,Zou:2019kzq,Meng:2020euv,Hu:2020nkg} with details
found in Appendix~\ref{scrutiny}, in this section we present explicit results
for some sample decays on the connection of the current algebra results with
the topological tensor invariants. We emphasize again that this connection
holds true in the SU(3) limit. The relation between the two approaches derived
in Sec.~\ref{current} is explicitly verified in these examples. We include at
least one decay of each of the classes of decays discussed in
Sec.~\ref{tables}. For each decay we include in the header the values of the
seven tensor invariants in the same sequence
$(I^-_1,I^-_2,I_3,I_4,\hat I_3,\hat I_4,I_5)$ as in the tables. For a better
perception we have underlined the nonvanishing contributions of the
topological tensor invariants.

%%%%%%%%%%%%%%%%%%%%%%%%%%%%%%%%%%%%%%%%%%%%%%%%%%%%%%%%%%%%%%%%%%%%%%%%%%%%%%%
\subsection{The CF decay $\Lambda_{c}^+ \to \Lambda^0\, \pi^+$ \quad
  {\small$12(I_i)=(-2,-2,-2,4,-2,4,1)$}}
%%%%%%%%%%%%%%%%%%%%%%%%%%%%%%%%%%%%%%%%%%%%%%%%%%%%%%%%%%%%%%%%%%%%%%%%%%%%%%%
In this case, all seven tensor invariants are non-zero. Together with the
SCS decay $\Lambda_{c}^+ \to \Lambda^0\, K^+$, this decay is the only one in
which $I_3=\hat I_3 \neq 0$. Since one has $I_3-\hat I_3=0$, one concludes
that in these decays the $W$-exchange contribution to the p.v.\ amplitude $A$
is zero. Nevertheless, there are also tree diagram contributions which will
contribute to the amplitude $A$. Expressed in terms of the topological tensor
invariants, the $W$-exchange contributions to the reduced amplitudes $\hat A$
and $\hat B$ are given by
\begin{eqnarray}\label{com1B}
\hat A^{\rm com} &=& (\underline{2I_3}+\underline{4I_5})
  - (\underline{2\hat I_3}+\underline{4I_5})=0-0,\nonumber\\
\hat B^{\rm pole}&=& (\underline{I_3}+\underline{2I_4}+\underline{6I_5})\,
  R_s(\Sigma^+)+(\underline{\hat I_3} +\underline{2\hat I_4}+\underline{6I_5})\,
  R_u(\Sigma_{c}^0)\ =\nonumber\\
  &=&1 \cdot R_s(\Sigma^0) + 1 \cdot R_u(\Sigma_{c}^0).
\end{eqnarray}
The same result is obtained by explicit summation over the intermediate
states. The intermediate states are $\Sigma^+$ in the $s$-channel and
$\Sigma_{c}^0$ in the $u$-channel. The intermediate states do not contribute to
the p.v.\ amplitude $A$ since $F_{\pi^+}$ is a conserved charge operator which
implies $I^f_{\Lambda^0\pi^+\Sigma^+}=0$ and
$I^f_{\Sigma_{c}^0\pi^+\Lambda_{c}^+}=0$. The contribution of the
intermediate states to the p.c.\ amplitude $B$ is given in terms of the flavor
factors multiplying the relavent mass ratio factors $R_s(\Sigma^0)$ and
$R_u(\Sigma_{c}^0)$. One has
\begin{eqnarray}
\hat B^{\rm pole}_s(\Sigma^0)&:&
I^{\rm CQM}_{\Lambda^0\pi^+\Sigma^+}I^{\rm pc}_{\Sigma^+\Lambda_{c}^+}=1,
  \nonumber\\
\hat B^{\rm pole}(u;\Sigma_{c}^0)&:&
I^{\rm pc}_{\Lambda^0\Sigma_{c}^0}
I^{\rm CQM}_{\Sigma_{c}^0\pi^+\Lambda_{c}^+}=1,
\end{eqnarray}
where $I^{\rm pc}_{\Sigma^+\Lambda_{c}^+}=-4/2\sqrt{6}$,
$I^{\rm pc}_{\Lambda^0\Sigma_{c}^0}=4/2\sqrt{6}$,
$I^{\rm CQM}_{\Lambda^0\pi^+\Sigma^+}=-6/2\sqrt{6}$, and
$I^{\rm CQM}_{\Sigma_{c}^0\pi^+\Lambda_{c}^+}=6/2\sqrt{6}$ (see
Tables~\ref{weakme2} and~\ref{strongme2} in Appendix~\ref{weak-strong}).
The mass ratios read
\begin{equation}
R_s(\Sigma^0)
  =\frac{m_{\Lambda^0}+m_{\Sigma^0}}{m_{\Lambda_{c}^0}-m_{\Sigma^0}}=2.11\qquad
R_u(\Sigma_{c}^0)
  =\frac{m_{\Lambda_{c}^0}+m_{\Sigma_{c}^0}}
  {m_{\Lambda^0}-m_{\Sigma_{c}^0}}=-3.54,
\end{equation}
Our results are in agreement with the results in Ref.~\cite{Zou:2019kzq} up to
an overall sign difference. Note that there is a contribution to
$\hat B^{\rm pole}$ in Eq.~(\ref{com1B}) from the topology III given by the
term proportional to $I_5$. We conclude that this term was in fact included in
the analysis of Ref.~\cite{Zou:2019kzq} despite the claim of the authors that
the contributions from the type III diagram were omitted.

%%%%%%%%%%%%%%%%%%%%%%%%%%%%%%%%%%%%%%%%%%%%%%%%%%%%%%%%%%%%%%%%%%%%%%%%%%%%%%%
\subsection{The CF decay $\Lambda_{c}^+ \to \Sigma^0 \pi^+$ \quad
  {\small$4\sqrt3(I_i)=(0,2,2,0,-2,4,1)$}}
%%%%%%%%%%%%%%%%%%%%%%%%%%%%%%%%%%%%%%%%%%%%%%%%%%%%%%%%%%%%%%%%%%%%%%%%%%%%%%%
The second decay we discuss is the CF decay $\Lambda_{c}^+ \to \Sigma^0 \pi^+$.
Again, this decay proceeds only by $W$-exchange. The asymmetry parameter has
been measured and is given by
$\alpha(\Lambda_{c} \to \Sigma^0 \pi^+)=-0.73\pm0.17\pm0.07$. This implies
that the $S$-wave amplitude must be nonvanishing. And in fact, a glance at
Table~\ref{TableCF} shows that one has a $s$-channel contribution while the
$u$-channel contribution vanishes since $2\hat I_3+4I_5=0$. Let us check this
in more detail. The $W$-exchange contributions to the reduced amplitudes
$\hat A$ and $\hat B$ are given by
\begin{eqnarray}\label{com1}
\hat A^{\rm com} &=& (\underline{2I_3}+\underline{4I_5})
  -(\underline{2\hat I_3}+\underline{4I_5}) \ =\ 2/\sqrt{3}-0(u),\nonumber\\
\hat B^{\rm pole}&=&(\underline{ I_3}+2 I_4+\underline{6I_5})\,R_s(\Sigma^+)
  +(\underline{\hat I_3}+\underline{2\hat I_4}+\underline{6I_5})
  \,R_u(\Sigma_{c}^0) \ =\nonumber\\
  &=&2/\sqrt{3}\,R_s(\Sigma^+)+3/\sqrt{3}\,R_u(\Sigma_{c}^0)
\end{eqnarray}
A direct summation over the intermediate state results in
\begin{eqnarray}
\hat A^{\rm com}(s;\Sigma^+)&:&
I^f_{\Sigma^0\pi^+\Sigma^+}I^{\rm pc}_{\Sigma^+\Lambda_{c}^+}
  =2/\sqrt{3},\nonumber\\ 
\hat A^{\rm com}(u;\Sigma_{c}^0)&:&
I^{\rm pc}_{\Sigma^0\Sigma_{c}^0}I^f_{\Sigma_{c}^0\pi^+\Lambda_{c}^+}
  =0,\nonumber\\
\hat B^{\rm pole}(s;\Sigma^+)&:&
I^{\rm CQM}_{\Sigma^0\pi^+\Sigma^+} I^{\rm pc}_{\Sigma^+\Lambda_{c}^+}
  =2/\sqrt{3},\nonumber\\
\hat B^{\rm pole}(u;\Sigma_{c}^0)&:&
I^{\rm pc}_{\Sigma^0\Sigma_{c}^0}
I^{\rm CQM}_{\Sigma_{c}^0\pi^+\Lambda_{c}^+}=3/\sqrt{3}
 \end{eqnarray}
where we have used
$I^f_{\Sigma^0\pi^+\Sigma^+}=-4/2\sqrt{2}$,
$I^{\rm pc}_{\Sigma^+\Lambda_{c}^+}=-4/2\sqrt{6}$,
$I^f_{\Sigma_{c}^0\pi^+\Lambda_{c}^+}=0$,
$I^{\rm CQM}_{\Sigma^0\pi^+\Sigma^+}=-4/2\sqrt{2}$ and
$I^{\rm CQM}_{\Sigma_{c}^0\pi^+\Lambda_{c}^+}=6/2\sqrt{6}$
from Tables~\ref{weakme2} and~\ref{strongme2} in Appendix~\ref{weak-strong}.

%%%%%%%%%%%%%%%%%%%%%%%%%%%%%%%%%%%%%%%%%%%%%%%%%%%%%%%%%%%%%%%%%%%%%%%%%%%%%%%
\subsection{The SCS decay $\Xi_{c}^0 \to p K^-$ \quad
  {\small$2\sqrt6(I_i)=(0,0,0,-2_a,0,0,-1_b)$}}
%%%%%%%%%%%%%%%%%%%%%%%%%%%%%%%%%%%%%%%%%%%%%%%%%%%%%%%%%%%%%%%%%%%%%%%%%%%%%%%
This decay is interesting from the point of view that there are altogether
four states that contribute as intermediate states. These are
$(\Lambda^0,\Sigma^0)$ in the $s$-channel and $(\Lambda_{c}^+,\Sigma_{c}^+)$
in the $u$-channel. The structure of the tensor invariants can be seen to be
identical for this decay and the decay $\Lambda_{c}^+ \to \Xi^0\,K^+$ treated
in the previous section. We anticipate that the treatment in terms of tensor
invariants will be a much simpler undertaking than the explicit summation of
the four contributing intermediate states. In our analysis we shall retain the
identification of the weak transitions in terms of the ($a$) and ($b$) type
contributions.

The reduced commutator term $\hat A^{\rm com}$ and pole term
$\hat B^{\rm pole}$ are given by
\begin{eqnarray}
\hat A^{\rm com} &=& (2I_3+4I_5)-(2\hat I_3+4I_5)
  \ =\ (-4_b/2\sqrt{6})-(-4_b/2\sqrt{6})=0, \nonumber \\
\hat B^{\rm pole}&=&
  (I_3+\underline{2I_4}+\underline{6I_5})\bar R_s(\Lambda^0,\Sigma^0)
  +(\hat I_3+2\hat I_4+\underline{6I_5}) R_u(\Sigma_{c}^0)\ = \nonumber\\
  &=&-(4_a+6_b)/2\sqrt{6}\,\bar R_s(\Lambda^0,\Sigma^0)
  -6_b/2\sqrt{6}\,R_u(\Sigma_{c}^0).
\end{eqnarray}
When taking the intermediate state route, one has to take into account that
$I^f_{\Sigma_{c}^+K^-\Xi_{c}^0}=0$ and
$I^{\rm CQM}_{\Lambda_{c}^+K^-\Xi_{c}^0}=0$ (see Appendix~\ref{weak-strong}).
For the relevant products of flavor factors one obtains
\begin{eqnarray}
\hat A^{\rm com}(s;\Lambda^0,\Sigma^0)&:&
I^f_{pK^-\Lambda^0}I^{\rm pc}_{\Lambda^0\,\Xi_{c}^0}
  +I^f_{pK^-\Sigma^0}I^{\rm pc}_{\Sigma^0\,\Xi_{c}^0}
  =\big(-(2_a+4_b)+2_a\big)/2\sqrt{6},\nonumber\\
\hat A^{\rm com}(u;\Lambda_{c}^+)\,&:&
I^{\rm pc}_{p\,\Lambda_{c}^+ }I^f_{\Lambda_{c}^+K^-\Xi_{c}^0}
        =(-4_b)/2\sqrt{6},\nonumber\\
\hat B^{\rm pole}(s;\Lambda^0,\Sigma^0)&:&
I^{\rm CQM}_{pK^-\Lambda^0}I^{\rm pc}_{\Lambda^0\Xi_{c}^0}
  +I^{\rm CQM}_{pK^-\Sigma^0}I^{\rm pc}_{\Sigma^0\Xi_{c}^0}
  =\big(-(3_a+6_b)-1_a\big)/2\sqrt{6},\nonumber\\
\hat B^{\rm pole}(u;\Sigma_{c}^+)&:&
I^{\rm pc}_{p\,\Sigma_{c}^+}I^{\rm CQM}_{\Sigma_{c}^+K^-\Xi_{c}^0}
  =-6_b/2\sqrt{6}.     
\end{eqnarray}
Keeping in mind that the two $S$-wave contributions have to be subtracted
while the two $P$-wave contributions have to be added, one finds agreement of
the two calculational routes. The net result is that the commutator
contribution to the decay $\Xi_{c}^0 \to p K^-$ vanishes.

%%%%%%%%%%%%%%%%%%%%%%%%%%%%%%%%%%%%%%%%%%%%%%%%%%%%%%%%%%%%%%%%%%%%%%%%%%%%%%%
\subsection{The CF decay $\Omega_{c}^0 \to \Xi^0 \bar{K}^0$ \quad
  {\small$2(I_i)=(-1,-1,0,0,-2,0,0)$}}
%%%%%%%%%%%%%%%%%%%%%%%%%%%%%%%%%%%%%%%%%%%%%%%%%%%%%%%%%%%%%%%%%%%%%%%%%%%%%%%
The only $W$-exchange contribution to this decay is from the tensor invariant
$\hat I_3=-1$ (see Table~\ref{TableOmegac}). Therefore, the decay proceeds
only via the $u$-channel, which becomes obvious from
\begin{eqnarray}
\hat A^{\rm com} &=&\,(2I_3+4I_5)-(\underline{2\hat I_3}+4I_5)
  \ =\ \,0(s)-(-2), \nonumber \\
\hat B^{\rm pole}&=&0(s)+(\underline{\hat I_3}+2\hat I_4+6I_5)\,
  R_u(\Xi_{c}^0,\Xi_{c}^{\prime\,0})
  \ =\ 0(s)-\bar R_u(\Xi_{c}^0,\Xi_{c}^{\prime\,0}).\qquad
\end{eqnarray}
where $0(s)$ stands for the absence of an $s$-channel contribution. The
interest in this decay is caused by the fact that both intermediate states
$\Xi_{c}^0$ and $\Xi_{c}^{'0}$ contribute to the $P$-wave amplitude
$B^{\rm pole}(u)$. In terms of the flavor invariants $I^f$, $ I^{\rm pc}$ and
$I^{\rm CQM}$ one obtains
\begin{eqnarray}
\hat A^{\rm com}(u;\Xi_{c}^{'0})&:&
I^{\rm pc}_{\Xi^0\Xi_{c}^{'0}}\,I^f_{\Xi_{c}^{'0}\bar K^0\Omega_{c}^0}
  \ =\ -2,\nonumber\\ 
\hat B^{\rm pole}(u;\Xi_{c}^0,\Xi_{c}^{\prime\,0})&:&
I^{\rm pc}_{\Xi^0\Xi_{c}^{0}}\,I^{\rm CQM}_{\Xi_{c}^{0}\bar K^0\Omega_{c}^0}
  +I^{\rm pc}_{\Xi^0\Xi_{c}^{'0}}\,
  I^{\rm CQM}_{\Xi_{c}^{'0}\bar K^0\Omega_{c}^0}\ =\ -1.\qquad
\end{eqnarray}
The results are in agreement with Ref.~\cite{Hu:2020nkg}. There is no need to
differentiate between the current algebra approach and the modified current
algebra result in this case since $I_5=0$.
  
In the limit of vanishing hyperfine splitting
$m_{\Xi_{c}^{0}}=m_{\Xi_{c}^{\prime 0}}$ one can sum the two $P$-wave
$u$-channel contributions to obtain
\begin{equation}\label{ABtest4}
\hat B^{\rm pole}=2(\underline{\hat I_3}+2\hat I_4+6I_5)
  \frac{m_{\Xi_0}-\bar m_{\Xi_{c}^0}}{m_{\Omega_{c}^0}-\bar m_{\Xi_{c}^0}}
  =-2\frac{m_{\Xi_0}-\bar m_{\Xi_{c}^0}}{m_{\Omega_{c}^0}-\bar m_{\Xi_{c}^0}},
\end{equation}
where $\bar m_{\Xi_{c}^{0}}$ denotes a suitable mass average of the two flavor
degenerate states $\Xi_{c}^0$ and $\Xi_{c}^{\prime\,0}$. The decay was analyzed
in Ref.~\cite{Fayyazuddin:1996iy}, again in the context of the current algebra
approach. In the first version of Ref.~\cite{Fayyazuddin:1996iy} there was the
claim that the contributions of the intermediate states $\Xi^0_{c}$ and
$\Xi^{\prime\,0}_{c}$ cancel each other for
$m_{\Xi_{c}^0}=m_{\Xi^{\prime\,0}_{c}}$. This statement was corrected later on
in an Erratum to Ref.~\cite{Fayyazuddin:1996iy} with a result in agreement with
Eq.~(\ref{ABtest4}). That the two contributions do not, in fact, cancel can be
easily checked by a mere visual inspection of the tensor invariants in
Table~\ref{TableOmegac}. We have provided this example to illustrate the power
of the current algebra approach when expressed in terms of topological tensor
invariants because the compactness of the tensor invariant expressions allows
one to easily trace mistakes in a current algebra result.
   
%%%%%%%%%%%%%%%%%%%%%%%%%%%%%%%%%%%%%%%%%%%%%%%%%%%%%%%%%%%%%%%%%%%%%%%%%%%%%%%
\subsection{The SCS decay $\Omega_{c}^0 \to \Sigma^0 \bar K^0$ \quad
  {\small$2\sqrt2(I_i)=(0,0,0,-2_a,0,0,-1_a)$}}
%%%%%%%%%%%%%%%%%%%%%%%%%%%%%%%%%%%%%%%%%%%%%%%%%%%%%%%%%%%%%%%%%%%%%%%%%%%%%%%
The decay proceeds only through $W$ exchange and via the ($a$)-type weak
interaction. The interest in this decay is based on the fact that up to
normalization factors the topological invariant structure is identical to the
one of the previous decay $\Lambda_{c}^+ \to \Xi^0\,K^+$. The same statement
holds true for the decays
\begin{eqnarray}
{\rm CF}&:& \Lambda_{c}^+ \to \Xi^0 K^+;\,\,\quad\quad \qquad \qquad
  \Xi_{c}^0 \to \Sigma^+ K^- \nonumber \\
{\rm SCS}&:& \Omega_{c}^0 \to \Sigma^+ K^-,\,\Sigma^0 \bar K^0;\,\quad\qquad
  \Xi_{c}^0 \to \Sigma^+ \pi^-,\,pK^- \nonumber \\ 
{\rm DCS}&:& \Xi_{c}^+ \to p\pi^0,\,n \pi^+;\quad\qquad\qquad
  \Xi_{c}^0 \to p\pi^-,\,n \pi^0
\end{eqnarray}
One obtains
\begin{eqnarray}
\hat A^{\rm com} &=&\,(2I_3+4I_5)-(2\hat I_3+4I_5)
 \ =\ (-4_b/2\sqrt{2})- (-4_b/2\sqrt{2})=0, \nonumber \\
\hat B^{\rm pole}&=&(I_3+\underline{2I_4}+\underline{6I_5})\bar R_s(\Xi^0)
  +(\hat I_3+2\hat I_4+\underline{6I_5})\bar R_u(\Xi_{c}^0,\Xi_{c}^{\prime\,0})
  \ =\nonumber\\
  &=&-(10_a)/2\sqrt{2}\,\bar R_s(\Xi^0)
  -6_a/2\sqrt{2}\,\bar R_u(\Xi_{c}^0,\Xi_{c}^{\prime\,0})
\end{eqnarray}
or
 \begin{eqnarray}
\hat A^{\rm com}(s;\Xi^0) &:&
  I^f_{\Sigma^0\bar K^0\Xi^0}I^H_{\Xi^0\Omega_{c}^0}
  =-4_a/2\sqrt{2} \nonumber \\
\hat A^{\rm com}(u;\Xi_{c}^0) &:&
  I^H_{\Sigma^0\Xi_{c}^0}I^f_{\Xi_{c}^0\bar K^0\Omega_{c}^0}
   =-4_a/2\sqrt{2} \nonumber \\
\hat B^{\rm pole}(s;\Xi^0) &:&
  I^{\rm CQM}_{\Sigma^0K^0\Xi^0}I^H_{\Xi^0\Omega_{c}^0}
  =-10_a/2\sqrt{2} \nonumber \\ 
\hat B^{\rm pole}(u;\Xi_{c}^0,\Xi_{c}^{'0}) &:&
  I^H_{\Sigma^0\Xi_{c}^0}I^{\rm CQM}_{\Xi_{c}^0\bar K^0\Omega_{c}^0}
  +I^H_{\Sigma^0\Xi_{c}^{'0}} I^{\rm CQM}_{\Xi_{c}^{'0}\bar K^0\Omega_{c}^0}
  =(-2_a-4_a)/2\sqrt{2},\qquad
\end{eqnarray}
where   
\begin{equation}
R_u(\Xi_{c}^0)=\frac{m_{\Omega_{c}^0}+m_{\Xi_{c}^0}}{m_{\Sigma^0}-m_{\Xi_{c}^0}}
  =4.83, \qquad
R_u(\Xi_{c}^{\prime\,0})=\frac{m_{\Omega_{c}^0}+m_{\Xi_{c}^{\prime\,0}}}
{m_{\Sigma^0}-m_{\Xi_{c}^{\prime\,0}}}=4.53.
\end{equation}   

%%%%%%%%%%%%%%%%%%%%%%%%%%%%%%%%%%%%%%%%%%%%%%%%%%%%%%%%%%%%%%%%%%%%%%%%%%%%%%%
\subsection{The $\Delta C=0$ SCS decays $\Xi_{c}^+ \to \Lambda_{c}^+\,\pi^0$
   and $\Xi_{c}^0 \to \Lambda_{c}^+\,\pi^-$\\
   {\small$12\sqrt2(I_i(\Xi_{c}^+ \to \Lambda_{c}^+\,\pi^0))
   =12(I_i(\Xi_{c}^0 \to \Lambda_{c}^+\,\pi^-))
   =(-5_{a'},4_{a'},-2_{b'},4_{b'},-8_{a'},4_{a'},1_{b'})$}}
%%%%%%%%%%%%%%%%%%%%%%%%%%%%%%%%%%%%%%%%%%%%%%%%%%%%%%%%%%%%%%%%%%%%%%%%%%%%%%%
The decays are contributed to by both the factorizing tree diagram and the
nonfactorizing $W$-exchange contributions. The tree diagram contributions
induced by the transition ($a'$) $s\to u;\,u\to d$ are purely p.v.\ $S$-wave
contributions, as follows from the light diquark transition $0^+\to 0^++0^-$
in the background field of the heavy charm quark. It is therefore interesting
to have a closer look at the structure of the $W$-exchange contributions.

Let us first state that the two decays are related by the $\Delta I=1/2$
rule as follows (see Table~\ref{deltaC=0})
\begin{equation}
\sqrt{2}\,{\cal M}(\Xi_{c}^+ \to \Lambda_{c}^+\,\pi^0)
  ={\cal M}( \Xi_{c}^0 \to \Lambda_{c}^+\,\pi^-)
\end{equation}
For the tree level contribution this comes about since we are using only the
${\cal H}_{\rm eff}({\cal O}_-)$ contribution. For the $W$-exchange
contributions the $\Delta I=1/2$ rule is a consequence of the KPW theorem.
  
Let us concentrate on the $W$-exchange contribution to the decay
$\Xi_{c}^0 \to \Lambda_{c}^+\,\pi^-$. We begin with the tensor invariant
representation. Since $2I_3+4I_5=0$ (see Table~\ref{deltaC=0}), the $S$-wave
$s$-channel contribution vanishes, i.e.\ one has $A^{\rm com}(s)=0$. The
nonvanishing $S$-wave $u$-channel and $P$-wave $(s,u)$-channel contributions
read
\begin{eqnarray}
\hat A^{\rm com}&=&0(s)-\big(\underline{2\hat I_3}+\underline{4 I_5}\big)
  \ =\ 0(s)-(-16_{a'}+4_{b'})/12, \nonumber \\
\hat B^{\rm pole}&=&
  \big(\underline{I_3}+\underline{2I_4}+\underline{6 I_5}\big)R_s(\Sigma_{c}^0)
  +\big(\underline{\hat I_3}+\underline{2\hat I_4}+\underline{6I_5}\big)
  R_u(\Xi_{c}^{\prime+})\ =\nonumber\\
  &=&12_{b'}/12\,R_s(\Sigma_{c}^0)+6_{b'}/12\,R_u(\Xi_{c}^{\prime+}),
\end{eqnarray}
where $2\hat I_3+4I_5=(-16_a+4_b)/12$, $I_3+2I_4+6I_5=12_b/12$ and
$\hat I_3+2\hat I_4+6I_5=6_b/12$ in agreement with the result of the
intermediate state route where
\begin{eqnarray}
\hat A^{\rm com}(u;\Xi_{c}^+)&:&
I^H_{\Lambda_{c}^+\Xi_{c}^+}I^f_{\Xi_{c}^+\pi^-\Xi_{c}^0}
  =(-16_{a'}+4_{b'})/12, \nonumber \\
\hat B^{\rm pole}(s;\Sigma_{c}^0)&:&
I^{\rm CQM}_{\Lambda_{c}^+\pi^-\Sigma_{c}^0}I^H_{\Sigma_{c}^0\Xi_{c}^0}
  =12_{b'}/12, \nonumber \\
\hat B^{\rm pole}(u;\Xi_{c}^{'+})&:&
I^H_{\Lambda_{c}^+\Xi_{c}^{'+}}I^{pc \,(\pi^-)}_{\Xi_{c}^{'+}\Xi_{c}^0}
  =6_{b'}/12.
\end{eqnarray}
Note that $P$-wave transitions are solely induced by the quark level
transition ($b'$) $c\to d;\,s\to c$. The transition ($a'$) $s\to u;\,u\to d$
does not contribute because of a quantum number mismatch for the $P$-wave
$s$-channel amplitude as exemplified by the fact that the $W$-exchange
topologies IIa and III do not admit a contribution ($a'$). The KPW theorem is
responsible for the absence of a contribution ($a'$) to the $P$-wave
$u$-channel amplitude. Contrary to what is stated in the
literature~\cite{Cheng:1992ff,Faller:2015oma,Cheng:2015ckx}, one obtains a
nonvanishing $P$-wave contribution from the transition ($b'$) which, in
addition, is considerably enhanced by the mass ratio factors
$R_s(\Sigma_{c}^0)=278.82$ and $R_u(\Xi_{c}^{\prime+})=-17.29$. Such a large
$P$-wave contribution is quite welcome since the present model calculations,
which are based on an assumed $S$-wave transition, are below or far below the
experimental rate measurement~\cite{Aaij:2020wtg}.

%%%%%%%%%%%%%%%%%%%%%%%%%%%%%%%%%%%%%%%%%%%%%%%%%%%%%%%%%%%%%%%%%%%%%%%%%%%%%%%
\subsection{The $\Delta C=0$ SCS decays $\Omega_{c}^0 \to \Xi_{c}^+\,\pi^-$ and
  $\Omega_{c}^0 \to \Xi_{c}^0\,\pi^0$\\
  {\small$2\sqrt6(I_i(\Omega_{c}^0 \to \Xi_{c}^+\,\pi^-))
  =4\sqrt3(I_i(\Omega_{c}^0 \to \Xi_{c}^0\,\pi^0))
  =(1_{a'},-2_{a'},2_{b'},-4_{b'},0,0,0)$}}
%%%%%%%%%%%%%%%%%%%%%%%%%%%%%%%%%%%%%%%%%%%%%%%%%%%%%%%%%%%%%%%%%%%%%%%%%%%%%%%
The decays are contributed to by both tree diagram and $W$-exchange
contributions. Since one now has a $D(1^+) \to D(0^+) + 0^-$ diquark
transition in the background field of the heavy charm quark, the tree
contribution is purely $P$-wave.

The $W$-exchange contributions are induced by the quark level transition ($b'$)
$c\to d;s \to c$ which is a pure $\Delta I=1/2$ transition for both
${\cal H}_{\rm eff}({\cal O}_-)$ transitions. Therefore, for the $W$-exchange
contributions one has the $\Delta I=1/2$ relation
\begin{equation}
\sqrt{2}\,{\cal M}(\Omega_{c}^0 \to \Xi_{c}^0 \pi^0)
   =-{\cal M}(\Omega_{c}^0 \to \Xi_{c}^+ \pi^-),
\end{equation}
regardless of whether one invokes the KPW theorem.
  
As Table~\ref{deltaC=0} shows, the topologies IIb and III do not contribute
as can be surmised from the fact that $\hat I_3=\hat I_4=I_5=0$, i.e.\ there
are no $u$-channel $W$-exchange contributions. One obtains
\begin{eqnarray}
\hat A^{\rm com}&=&\big(\underline{2I_3}+4 I_5\big)
  -\big(2\hat I_3+4 I_5\big)\ =\ 4_{b'}/2\sqrt{6}-0(u), \nonumber \\
\hat B^{\rm pole}&=&\big(\underline{I_3}+\underline{2 I_4}+6I_5\big)\,
  R_s(\Xi_{c}^{\prime\,0})+0(u)
  \ =\ -6_{b'}/2\sqrt{6}\,R_s(\Xi_{c}^{\prime\,0})+0(u),\qquad
\end{eqnarray}
where $2I_3=4_b/2\sqrt{6}$ and $I_3+2I_4=-6_b/2\sqrt{6}$. It is apparent that
the $W$-exchange contributions generate an $S$-wave contribution through the
topology IIa.

Let us confirm that we obtain the same result for the intermediate state
representation. The $s$-channel intermediate states are $\Xi_{c}^0$ and
$\Xi_{c}^{\prime\,0}$, resp., since
$I^f_{\Xi_{c}^+\pi^-\Xi_{c}^{\prime\,0}}=0$ and
$I^{\rm CQM}_{\Xi_{c}^+\pi^-\Xi_{c}^{0}}=0$. One therefore has
\begin{eqnarray}
\hat A^{\rm com}(s;\Xi_{c}^0)&:&
I^f_{\Xi_{c}^+\pi^-\Xi_{c}^0}\,I^{\rm pc}_{\Xi_{c}^0\Omega_{c}^0}
  =4_{b'}/2\sqrt{6}\ \nonumber \\
\hat B^{\rm pole}(s;\Xi_{c}^{\prime\,0})&:&
I^{\rm CQM}_{\Xi_{c}^+\pi^-\Xi_{c}^{'0}}\,
  I^{\rm pc}_{\Xi_{c}^{\prime\,0}\Omega_{c}^0}=-6_{b'}/2\sqrt{6}.
\end{eqnarray}
Again, the $P$-wave contribution is enhanced by the mass factor
$R_s(\Xi_{c}^{\prime\,0})=43.51$, but not as much as in the previous case
$\Xi_{c}^0 \to \Lambda_{c}^+\,\pi^-$.

%%%%%%%%%%%%%%%%%%%%%%%%%%%%%%%%%%%%%%%%%%%%%%%%%%%%%%%%%%%%%%%%%%%%%%%%%%%%%%%
\subsection{\bf The CF decay $\Xi_{cc}^{++} \to \Xi_{c}^{+} \pi^+$ \quad 
  {\small$2\sqrt6(I_i)=(-2,1,0,0,-4,4,0)$}}
%%%%%%%%%%%%%%%%%%%%%%%%%%%%%%%%%%%%%%%%%%%%%%%%%%%%%%%%%%%%%%%%%%%%%%%%%%%%%%%
The decay belongs to the class of decays
$B_{cc}({\bf 3})\to B_{c}(\overline{\bf 3})+M({\bf 8})$. One has
\begin{eqnarray}
\hat A^{\rm com}&=&0(s)-(\underline{2\hat I_3} +4I_5)=-8/2\sqrt{6},
  \nonumber \\
\hat B^{\rm pole}&=&0(s)+\big(\underline{\hat I_3}+\underline{2\hat I_4}
  +6I_5\big)\,R_u(\Xi_{cc}^{+})=4/2\sqrt{6}\,R_u(\Xi_{cc}^{+}) 
\end{eqnarray}
or
\begin{eqnarray}
\hat A^{\rm com}(u;\Xi_{cc}^{+})&:&
I^{\rm pc}_{\Xi_{c}^+\Xi_{cc}^+}I^f_{\Xi_{cc}^+\pi^+\Xi_{cc}^{++}}
  =(\tfrac{8}{2\sqrt{6}})(-1)=-8/2\sqrt{6}, \nonumber \\
\hat B^{\rm pole}(u;\Xi_{cc}^+)&:&
I^{\rm pc}_{\Xi_{c}^+\Xi_{cc}^+}I^{\rm CQM}_{\Xi_{cc}^+\pi^+\Xi_{cc}^{++}}
  =(\tfrac{8}{2\sqrt{6}})(\tfrac{1}{2})=4/2\sqrt{6}.
\end{eqnarray}
  
%%%%%%%%%%%%%%%%%%%%%%%%%%%%%%%%%%%%%%%%%%%%%%%%%%%%%%%%%%%%%%%%%%%%%%%%%%%%%%%
\subsection{The CF decay $\Xi_{cc}^+ \to \Xi_{c}^{+} \pi^0$ \quad
  {\small$4\sqrt2(I_i)=(0,0,4,-2,-4,4,0)$}}
%%%%%%%%%%%%%%%%%%%%%%%%%%%%%%%%%%%%%%%%%%%%%%%%%%%%%%%%%%%%%%%%%%%%%%%%%%%%%%%
This decay also belongs to the class
$B_{cc}({\bf 3})\to B_{c}({\overline{\bf 3}})+M({\bf 8})$. The decay is not
related to the previous decay $\Xi_{cc}^{++} \to \Xi_{c}^{+} \pi^+$ by isospin
symmetry, as a comparison of the tensor invariants shows. There are, in fact,
two reduced isospin amplitudes describing the decays $\Xi_{cc}\to\Xi_{c}\pi$. 
Analyzing the flavor flow in diagram III one concludes that the topological
invariant $I_5$ vanishes. From the general analysis in Sec.~\ref{tables} we
know that $I_3 +2I_4=0$ is a general result for this class of decays. In the
language of the current algebra approach this implies that the $P$-wave
$s$-channel contribution vanishes. Candidates for the intermediate $s$-channel
states are $\Xi_{c}^+$ and $\Xi_{c}^{\prime+}$. Still, $\Xi_{c}^{\prime+}$
does not contribute since the weak transition
$\langle\Xi_{c}^{\prime+}|H^{\rm pc}|\Xi_{cc}^+\rangle$ vanishes in the
SU(3) limit due to the KPW theorem. In terms of the topological tensor
invariants one has
\begin{eqnarray}
\hat A^{\rm com}&=&(\underline{2I_3}+4I_5)-(\underline{2\hat I_3}+4I_5)
 \ =\ 8/4\sqrt{3}-(-8/4\sqrt{3}), \nonumber \\
\hat B^{\rm pole}&=&(\underline{I_3}+\underline{2I_4}+6I_5)\,R_s(\Xi_{c}^+)
  +(\underline{\hat I_3}+\underline{2\hat I_4} +6I_5)\,R_u(\Xi_{cc}^+)
  \ =\ 4/4\sqrt{3}R_u(\Xi_{cc}^+)\qquad
\end{eqnarray}
with $2(\underline{I_3}-\underline{\widehat I_3})=4/\sqrt{3}$ and
$\underline{\hat I_3}+\underline{2\hat I_4}+6I_5=1/\sqrt{3}$. The same result
is obtained by the intermediate state analysis where one has
\begin{eqnarray}
\hat A^{\rm com}(s;\Xi_{c}^+)&:& 
I^f_{\Xi_{c}^+\pi^0\Xi_{c}^+}I^{\rm pc}_{\Xi_{c}^+\Xi_{cc}^+}
  =8/4\sqrt{3}, \nonumber \\
\hat A^{\rm com}(u;\Xi_{c}^{++})&:& 
I^{\rm pc}_{\Xi_{c}^+\Xi_{cc}^+}I^f_{\Xi_{cc}^+\pi^0\Xi_{cc}^+}
  =-8/4\sqrt{3}, \nonumber \\
\hat B^{\rm CQM}(u;\Xi_{cc}^+)&:&
I^{\rm pc}_{\Xi_{c}^+\Xi_{cc}^+}I^{\rm CQM}_{\Xi_{cc}^+\pi^0\Xi_{cc}^+}
  =4/4\sqrt{3}.
\end{eqnarray}

%%%%%%%%%%%%%%%%%%%%%%%%%%%%%%%%%%%%%%%%%%%%%%%%%%%%%%%%%%%%%%%%%%%%%%%%%%%%%%%
\subsection{The CF decay $\Xi^+_{cc}\to \Sigma_{c}^{++}K^-$ \quad
  {\small$2(I_i)=(0,0,0,2,0,0,0)$}}
%%%%%%%%%%%%%%%%%%%%%%%%%%%%%%%%%%%%%%%%%%%%%%%%%%%%%%%%%%%%%%%%%%%%%%%%%%%%%%%
This decay belongs to the class $B_{cc}({\bf 3}) \to B_{c}({\bf 6})+P({\bf 8})$.
From the discussion in Sec.~\ref{tables} we know that
$I_3=\hat I_3=\hat I_4=I_5=0$ for this class of decays. The only nonvanishing
contribution is $I_4=1$, i.e.\ there is no factorizing contribution to this
decay. When stated in terms of the current algebra approach, this implies that
the only nonvanishing contribution to this decay is the $P$-wave $s$-channel
contribution.

We verify this in explicit form using the current algebra representation.
First note that there is no candidate for the $u$-channel intermediate state.
As concerns the $s$-channel, due to the KPW theorem $\Xi_{c}^{\prime+}$ does
not contribute as intermediate state. On the other hand $\Xi_{c}^+$ does not
contribute to the $S$-wave $s$-channel since
$I^f_{\Sigma_{c}^{++}K^-\Xi_{c}^+}=0$. One thus remains with the
$P$-wave $s$-channel contribution $B^{\rm pole}(s)$. One has
\begin{equation}
\hat B^{\rm pole}\ =\ (I_3+\underline{2I_4}+ 6I_5)\,R_s(\Xi_{c}^+)+0(u)
  \ =\ 2R_s(\Xi_{c}^+)+0(u),
\end{equation}
where $I_3+\underline{2I_4}+6I_5=2$. In terms of the intermediate state path
one has
\begin{eqnarray}
\hat B^{\rm pole}(s,\,\Xi_{c}^+)&:&
I^{\rm CQM}_{\Sigma_{c}^{++}K^-\Xi_{c}^+}\,
I^{\rm pc}_{\Xi_{c}^+\Xi_{cc}^{+}}=2
\end{eqnarray}
in accord with the result in terms of tensor invariants. This result is in
agreement with the result in Ref.~\cite{Cheng:2020wmk}. Note that there is no
need fo modify the current algebra approach in this case since $I_5=0$.

%%%%%%%%%%%%%%%%%%%%%%%%%%%%%%%%%%%%%%%%%%%%%%%%%%%%%%%%%%%%%%%%%%%%%%%%%%%%%%%
\subsection{The CF decay $\Xi_{cc}^{++} \to \Sigma^+\,D^+$ \quad
  {\small$2(I_i)=(0,0,0,0,0,-2,0)$}}
%%%%%%%%%%%%%%%%%%%%%%%%%%%%%%%%%%%%%%%%%%%%%%%%%%%%%%%%%%%%%%%%%%%%%%%%%%%%%%%
This decay belong to the class
$B_{cc}({\bf 3})\to B({\bf 8})+M(\overline{\bf 3})$. The current algebra
approach applies to the $P$-waves only where we can check the structure of the
$1/2^+$ pole contributions. As Table~\ref{TableBccBD} shows, there are only
$u$-channel pole contributions since the only nonvanishing tensor invariant is
$\hat I_4$. This agrees with the observation that, from the flavor flow, there
can be no intermediate $s$-channel contributions. One has
\begin{eqnarray}
\hat B^{\rm pole}(u;\Sigma_{c}^+,\Lambda_{c}^+)&=&
  (\hat I_3+\underline{2\hat I_4}+I_5)\,\bar R_u(\Sigma_{c}^+,\Lambda_{c}^+)
  \ =\ -2\,\bar R_u(\Sigma_{c}^+,\Lambda_{c}^+)
\end{eqnarray}
or
\begin{eqnarray}
\hat B^{\rm pole}(u;\Sigma_{c}^+,\Lambda_{c}^+)&:&
I^{\rm pc}_{\Sigma^+\Sigma_{c}^+}I^{\rm CQM}_{\Sigma_{c}^+D^+\Xi_{cc}^{++}}
  +I^H_{\Sigma^+\Lambda_{c}^+}I^{\rm CQM}_{\Lambda_{c}^+D^+\Xi_{cc}^{++}}
  =1/2-5/2=-2.
\end{eqnarray}

%%%%%%%%%%%%%%%%%%%%%%%%%%%%%%%%%%%%%%%%%%%%%%%%%%%%%%%%%%%%%%%%%%%%%%%%%%%%%%%
\subsection{The CF decay $\Xi_{cc}^{+} \to \Sigma^+\,D^0$ \quad
  {\small$2(I_i)=(0,0,0,0,0,0,1)$}}
%%%%%%%%%%%%%%%%%%%%%%%%%%%%%%%%%%%%%%%%%%%%%%%%%%%%%%%%%%%%%%%%%%%%%%%%%%%%%%%
This decay belongs to the same class as the previous one,
$B_{cc}({\bf 3}) \to B({\bf 8})+M(\overline{\bf 3})$. Since $I^\pm_{1.2}=0$,
there are no tree diagram contributions. The only nonvanishing tensor
invariant is $I_5$. This implies that one has both $s$- and $u$-channel pole
contributions. The current algebra approach only applies to the $P$-wave pole
contribution where we can check on the structure of the $1/2^+$ pole
contributions. The $s$-channel intermediate state is $\Xi_{c}^+$. In the
$u$-channel both $\Lambda_{c}^+$ and $\Sigma_{c}^+$ contribute as intermediate
states. The structure of the $P$-wave contributions is given by
\begin{eqnarray}
\hat B^{\rm pole}(s;\Xi_{c}^+)&=&(I_3+2I_4+\underline{6I_5})\,R_s(\Xi_{c}^+)
  \ =\ 3R_s(\Xi_{c}^+), \nonumber \\
\hat B^{\rm pole}(u;\Lambda_{c}^+,\Sigma_{c}^+)&=&
  (\hat I_3+2\hat I_4+\underline{6I_5})\,R_u(\Lambda_{c}^+,\Sigma_{c}^+)
  \ =\ 3R_u(\Lambda_{c}^+,\Sigma_{c}^+).\qquad
\end{eqnarray}
Since $I^{\rm CQM}_{\Sigma^+D^0\Xi_{c}^{+}}I^H_{\Xi_{c}^+\Xi_{cc}^+}=3$ and 
$I^{\rm CQM}_{\Lambda_{c}^+D^0\Xi_{cc}^{+}}+I^H_{\Sigma^+\Sigma_{c}^+}
I^{\rm CQM}_{\Sigma_{c}^+D^0\Xi_{cc}^{+}}=1/2+5/2=3$, the result obtained
via intermediate states agrees with this result. Note that the modified
current algebra approach predicts that the decay
$\Xi_{cc}^{+} \to \Sigma^+\,D^0$ vanishes.

%%%%%%%%%%%%%%%%%%%%%%%%%%%%%%%%%%%%%%%%%%%%%%%%%%%%%%%%%%%%%%%%%%%%%%%%%%%%%%%
\section{\label{features}Some general features of the topological tensor
  and current algebra approaches}
%%%%%%%%%%%%%%%%%%%%%%%%%%%%%%%%%%%%%%%%%%%%%%%%%%%%%%%%%%%%%%%%%%%%%%%%%%%%%%%
The tables in Sec.~\ref{tables} contain a wealth of physics information on the
nonleptonic decays of charm baryons. Just by visual inspection one can
identify many interesting patterns of physical relevance concerning the
nonleptonic charm baryon decays. We divide the general remarks into two parts.
In the first part we list general features of the various charm baryon decays
without using any dynamical input. In the second part we identify some general
features in the tables using elements of the current algebra approach. We
begin with by listing decays that have all zero entries in the tables, i.e.\
their decay rates are predicted to be zero. They are all DCS decays and read
\begin{eqnarray}
\Gamma(\Omega_{c}^0 \to \Lambda^0\,\pi^0,\Sigma^0\,\eta(\eta'))&=&0,
  \nonumber\\
\Gamma(\Omega_{cc}^+ \to \Lambda_{c}^+\,\pi^0,\,\Sigma_{c}^+\,\eta(\eta'))&=&0,
  \nonumber\\
\Gamma(\Omega_{cc}^+ \to \Sigma^0\,D_s^+)&=&0.
\end{eqnarray}
A corresponding prediction is given in the following.
Note that there are no tree contributions to any of these decays which implies
that one only has transitions induced by the effective Hamiltonian
${\cal H}_{\rm eff}({\cal O}_-)$. According to Table~\ref{Heff} the DCS
Hamiltonian ${\cal H}_{\rm eff}({\cal O}_-)$ from Eq.~(\ref{DCS}) transforms
as an isosinglet ($\Delta I=0$). From the isospin assignments of the charm
baryons in Table~\ref{Tablemass} and the fact that $I(D_s^{(\ast)+})=0$ one
can see that all the listed decays are isospin forbidden. A related derivation
of the vanishing of the decay $\Omega_{cc}^+ \to \Sigma^0\,D_s^{\ast\,+}$ can
also be obtained from a topological point of view and the KPW theorem. Both
the isospin forbidden decay $\Omega_{cc}^+ \to \Sigma^0\,D_s^{\ast\,+}$ and
the isospin allowed decay $\Omega_{cc}^+ \to \Lambda^0\,D_s^{\ast\,+}$ proceed
via the topological diagrams IIb and III in which the final state quark pair
$[du]$ emerging from the weak vertex is antisymmetric. It then follows that
the decay $\Omega_{cc}^+ \to \Sigma^0\,D_s^{\ast\,+}$ is forbidden while the
decay $\Omega_{cc}^+ \to \Lambda^0\,D_s^{\ast\,+}$ is allowed. In view of this
observation the result $\Gamma(\Omega_{cc}^+ \to \Sigma^0\,D_s^{\ast\,+})/
\Gamma(\Omega_{cc}^+ \to \Lambda^0\,D_s^{\ast\,+})=16.71$ obtained in
Ref.~\cite{Li:2020qrh} is rather puzzling.

We discuss a sample of linear relations between pairs of partner amplitudes
that follow from the tables which are based on SU(3) invariance. These
relations can also be derived by making use of the SU(2) isospin ($I$-spin),
$U$-spin and $V$-spin subgroups of SU(3). The identification of such bilinear
relations in terms of their $I$-spin, $U$-spin or $V$-spin origin may
sometimes be helpful from the mnemonic point of view. We discuss a few
examples for each of the $I$-spin, $U$-spin and $V$-spin subgroup relations
where we restrict ourselves to bilinear relations between pairs of particle
decay amplitudes, leaving out triangular relations between three decay
amplitudes. We thus concentrate on bilinear amplitude relations which are
governed by a single $I$-spin, $U$-spin or $V$-spin reduced matrix element. 

\subsection{$I$-spin sum rules}
As an example we take the CF relation
\begin{equation}\label{I-spin}
{\cal A}(\Lambda_{c}^+ \to \Sigma^+ \pi^0)
  = -{\cal A}(\Lambda_{c}^+ \to \Sigma^0 \pi^+)
\end{equation}
following from Table~\ref{TableCF}. The relation~(\ref{I-spin}) can also be
derived from the $I$-spin subgroup using the isospin decomposition
$0 \to (\Delta I=1)\otimes1\otimes1$.\footnote{Note that in this section we
prefer to use the angular momentum notation instead of the notation in
irreducible representations of SU(3) which are characterized by their
multiplicity (in boldface). In terms of these irreducible representations, the
decomposition at hand reads ${\bf 1}\to{\bf 3}\otimes{\bf 3}\otimes{\bf 3}$.}
We have used a calligraphic notation ${\cal A}$ in Eq.~(\ref{I-spin}) where
${\cal A}$ stands either for the p.v.\ amplitude $A$ or for the p.c.\
amplitude $B$. If one neglects isospin mass breaking effects, this leads to
the equality of the two rates
\begin{equation}
\Gamma(\Lambda_{c}^+\to\Sigma^+\pi^0)=\Gamma(\Lambda_{c}^+\to\Sigma^0\pi^+).
\end{equation}
There are many such isospin relations which can be read off from the tables.
We do not list them all except for an interesting prediction on the vanishing
of amplitudes following from $I$-spin symmetry. This prediction comes about by
noticing that the DCS transition $c\to d;\,s \to u$ is an isospin scalar with
$\Delta I=0$. As a result one has three DCS decays that are altogether isospin
forbidden. These are
\begin{equation}
\Gamma(\Omega_{c}^0 \to \Lambda^0\,\pi^0)=0,\quad
\Gamma(\Omega_{cc}^+ \to \Lambda_{c}^+\,\pi^0)=0,
\end{equation}
with an isospin decomposition $0 \to (\Delta I=0)\otimes 0\otimes 1$ or
$0 \to (\Delta I=0)\otimes 1 \otimes 0$. Of physical relevance are also the
related six forbidden decays involving the vector mesons $\omega$, $\phi$ and
$D_s^{\ast\,+}$. First of all, these are the five $\Delta C=1$ DCS decays
\begin{equation}
\Gamma(\Omega_{c}^0 \to \Sigma^0\,\eta_8,\,\Sigma^0\,\eta_1)=0,\quad
\Gamma(\Omega_{cc}^+ \to \Sigma_{c}^+\,\eta_8,\,\Sigma_{c}^+\,\eta_1)=0.
\end{equation}
Finally, the $\Delta C=2$ DCS decay $\Omega_{cc}^+ \to \Sigma^0\,D_s^+$ is
allowed from the topological diagram point of view IIb and III but forbidden
by the KPW theorem.

\subsection{$U$-spin sum rules}
First of all, note that the three effective CF, SCS and DCS Hamiltonians
belong to the same $\Delta U=1$ multiplet with $\Delta U_3$ quantum numbers
$+1$, $0$, and $-1$, respectively (cf.\ Table~\ref{Heff} and
Fig.~\ref{tripsex}). Again we shall only discuss bilinear $U$-spin sum rule
relations. As an example we consider the four decays
$\Xi_{c}^0 \to \Sigma^+ K^+$ (CF), $\Xi_{c}^0 \to \Sigma^+ \pi^-$ (SCS),
$\Xi_{c}^0 \to p\,K^-$ (SCS) and $\Xi_{c}^0 \to p\,\pi^-$ (DCS). There is only
one $U$-spin reduced matrix element for all four decays, as can be seen from
the decomposition of the $U$-spin product
$0\to(\Delta U=1)\otimes 1/2\otimes 1/2$. Accordingly, one has the amplitude
relations
\begin{equation}\label{Uspin}
 \frac{{\cal A}(\Xi_{c}^0 \to \Sigma^+ K^-)}{c^2}
   = - \frac{{\cal A}(\Xi_{c}^0 \to \Sigma^+ \pi^-)}{cs}
   = - \frac{{\cal A}(\Xi_{c}^0 \to p\,K^-)}{cs}
   = - \frac{{\cal A}(\Xi_{c}^0 \to p\,\pi^-)}{s^2},
\end{equation}
where we have divided the amplitudes by the relevant Cabibbo angle factors
(one has $V_{us}=\lambda=s=\sin\theta_C\approx 0.224$ and
$c=\cos\theta_C\approx 0.975$). The relations~(\ref{Uspin}) are in agreement
with the SU(3) analysis in the respective tables. More $U$-spin relations
involving also more than two decay amplitudes can be found in
Refs.~\cite{Korner:1978tc,Jia:2019zxi}.

\subsection{$V$-spin sum rules}
Again we shall only discuss bilinear $V$-spin sum rule relations. For example,
using Table~\ref{TableCF} for the CF decays
$B_{c}(\overline{\bf 3}) \to B({\bf 8})+ M({\bf 8})$ one reads off 
\begin{equation}\label{Vspin1}
{\cal A}(\Lambda_{c}^+ \to p \,\bar K^0)
  = - {\cal A}(\Xi_{c}^0 \to \Xi^- \,\pi^+),\quad
{\cal A}(\Xi_{c}^+ \to \Xi^0\,\pi^+)
  = - {\cal A}(\Xi_{c}^+ \to \Sigma^+\,\bar K^0) 
\end{equation}
which also follow from $V$-spin symmetry via the decomposition of the $V$-spin
products $1/2 \to (\Delta V=0) \otimes 1 \otimes 1/2$ and
$0 \to (\Delta V=0) \otimes 1/2 \otimes 1/2$. In the same way one finds the
$V$-spin CF amplitude relation
\begin{equation}\label{Vspin2}
{\cal A}(\Lambda_{c}^+ \to \Xi^0 \, K^+)
  = - {\cal A}(\Xi_{c}^0 \to \Sigma^+ \,K^-).
\end{equation}
The $V$-spin sum rules~(\ref{Vspin1}) and~(\ref{Vspin2}) have been listed
before in Ref.~\cite{Korner:1978tc} but have been missed in
Ref.~\cite{Jia:2019zxi}.

\subsection{Combined sum rules}
One can then combine the $U$-spin sum rule~(\ref{Uspin}) and the $V$-spin
sum rule~(\ref{Vspin2}) into one sum rule involving the five decays in
Eqs.~(\ref{Uspin}) and~(\ref{Vspin2}), as will be discussed further on.

The $U$-spin and $V$-spin amplitude relations do not directly translate into
the corresponding rate relations since the rate formula involves different
kinematical factors for the p.v.\ $|A|^2$ and p.c.\ $|B|^2$ contributions
which can differ considerably from one another, as can be seen by the rate
expression (cf.\ Appendix~\ref{spinkinematics})
\begin{equation}\label{rate}
\Gamma=\frac{p}{4\pi m_i^2}\Big(Q_+|A|^2+ Q_-|B|^2\Big)
  =\frac{p}{32\pi m_i^2}
  \Big(|H_{1/2\,0}^{\rm pv}|^2+|H_{1/2\,0}^{\rm pc}|^2\Big)
\end{equation}
($Q_\pm=(m_f\pm m_i)^2-m_k^2$, $p=\sqrt{Q_+Q_-}/(2m_i)$). In the last
expression of Eq.~(\ref{rate}) we have written the rate in terms of the two
helicity amplitudes of the process. As discussed in the following, it can make
a big difference for the SU(3) rate predictions depending on whether one
postulates SU(3) invariance for the invariant amplitudes $A$ and $B$ or for
the helicity amplitudes $H_{1/2\,0}^{\rm pv}$ and $H_{1/2\,0}^{\rm pc}$. The
SU(3) analysis in Refs.~\cite{Geng:2018plk,Geng:2018bow,Jia:2019zxi} has been
done using the helicity amplitude option while the SU(3) analysis in
Refs.~\cite{Verma:1995dk,Sharma:1996sc,Geng:2019xbo} is based on the invariant
amplitude option. The authors of Ref.~\cite{Lu:2016ogy} take an extreme view
and postulate SU(3) invariance for the rates or equivalently branching
fractions, thereby disregarding all kinematic factors. This may be justified
for isospin related decays but not for decays where the final states are
composed of differing hypercharges. Savage and Springer advocate the use of an
average of the kinematical $S$-wave and $P$-wave factors $Q_+$ and $Q_-$
when comparing rates to SU(3) predictions~\cite{Savage:1989qr}. 

\subsection{More sum rules}
Returning to the sum rules, more sum rules can be obtained by visually
scanning the tables for decays with repeating patterns of the values of
the tensor invariants. For example, for the pattern $(0,0,-2,0,2,-4,-1)$ one
has the amplitude relation
\begin{equation}
\sqrt{2}{\cal A}(\Lambda_{c}^+ \to \Sigma^0\,\pi^+)
  = - \sqrt{2}{\cal A}(\Lambda_{c}^+ \to \Sigma^+\,\pi^0)
  = - {\cal A}(\Xi_{c}^0 \to \Xi^0\,K^0)
  = {\cal A}(\Xi_{c}^0 \to n\,\bar K^0).
\end{equation}
The four listed decays proceed via $W$-exchange alone.

The four decays $\Lambda_{c}^+\to n\,\pi^+$, $\Lambda_{c}^+\to\Lambda^0\,\pi^+$,
$\Lambda_{c}^+\to\Sigma^0\,\pi^+$ and $\Lambda_{c}^+\to p\,K^-$ follow from the
pattern $(x,x,x,x,x,-4,-1)$. The tensor invariant relation $\hat I_4=4 I_5$
holds in addition to the four inherent relations listed in Eq.~(\ref{linrel1}).
This implies that there are only two topological invariants each for the
amplitudes $A$ and $B$ describing the four decays. For definiteness we take
the topological invariants $\mathbfcal{T}_3$ and $\mathbfcal{T}_5$. In this
way one can then derive the sum rule
\begin{equation}
   |{\cal A}|^2(\Lambda_{c}^+ \to n\,\pi^+)/s^2=
   3\,|{\cal A}|^2(\Lambda_{c}^+ \to \Lambda^0\,\pi^+)
   +|{\cal A}|^2(\Lambda_{c}^+ \to \Sigma^0\,\pi^+)
   -|{\cal A}|^2(\Lambda_{c}^+ \to p\,\bar K^0)
\end{equation}
involving only squared magnitudes of the amplitudes ${\cal A}=A,B$. A
corresponding sum rule was listed in Eq.~(27) of Ref.~\cite{Lu:2016ogy}.
However, the authors of Ref.~\cite{Lu:2016ogy} have written their sum rule in
terms of rates, or equivalently, in terms of branching fractions assuming
SU(3) to hold for the rates. In view of the widely diverging kinematical
factors in the rate expression Eq.~(\ref{rate}) the validity of their sum rule
may be called into question.

Further sum rules are obtained by visually scanning the tables for decays
with repeating patterns of the values of tensor invariants. One of the
patterns is $(0,0,0,-2,0,0,-1)/N_j$, where $N_j$ is the product of
normalization factors of the flavor wave functions in the decay process
$(B_{c}\to B+M)_j$. Collecting the decays with this pattern, we leave out
decays related to isospin. Left are the four decays $\Xi_{c}^0\to\Sigma^+K^-$,
$\Xi_{c}^0\to\Sigma^+\pi^-$, $\Xi_{c}^0\to p\,K^-$ and $\Xi_{c}^0\to p\,\pi^-$
listed in Eq.~(\ref{Uspin}), and the decay $\Lambda_{c}^+ \to \Xi^0 \, K^+$ (CF)
listed in Eq.~(\ref{Vspin2}). The decay $\Omega_{c}^0 \to \Sigma^+ K^-$ (SCS)
also possesses the aforementioned flavor pattern and will be included in our
analysis by appealing to SU(6) symmetry (as discussed in
Sec.~\ref{topotensor}), even though the decay belongs to the class
$B_{c}({\bf 6}) \to B({\bf 8})+M({\bf 8})$ and not to the class
$B_{c}(\overline{\bf 3}) \to B({\bf 8})+M({\bf 8})$ as the other five decays.
The SU(3) analysis can then be done by relating the amplitudes of these six
decays by the inverse of the normalization factors $1/N_j$. There are no
tree-graph contributions to these decays and, adding a bit of dynamics from
the current algebra approach, one expects that the p.v.\ amplitude $A$
vanishes since $I_3=\hat I_3=0$ in these cases. This observation will be the
basis of our following analysis where we assume that the decays proceed via
p.c.\ $P$-wave $W$-exchange contributions and where we assume that SU(3)
invariance holds for the dimensionless amplitude $B$.

\subsection{Spin kinematics}
When comparing rates, one has to take into account the spin kinematical
$P$-wave rate factor which, according to Eq.~(\ref{rate}) reads
$p\,Q_-/m_i^2=Q_-\sqrt{Q_+Q_-}/2m_i^3$. One can then write down predictions
for the six relative rates which read
\begin{eqnarray}\label{rateSU3ia}
 \Gamma_{\Lambda_{c}^+ \to \Xi^0\,K^+}&:&
 \Gamma_{\Xi_{c}^0 \to \Sigma^+ K^-}:
 \Gamma_{ \Xi_{c}^0 \to \Sigma^+ \pi^-}:
 \Gamma_{\Xi_{c}^0 \to p\,K^-}: 
 \Gamma_{ \Omega_{c}^0 \to \Sigma^+ K^-}:
 \Gamma_{ \Xi_{c}^0 \to p\,\pi^-}\ =\nonumber\\
 = \qquad 1\,c^4\quad&:& 2.25\, c^4 \,\quad :\,2.83\,c^2s^2 \, \,\,:
 \,\,3.86\, c^2 s^2\,\, :\, 19.22\,c^2 s^2 \,\, : \,\,\,4.54\, s^4 \qquad
\end{eqnarray}
If instead one assumes that SU(3) holds for the dimensionful helicity
amplitudes (see Appendix~\ref{spinkinematics}) the corresponding kinematical
factor is $\Gamma\sim p/m_i^2|H^{\rm pc}|^2$, and one arrives at the relative
rate prediction
\begin{eqnarray}\label{rateSU3ha}
 \Gamma_{\Lambda_{c}^+ \to \Xi^0\,K^+}&:&
 \Gamma_{\Xi_{c}^0 \to \Sigma^+ K^-}:
 \Gamma_{ \Xi_{c}^0 \to \Sigma^+ \pi^-}:
 \Gamma_{\Xi_{c}^0 \to p\,K^-}: 
 \Gamma_{ \Omega_{c}^0 \to \Sigma^+ K^-}:
 \Gamma_{ \Xi_{c}^0 \to p\,\pi^-}\ =\nonumber\\
= \qquad 1\,c^4\quad&:& 1.13\, c^4 \,\quad :\,1.23\,c^2s^2 \, \,\,:
 \,\,1.29\, c^2 s^2\,\, :\, 6.70\,c^2 s^2 \,\, : \,\,\,1.38\, s^4 \qquad
\end{eqnarray} 
It is apparent that the SU(3) symmetry predictions diverge widely depending
on whether one postulates SU(3) symmetry for the invariant amplitudes or for
the helicity amplitudes. As the numbers in Eqs.~(\ref{rateSU3ia})
and~(\ref{rateSU3ha}) show, the difference in the predicted rate ratios can
amount up to a factor of $3.3$ for the rate ratio
$\Gamma_{ \Xi_{c}^0 \to p\,\pi^-}/\Gamma_{\Lambda_{c}^+ \to \Xi^0\,K^+}$. The
issue of whether to assume SU(3) invariance for the invariant or helicity
ampltitudes cannot be decided on a purely theoretical basis except for an
aesthetic argument favoring the invariant amplitude option because the
invariant amplitudes carry no mass dimension. In the long run only experiment
can decide which of the two options has to be favored. 

\subsection{Exclusive $W$-exchange contributions}
The hope is that from these decays one can glimpse a hint of the dynamics
governing the $W$-exchange contributions without interference from the tree
diagram contributions. Apart from the six decays listed in
Eq.~(\ref{rateSU3ia}) there are a large number of decays that proceed via
$W$-exchange contributions alone. Some examples are the CF decays
$\Lambda_{c}^+ \to \Sigma^0\,\pi^+$ and
$\Xi_{c}^0 \to \Xi^0\,\pi^0$ in Table~\ref{TableCF}. All of the decays
$B_{cc} \to B+D$ in Table~\ref{TableBccBD} have solely $W$-exchange
contributions.

\subsection{Tree diagrams alone}
While theoretical progress on the dynamics of the $W$-exchange contributions
has been rather slow, there exist a large body of literature on the treatment
of the tree diagram contributions to charm baryon decays. The tree diagram
contributions are related to the current induced charm baryon to light baryon
transitions which have been studied in a large number of papers using a
variety of theoretical models, including also lattice calculations. Among the
models are
\begin{itemize}
\item[i)]the relativistic quark
model~\cite{Hussain:1990ai,Hussain:1990uu,Hussain:1992rb}
\item[ii)]the MIT bag model~\cite{PerezMarcial:1989yh}
\item[iii)]the covariant confined quark
model~\cite{Gutsche:2014zna,Gutsche:2015rrt,Gutsche:2018utw,Gutsche:2019iac}
\item[iv)]the relativistic quark--diquark
model~\cite{Faustov:2016yza,Faustov:2019ddj}
\item[v)]the semirelativistic quark model~\cite{Pervin:2005ve,Pervin:2006ie} 
\item[vi)]SU(3) based analysis~\cite{Lu:2016ogy,Geng:2018plk}
\item[vii)]light-cone sum
rules~\cite{Liu:2009sn,Azizi:2009wn,Azizi:2011mw,Duan:2020xcc}
\item[viii)]the light front quark model~\cite{Zhao:2018zcb}
\item[ix)]QCD sum rules~\cite{Zhao:2021sje} 
\item[x)]lattice calculations~\cite{Zhang:2021oja}
\end{itemize}
For the single charm baryon decays one finds only one decay which proceeds via
the tree diagram alone: the well-measured SCS decay
$\Lambda_{c} \to p\,\phi$~\cite{Alexander:1995hd,Abe:2001mb,Ablikim:2016tze}
listed in Tables~\ref{TableSCSa} and~\ref{TableSCSb}. This decay has gained a
certain prominence since it allows one to fix the effective number of colors
$N_{c}({\rm eff})$ coming into play when calculating the nonleptonic decay
amplitude from the current induced transition amplitudes.

Contrary to the single charm decays, there are a large number of double charm
decays from the class $B_{cc}(3)\to B_{c}(6)+M(8)$ in Table~\ref{Tablecc3,6,8}
which decay via the tree diagram alone. From the CF decays of this class these
are the four decays $\Xi_{cc}^{++} \to \Sigma_{c}^{++}\,\bar K^0$,
$\Xi_{cc}^{++} \to \Xi_{c}^{\prime \,+}\,\pi^+$,
$\Omega_{cc}^+ \to \Omega_{c}^0\,\pi^+$ and
$\Omega_{cc}^+ \to \Xi_{c}^{\prime \,+}\,\bar K^0$. From the flavor flow of
diagram IIa one can check  that these four decays cannot proceed via diagram
IIa. As concerns the SCS and DCS decays, one finds a number of decays of this
class in Table~\ref{Tablecc3,6,8} which we do not list separably.

\subsection{Weak $1\to 3$ quark decays}
The topological diagrams in Fig.~\ref{topo} can be divided up into two
classes. Diagrams Ia, Ib and IIb are characterized by weak $1\to 3$ short
distance quark transitions while diagrams IIa and III are governed by $2\to 2$
short distance quark transitions. Diagrams IIa and III both involve the
creation of a quark pair from the vacuum. Quantum mechanically such a
situation can be handled by the $^3P_0$ model which, however, brings in a
considerable amount of uncertainties in terms of model parameters and an
unknown energy dependence of the quark pair creation process. The covariant
confined quark model (CCQM) is a quantum field theoretical model where the
topological diagrams are interpreted as Feynman loop integrals with nonlocal
quark particle interactions. In principle, a field theoretical model as the
CCQM should also be able to describe transitions involving quark pair
creation. However, as it turns out, the CCQM calculation of diagrams IIa and
III becomes unreliable because the energy release in charm baryon decays is
quite large and the energetic light quark corresponding to one of the light
quarks in the created pair probes the nonlocal Gaussian wave functions in a
problematic region. The CCQM does, however, provide reliable results for
diagrams Ia, Ib and IIb as exemplified by the recent calculation of the decay
$\Xi_{cc}^{++} \to \Xi_{c}^+ \,\pi^+$ which obtains contributions only from
diagrams Ia, Ib and IIb. Therefore, it is interesting to provide a list of
decays that decay via the rearrangement diagrams Ia, Ib and IIb. It would be
interesting to calculate this class of decays either in the quantum mechanical
model or in the quantum field theoretical CCQM model.

As an inspection of the tables shows, there are many decays that belong to
this class. Listing again only CF decays, we find the quark rearrangement
transitions ($I_3,I_4,I_5=0$)
\begin{equation}
\Xi_{c}^+\to\Sigma^+\bar K^0,\Xi^0\pi^+,\quad
\Omega_{c}\to\Xi^0\bar K^0,\quad
\Xi_{cc}^+\to\Xi_{c}^+\pi^+,\Sigma^+D^+,\quad
\Omega_{cc}^+\to\Xi_{c}^+\bar K^0.
\end{equation}

\subsection{No $S$-wave contributions from $W$-exchange}
According to the current algebra analysis, the $W$-exchange contribution to
the p.v.\ amplitude $A$ is proportional to $I_3-\hat I_3$, i.e.\
$A(\mbox{$W$-exchange})\sim I_3-\hat I_3$. The vanishing of
$A(\mbox{$W$-exchange})$ can be realized in two ways. One can have 
i) $I_3 =\hat I_3,\, I_3 \neq 0$ or ii) $I_3 =\hat I_3=0$. The tables show
that the first case $I_3 =\hat I_3,\, I_3 \neq 0$ is rather rare. We found
only two decays with this signature, namely the CF decay
$\Lambda_{c}^+ \to \Lambda^0 \,\pi^+$ and the SCS decay
$\Lambda_{c}^+ \to \Lambda^0 \,K^+$. Both of these decays also have tree graph
contributions which will also contribute to the $S$-wave amplitude. Therefore,
concerning the the vanishing of the p.v.\ amplitude $A$ and thus about the
vanishing of the asymmetry parameter, one cannot say anything general about
these two decays. The second case $I_3=\hat I_3=0$ occurs rather frequently.
Among them are the six decays listed in Eq.~(\ref{rateSU3ia}) including their
isospin partners. There are five DCS decays of the $\Omega_{c}^0$ charm baryon
with the signature $I_3=\hat I_3=0$. These are
\begin{eqnarray}
\mbox{DCS}&:&\Omega_{c}^0 \to \Sigma^+\pi^-,\,\Sigma^0\pi^0,\,\Sigma^-\pi^+,\,
pK^-,\,n \bar K^0
\end{eqnarray}
Since there are no tree graph contributions to these decays, within the
current algebra approach the five decays are thus predicted to have zero
asymmetry.

All of the double charm baryon decays
$B_{cc}({\bf 3}) \to B_{c}({\bf 6}) + M({\bf 8})$ listed in
Table~\ref{Tablecc3,6,8} have the signature $I_3=\hat I_3=0$. Of particular
interest are those decays that have no tree contributions. These are
\begin{eqnarray}
\mbox{CF}&:&\Xi_{cc}^+ \to \Sigma_{c}^{++} K^-,\, \Omega_{c}^0 K^+, \nonumber \\
\mbox{SCS}&:&\Omega_{cc}\to \Xi_{c}^0 \pi^+, \nonumber \\
\mbox{DCS}&:&\Omega_{cc}\to \Sigma_{c}^+ \pi^0. 
\end{eqnarray}
The asymmetry parameters of these decays are predicted in the current algebra
approach to be zero.

As concerns the p.c.\ amplitude $B$, one cannot find a single example in the
tables for which $B(\mbox{$W$-exchange})=0$ holds fully nontrivially. The
signature would be $I_3+2I_4+6I_5=0$ {\em and\/} $\hat I_3+2\hat I_4+6I_5=0$.
Independently, each of these signatures is found in many decays where the other
signature is not given. Still, there are a few examples in which both
conditions are satisfied, though the second only in the trivial way
$\hat I^3=\hat I^4=I^5=0$. Except for a single DCS example
$\Omega_{c}^0\to\Lambda^0\eta_\omega$ from Table~\ref{TableOmegac}, all the
examples are from Table~\ref{cc3,bar3,8}. We distinguish between cases with
and without tree contributions.
\begin{itemize}
\item[i)]The pattern $(0,0,+4,-2,0,0,0)$ or multiples of it, indicating a
  transition without tree contributions, is found in the CF decay
  $\Xi_{cc}^+\to\Xi_{c}^+\eta_\phi$, and in the DCS decays
  $\Omega_{c}^0\to\Lambda^0\eta_\omega$ and
  $\Omega_{cc}^+\to\Lambda_{c}^+\eta_\omega$, with the common property that
  the meson is an $\eta$.
\item[ii)]The pattern $(-2,+1,+4,-2,0,0,0)$ is found in the SCS decay
  $\Omega_{cc}^+\to\Xi_{c}^+\eta_\omega$.
\item[iii)]The pattern $(+2,-1,+4,-2,0,0,0)$ or multiples of it, indicating a
transition with tree contributions, is found in the CF decays
  $\Xi_{cc}^+\to\Xi_{c}^0\pi^+$ and $\Xi_{cc}^+\to\Lambda_{c}^+\bar K^0$, in
  the SCS decays $\Xi_{cc}^+\to\Xi_{c}^0K^+$, $\Omega_{cc}^+\to\Xi_{c}^+\pi^0$
  and $\Omega_{cc}^+\to\Xi_{c}^0\pi^+$, and in the DCS decays
  $\Omega_{cc}^+\to\Xi_{c}^0K^+$ and $\Omega_{cc}^+\to\Xi_{c}^+K^0$, i.e.\ for
  decays into pions and kaons.
\end{itemize}

%%%%%%%%%%%%%%%%%%%%%%%%%%%%%%%%%%%%%%%%%%%%%%%%%%%%%%%%%%%%%%%%%%%%%%%%%%%%%%%
\section{\label{summary}Summary and Outlook}
%%%%%%%%%%%%%%%%%%%%%%%%%%%%%%%%%%%%%%%%%%%%%%%%%%%%%%%%%%%%%%%%%%%%%%%%%%%%%%%
We have collected and organized a wealth of material on 196 nonleptonic
$B_i(1/2^+)\to B_f(1/2^+)+M_k(0^-)$ decays of single and double charm baryons
which else is scattered among many papers in the literature. We have collected
some group theoretical material on the symmetry groups SU(2)$_I$, SU(2)$_U$,
SU(2)$_V$, SU(3) and SU(6) needed in the analysis of the decays. We have
presented the results of calculating the values of the seven topological
tensor invariants for each of the 196 decays in a number of tables. Without
having to do any explicit numerical calculation, the information contained in
the tables leads to a number of important observations on the structure of the
nonleptonic decays which we explicate. In the second part of the paper we have
discussed a dynamical approach based on current algebra and the pole model,
the results of which can be represented very compactly in terms of the
topological tensor invariants introduced in the first part of the paper. The
compact result was achieved by performing the sum over the intermediate baryon
ground states inherent to the current algebra approach, using a completeness
relation. We have critically examined a modified version of the current
algebra approach introduced recently in the literature. From the experimental
point of view, the key experiment to discard or to keep this new modified
approach is to measure the asymmetry parameter in the decay
$\Lambda_{c}^+ \to \Xi^0\,K^+$ and, for that matter, to measure the asymmetry
parameter in all the decays listed in Eq.~(\ref{rateSU3ia}) plus their isospin
partners. While the asymmetry parameters in these decays are predicted by the
modified current algebra approach to be large, in the standard current algebra
approach they are predicted to be zero ore close to zero. Our results on the
topological tensor invariants are useful also for other dynamical calculations
of charm baryon decays such as constituent quark model approaches. In this
context we are planning to refine the covariantized constituent quark model
calculation of Ref.~\cite{Korner:1992wi} and extend it to include also the SCS
and DCS decays of single charm baryons.

An overall picture is difficult to obtain, as the experiments are on the way.
Therefore, no definite decision can be made on whether and to what extend our
estimates are compatible with the current data, except for the examples we
have exposed. Still, we hope that the material presented in this paper will
aid and provide extra stimulus for the experimental search of the many missing
decay modes of charm baryon decays including a measurement of their absolute
branching ratios -- and, what is of utmost importance, to measure the
asymmetry parameters in their decays. We are thus looking forward to new
experimental results on charm baryon decays from Belle II, BESIII and LHCb
which should be forthcoming in the near future.

\subsection*{Acknowledgments}
The research presented in this paper was begun while JGK was visiting the
Quaid-y-Azam University in Islamabad in 2006. His thanks go to Jamil Aslam,
Alimjian Kadeer and the late Faheem Hussain$^\dagger$ for their participation
in the early stages of this work. We acknowledge fruitful discussions with
M.A.~Ivanov and V.E.~Lyubovitskij. The research was supported in part by the
European Regional Development Fund under Grant No.~TK133. SG acknowledges
support by the PRISMA+ Cluster of Excellence at the University of Mainz. The
almost finished manuscript was finalized by SG in thankful remembrance of his
deceased collaborator and friend JGK.

\begin{appendix}

\section{\label{tensors}Representations of tensors}
\setcounter{equation}{0}\def\theequation{A\arabic{equation}}
For the tensor representations of the meson and baryon flavor wave functions
we use the phase convention of the second edition of
Lichtenberg~\cite{Lichtenberg:1978pc} (differing from the first edition).

\subsection{Flavor space wave functions for the mesons}
For the meson flavor wave functions we use Table~12.7 in
Ref.~\cite{Lichtenberg:1978pc},
\begin{eqnarray}
\pi^+&:& M^{1}{}_{2}=-1\,,\nonumber\\
\pi^-&:& M^{2}{}_{1}=1\,,\nonumber\\
\pi^0&:& M^{1}{}_{1}=-M^{2}{}_{2}=1/\sqrt{2}\,,\nonumber\\
K^0&:&   M^{2}{}_{3}=1\,,\nonumber\\
\bar{K}^0&:&   M^{3}{}_{2}=-1\,,\nonumber\\
K^+&:&   M^{1}{}_{3}=1\,,\nonumber\\
K^-&:& M^{3}{}_{1}=1\,,\nonumber\\
\eta_\omega&:& M^{1}{}_{1}=M^{2}{}_{2}=1/\sqrt{2}\,,\nonumber\\
\eta_\phi&:& M^{3}{}_{3}=1\,,\nonumber\\
\eta_8&:&M^{1}{}_{1}=M^{2}{}_{2}=-\tfrac12M^{3}{}_{3}=1/\sqrt{6}\,,\nonumber\\
\eta_1&:&M^{1}{}_{1}=M^{2}{}_{2}=M^{3}{}_{3}=1/\sqrt{3}\,,\nonumber\\
D^+&:&M^{4}{}_{2}=-1\,, \quad D^-\,:\,M^{2}{}_{4}=-1\,,\nonumber\\
D^0&:&M^{4}{}_{1}=1 \,,\,\,\quad \bar D^0\,:\,M^{1}{}_{4}=-1\,,\nonumber\\
D_s^+&:&M^{4}{}_{3}=1\,, \,\,\quad D_s^-\,:\,M^{3}{}_{4}=-1\,.
\end{eqnarray}
By employing the conjugate representation $M_a^b:=\bar M_a{}^b=M^a{}_b$ of the
flavor space wave functions in the main text we take into account that the
outgoing mesons are considered as incoming antiparticles to the interaction.
The matrix $M=M_k$ is identical to the conserved vector charge operator of the
pseudoscalar meson $k$ given in the first column of this list.

\subsection{Flavor space wave functions for the baryons}
For the baryon flavor wave function use the second set listed in Table~12.4
labelled as ``Octet 2'' and the second set listed in Table~12.5 labelled as
``Second $20_M$'' in Ref.~\cite{Lichtenberg:1978pc}. Instead of the ordering
$B_{[ab]c}$ we use the ordering $B_{c[ab]}$ such that our flavor wave
functions are antisymmetric in the last two indices.
\begin{itemize}
\item The light baryon octet:
\begin{eqnarray}
p(uud)&:&\quad B_{112}=-B_{121}=1/\sqrt{2}\nonumber\\
n(udd)&:&\quad B_{212}=-B_{221}=1/\sqrt{2}\nonumber\\
\Sigma^{+}(uus)&:&\quad B_{113}= -B_{131}=1/\sqrt{2}\nonumber\\
\Sigma^{0}(uds)&:&\quad B_{123}=-B_{132}=B_{213}=-B_{231}=1/2\nonumber\\
\Sigma^{-}(dds)&:&\quad B_{223}=-B_{232}=1/\sqrt{2}\nonumber\\
\Lambda(uds)&:&\quad B_{132}=-B_{123}=B_{213}=-B_{231}=\tfrac{1}{2}B_{312}
  =-\tfrac{1}{2}B_{321}=1/\sqrt{12}\nonumber\\
\Xi^{0}(uss)&:&\quad B_{313}=-B_{331}=1/\sqrt{2}\nonumber\\
\Xi^{-}(dss)&:&\quad B_{323}=-B_{332}=1/\sqrt{2}
\end{eqnarray}
\item The $C=1$ charm baryon antitriplet:
\begin{eqnarray}
\Lambda_{c}^{+}(udc)&:&\quad B_{142}=-B_{124}=B_{214}=-B_{241}
  =\tfrac{1}{2}B_{412}=-\tfrac{1}{2}B_{421}=1/\sqrt{12}\nonumber\\
\Xi^{+}_{c}(usc)&:&\quad B_{143}=-B_{134}=B_{314}=-B_{341}
  =\tfrac{1}{2}B_{413}=-\tfrac{1}{2}B_{431}=1/\sqrt{12}\nonumber\\
\Xi^{0}_{c}(dsc)&:&\quad B_{243}=-B_{234}=B_{324}=-B_{342}
  =\tfrac{1}{2}B_{423}=-\tfrac{1}{2}B_{432}=1/\sqrt{12}\qquad
\end{eqnarray}
\item The $C=1$ charm baryon sextet:
\begin{eqnarray}
\Sigma^{++}_{c}(uuc)&:&\quad B_{114}=-B_{141}=1/\sqrt{2}\nonumber\\
\Sigma^{+}_{c}(udc)&:&\quad B_{214}=B_{124}=-B_{142}=-B_{241}=1/2\nonumber\\
\Sigma^{0}_{c}(ddc)&:&\quad B_{224}=-B_{242}=1/\sqrt{2}\nonumber\\
\Xi^{\prime+}_{c}(usc)&:&\quad B_{314}=B_{134}=-B_{341}=-B_{143}=1/2\nonumber\\
\Xi^{\prime\,0}_{c}(dsc)&:&\quad B_{324}=B_{234}=-B_{342}=-B_{243}=1/2
  \nonumber\\
\Omega^{0}_{c}(ssc)&:&\quad B_{334}=-B_{343}=1/\sqrt{2}
\end{eqnarray}
\item The $C=2$ double charm baryon triplet:
\begin{eqnarray}
\Xi^{++}_{cc}(ucc)&:&\quad B_{414}=-B_{441}=1/\sqrt{2}\nonumber\\
\Xi^{+}_{cc}(dcc)&:&\quad B_{424}=-B_{442}=1/\sqrt{2}\nonumber\\
\Omega^{+}_{cc}(scc)&:&\quad B_{434}=-B_{443}=1/\sqrt{2}
\end{eqnarray}
\end{itemize}
The flavor space wave functions of the conjugate ${\bf 20'}$ states
are given by $\bar B_{ijk}=B^{ijk}$.

%%%%%%%%%%%%%%%%%%%%%%%%%%%%%%%%%%%%%%%%%%%%%%%%%%%%%%%%%%%%%%%%%%%%%%%%%%%%%%%
\subsection{The weak transition tensor $H^{[ij]}_{[kl]}({\cal O_-})$}
%%%%%%%%%%%%%%%%%%%%%%%%%%%%%%%%%%%%%%%%%%%%%%%%%%%%%%%%%%%%%%%%%%%%%%%%%%%%%%%
Early foundations for the properties in this section and in Table~\ref{Heff}
are found in Ref.~\cite{Korner:1970xq}.
\begin{itemize}
\item
The $\Delta C=1$ effective Hamiltonian
\begin{eqnarray}
 CF:\quad c\to s,\quad d\to u:
 &&H^{[42]}_{[31]}=-H^{[24]}_{[31]}= -H^{[42]}_{[13]}
 = H^{[24]}_{[13]}=1
 \nonumber\\
\mbox{SCS:($a$)}\quad c\to s,\quad s\to u:
&&H^{[43]}_{[31]}=-H^{[34]}_{[31]}=-H^{[43]}_{[13]}=H^{[34]}_{[13]}=1
\nonumber\\
\mbox{($b$)}\quad c\to d,\quad d\to u:
&&H^{[42]}_{[21]}=-H^{[24]}_{[21]}=-H^{[42]}_{[12]}=H^{[24]}_{[12]}=1
\nonumber\\
\mbox{DCS:}\quad c\to d,\quad s\to u:
 &&H^{[43]}_{[21]}=-H^{[34]}_{[21]}=-H^{[43]}_{[12]}=H^{[34]}_{[12]}=1
\end{eqnarray}
\item The $\Delta C=0$ effective Hamiltonian
\begin{eqnarray}
\mbox{SCS:($a'$)}\quad  s\to u,\quad u\to d:
&&H^{[31]}_{[12]}=-H^{[13]}_{[12]}=-H^{[31]}_{[21]}=H^{[13]}_{[21]}=1
\nonumber\\
\mbox{($b'$)}\quad c\to d,\quad s\to c:
 &&H^{[43]}_{[24]}=-H^{[34]}_{[24]}=-H^{[43]}_{[42]}=H^{[34]}_{[42]}=1
\end{eqnarray}
\item The $\Delta S=1$ hyperon decay effective Hamiltonian
\begin{equation}\label{H}
\mbox{SCS:}\quad s\to u,\quad u\to d:\quad
 H^{[31]}_{[12]}=-H^{[13]}_{[12]}= -H^{[31]}_{[21]}=H^{[13]}_{[21]}=1 
\end{equation}
\end{itemize}

%%%%%%%%%%%%%%%%%%%%%%%%%%%%%%%%%%%%%%%%%%%%%%%%%%%%%%%%%%%%%%%%%%%%%%%%%%%%%%%
\section{\label{complete}Derivation of the completeness relation}
%%%%%%%%%%%%%%%%%%%%%%%%%%%%%%%%%%%%%%%%%%%%%%%%%%%%%%%%%%%%%%%%%%%%%%%%%%%%%%%
\setcounter{equation}{0}\def\theequation{B\arabic{equation}}
The completeness relation~(\ref{completeness}) can be derived by writing down
a sixth rank tensor $T^{bcd}_{kmn}$ built from a linear combination of
the products of three $\delta$-functions which is separately antisymmetric
under the exchange of the tensor labels $c\leftrightarrow d$ and
$m\leftrightarrow n$. One has 
\begin{equation}\label{completeness1}
\sum_{\ell} B^\ell_{k[mn]}B_\ell^{b[cd]}
  =a(\delta_k^b \delta_m^c \delta_n^d- \delta_k^b \delta_m^d \delta_n^c)
  +b\Big((\delta_m^b \delta_n^c \delta_k^d -\delta_m^b \delta_n^d \delta_k^c) 
  +(\delta_n^b \delta_k^c \delta_m^d -\delta_n^b \delta_k^d \delta_m^c)\Big).
\end{equation}
By contracting~(\ref{completeness1}) again with $B^{\ell'}_{b[cd]}$ and using
$B^{\ell'}_{m[nk]}+B^{\ell'}_{n[km]}=- B^{\ell'}_{k[mn]}$ one obtains
$a-b=1/2$. The coefficients $a$ and $b$ can be determined by contracting
Eq.~(\ref{completeness1}) with $\delta_b^k \delta_{c}^m \delta_d^n$ which gives
\begin{equation}
\frac13 N(N^2-1)=a(N^3-N^2)+2b(N-N^2) 
\end{equation}
due to the normalization and orthogonality relation~(\ref{orthogonality})
with the solution $b=-a/2$. One thus has $a=2/6$ and $b=-1/6$ as in the
completeness relation~(\ref{completeness}). 

\begin{figure}\begin{center}
\epsfig{figure=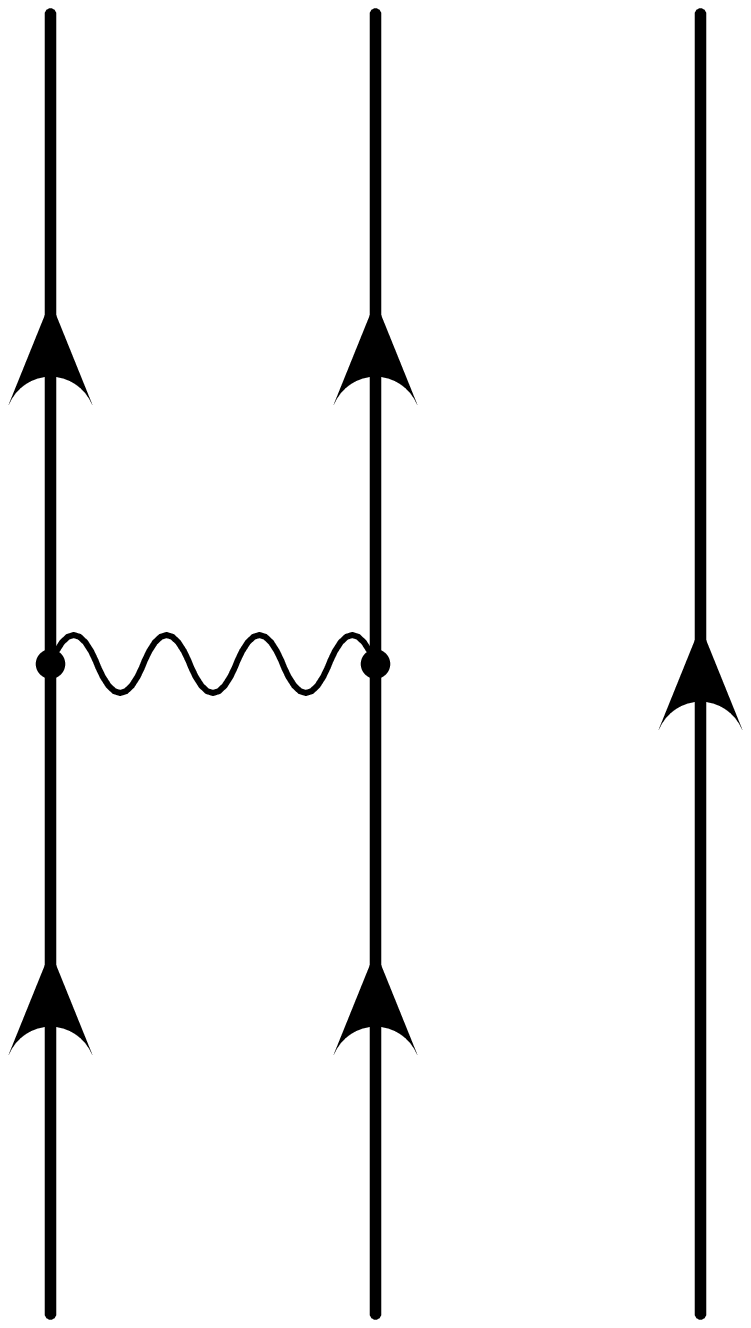, scale=0.25}\qquad \qquad \qquad
\epsfig{figure=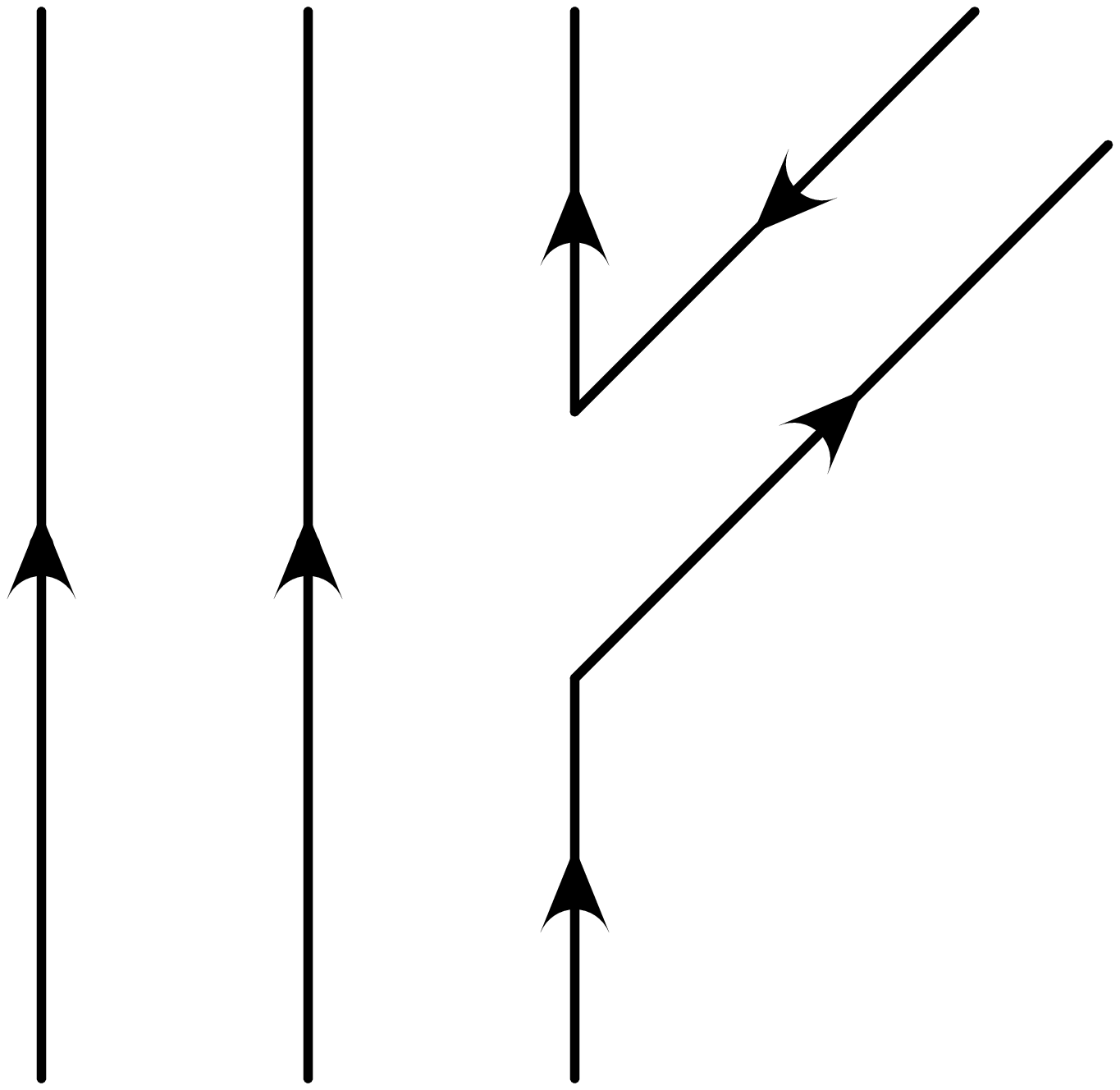, scale=0.25}\end{center}
\caption{\label{weakstrong} weak transition (left) and strong
  transition (right)}
\end{figure}

%%%%%%%%%%%%%%%%%%%%%%%%%%%%%%%%%%%%%%%%%%%%%%%%%%%%%%%%%%%%%%%%%%%%%%%%%%%%%%%
\section{\label{weak-strong}SU(3) structure of weak and strong\\ transition
  matrix elements}
%%%%%%%%%%%%%%%%%%%%%%%%%%%%%%%%%%%%%%%%%%%%%%%%%%%%%%%%%%%%%%%%%%%%%%%%%%%%%%%
\setcounter{equation}{0}\def\theequation{C\arabic{equation}}
In this Appendix we present explicit values for the weak and strong matrix
elements needed in the current algebra analysis of Sec.~\ref{sample} and
Appendix~\ref{scrutiny} and depicted in Fig.~\ref{weakstrong}. We begin with a
discussion of the weak transition matrix elements.

\begin{table}
\caption{\label{weakme}
  SU(3) properties of the weak transition matrix elements}
\begin{center}
\begin{tabular}{cccl}
\hline \hline \\[-11pt]
$\phantom{\Delta C=1}$
& matrix element & direct product & $N_{\rm SU(3)}$ \\[3pt] \hline \\[-11pt]
%%% decay 1
$\Delta C=1$
& $\langle B({\bf 8})|H^{\rm pc}({\bf 6}) |B_{c}(\overline{\bf 3})\rangle$
& $\overline{\bf 3}\otimes\overline{\bf 6}\otimes{\bf 8}$& 1 \\[3pt]
%%% decay 2
$\phantom{\Delta C=1}$
& $\langle B{(\bf 8})|H^{\rm pc}({\bf 6})|B_{c}({\bf 6})\rangle$
& ${\bf 6}\otimes\overline{\bf 6}\otimes{\bf 8}$ & 1 \\[3pt]
%%% decay 3
$\phantom{\Delta C=1}$
& $\langle B_{c}(\overline{\bf 3})|H^{\rm pc}({\bf 6})|B_{cc}({\bf 3})\rangle$
& ${\bf 3}\otimes\overline{\bf 6}\otimes{\bf 3}$ & 1 \\[3pt]
%%% decay 4
$\phantom{\Delta C=1}$
& $\langle B_{c}({\bf 6)}|H^{\rm pc}({\bf 6})|B_{cc}({\bf 3})\rangle$
& ${\bf 3}\otimes\overline{\bf 6}\otimes\overline{\bf 6}$ & 0 \\[3pt]
%%% decay 5
$\Delta C=0$
& $\langle B_{c}(\overline{\bf 3})|H^{\rm pc}({\bf 8})
  |B_{c}(\overline{\bf 3})\rangle$
& $\overline{\bf 3}\otimes{\bf 8}\otimes{\bf 3}$ & 1\quad ($a',b'$) \\[3pt]
%%% decay 6
$\phantom{\Delta C=0}$
& $\langle B_{c}({\bf 6})|H^{\rm pc}({\bf 8})|B_{c}(\overline{\bf 3})\rangle$
& $\overline{\bf 3}\otimes{\bf 8}\otimes\overline{\bf 6}$ & 1 \quad ($b'$)
\\[3pt]
%%% decay 7
$\phantom{\Delta C=0}$
& $\langle B_{c}(\overline{\bf 3})|H^{\rm pc}({\bf 8})|B_{c}({\bf 6})\rangle$
& ${\bf 6}\otimes{\bf 8}\otimes{\bf 3}$ & 1 \quad ($b'$) \\[3pt]
%%% decay 8
$\phantom{\Delta C=0}$
& $\langle B_{c}({\bf 6})|H^{\rm pc}({\bf 8})|B_{c}({\bf 6})\rangle$
& ${\bf 6}\otimes{\bf 8}\otimes\overline{\bf 6}$ & 1 \quad ($b'$)
\\[3pt]
\hline\hline
\end{tabular}
\end{center}
\end{table}

\subsection{Weak matrix elements}
In Table~\ref{weakme} we list the SU(3) decomposition of the four classes
of decays each for the $\Delta C=1$ and $\Delta C=0$ decays. The SU(3)
analysis shows that all matrix elements but one are allowed. The forbidden
transition is the $\Delta C=1$ transition
$\langle B_{c}({\bf 6)}|H^{\rm pc}({\bf 6})|B_{cc}({\bf 3})\rangle$ which
vanishes in the SU(3) limit since the direct product
${\bf 3}\otimes\overline{\bf 6}\otimes\overline{\bf 6}$ does not contain the
unit representation. All the other transitions are governed by one SU(3)
reduced matrix element. While the $\Delta C=0$ quark transition ($b'$)
$c \to d;\,s \to c$ contributes to all four possible matrix elements, the
quark transition ($a'$) $s \to u;\, u\to d$ contributes only to the matrix
element $\langle B_{c}(\overline{\bf 3})|H^{\rm pc}({\bf 8})|
B_{c}(\overline{\bf 3})\rangle$ because of the KPW theorem.\footnote{We do not
agree with the $\Delta C=0$ analysis of Cheng et al.~\cite{Cheng:1992ff,%
Cheng:2015ckx} who assert that the nondiagonal matrix elements
$\langle B_{c}({\bf 6})|H^{\rm pc}({\bf 8})|B_{c}(\overline{\bf 3})\rangle$ and
$\langle B_{c}(\overline{\bf 3})|H^{\rm pc}({\bf 8})|B_{c}({\bf 6})\rangle$ are
zero.} This pattern is confirmed by the tensor analysis. The only possible
connected tensor invariant is given by the contraction
\begin{equation}\label{pc1}
I^{\rm pc}=B^{a[bc]}B_{a[b'c']}H_{[b\,c\,]}^{[b'c']}.
\end{equation}
The notation $I^{\rm pc}$ is a short-hand notation for
$I^{\rm pc}_{B_fB_i}=\langle B_f|H|B_i\rangle$.
That the tensor invariant $I^{\rm pc}$ vanishes for the transition
$B_{cc}({\bf 3}) \to H({\bf 6})+ B_{c}({\bf 6})$ can be understood by realizing
that the flavor of the noninteracting quark line $a$ must be charmed. But then
the quark lines $b$ and $c$ that go into the final baryon must be light and
symmetric $\{bc\}$ in the ${\bf 6}$ representation. This clashes with the
antisymmetry of the flavor wave function $B^{a[bc]}$ of the final state.

\begin{table}[ht]
\caption{\label{weakme1}
  $\Delta C=1$ values of the weak transition matrix element
  $I^{\rm pc}=\langle B_f|H^{\rm pc}({\bf 6})|B_i\rangle$.
  The suffix labelling ($a$) and ($b$) for the SCS decays is explained in the
  caption of Table~\ref{TableOmegac}.}
\begin{center}
\begin{tabular}{lcclcc}
\hline \hline \\[-11pt]
$\phantom{CF}$   & matrix element & $I^{\rm pc}$ &
$\phantom{SCS}$   & matrix element & $I^{\rm pc}$ \\[3pt] \hline \\[-11pt]
CF
& $ 2\sqrt{6}\langle\Sigma^+|H^{\rm pc}({\bf 6})|\Lambda_{c}^+\rangle$& $-4$ &
SCS
& $ 2\sqrt{6}\langle p|H^{\rm pc}({\bf 6})|\Lambda_{c}^+\rangle$& $+4_b$ \\[3pt]
$\phantom{CF}$&
$2\sqrt{6}\langle\Xi^0|H^{\rm pc}({\bf 6})|\Xi_{c}^0\rangle$& $+4$ &
$\phantom{SCS}$&
$2\sqrt{6}\langle\Sigma^+|H^{\rm pc}({\bf 6})|\Xi_{c}^+\rangle$& $-4_a$ \\[3pt]
$\phantom{CF}$&
$2\sqrt{2}\langle\Sigma^+|H^{\rm pc}({\bf 6})|\Sigma_{c}^+\rangle$& $+4$ &
$\phantom{SCS}$ &
$4\sqrt{3}\langle\Sigma^0|H^{\rm pc}({\bf 6})|\Xi_{c}^0\rangle$& $-4_a$ \\[3pt]
$\phantom{CF}$&
$2\sqrt{2}\langle\Sigma^0|H^{\rm pc}({\bf 6})|\Sigma_{c}^0\rangle$& $+4$ &
$\phantom{SCS}$&
$12\langle\Lambda^0|H^{\rm pc}({\bf 6})|\Xi_{c}^0\rangle$& $-4_a-8_b$ \\[3pt]
$\phantom{CF}$&
$2\sqrt{6}\langle\Lambda^0|H^{\rm pc}({\bf 6})|\Sigma_{c}^0\rangle$& $+4$ &
$\phantom{SCS}$&
$2\sqrt{2}\langle p|H^{\rm pc}({\bf 6})|\Sigma_{c}^+\rangle$& $-4_b$ \\[3pt]
$\phantom{CF}$&
$2\sqrt{2}\langle\Xi^0|H^{\rm pc}({\bf 6})|\Xi_{c}^{\prime\,0}\rangle$& $+4$ &
$\phantom{SCS}$&
$4\langle\Sigma^0|H^{\rm pc}({\bf 6})|\Xi_{c}^{\prime\,0}\rangle$& $+4_a$\\[3pt]
$\phantom{CF}$&
$2\sqrt{6}\langle\Xi_{c}^+|H^{\rm pc}({\bf 6})|\Xi_{cc}^+\rangle$& $+8$ &
$\phantom{SCS}$&
$4\sqrt{3}\langle\Lambda^0|H^{\rm pc}({\bf 6})|\Xi_{c}^{\prime\,0}\rangle$&
  $+4_a-8_b$\\[3pt]
$\phantom{CF}$&
$2\sqrt{2}\langle\Xi_{c}^{\prime+}|H^{\rm pc}({\bf 6})|\Xi_{cc}^+\rangle$& $0$ &
$\phantom{SCS}$&
$2\langle\Xi^0|H^{\rm pc}({\bf 6})|\Omega_{c}^0\rangle$& $+4_a$ \\[3pt]
\hline\hline
\end{tabular}
\end{center}
 \end{table}

We will not provide an exhaustive list of weak matrix elements. Instead, in
Tables~\ref{weakme1} and~\ref{weakme2} we list all the weak matrix elements
that appear in the sample decays discussed in Sec.~\ref{sample} and
Appendix~\ref{scrutiny}. In Table~\ref{weakme1} we confirm the vanishing of
the matrix element $\langle\Xi_{c}^{\prime+}|H^{\rm pc}|\Xi_{cc}^+\rangle$ by
explicit calculation.

\subsection{The $\Delta Y=1$ hyperon decay}

For the $\Delta Y=1$ hyperon decays there are two reduced SU(3) matrix
elements as can be seen by the reduction of the direct product
${\bf 8}\otimes{\bf 8}\otimes{\bf 8}\to 2\cdot{\bf 1}\oplus\ldots$. It is
common practise to characterize the two couplings by their symmetry structure
which are labelled by $d$ (symmetric coupling) and $f$ (antisymmetric
coupling). As we shall see in a moment, the invariant~(\ref{pc1}) has a $d/f$
structure of $d/f=-1$. In addition to the tensor invariant~(\ref{pc1}) one has
a second tensor invariant given by the contraction
$B^{j[bc]}B_{i[bc]}H_{[rj]}^{[ir]}$. This second invariant does not contribute
to the $\Delta C=1$ transitions because i) there are no two same light quarks
in the effective Hamiltonian for the CF and DCS transitions and ii) the two
contributions from the internal contractions ($a$) $(s \bar s) $ and ($b$)
$(d\bar d)$ cancel in the SCS contributions.

\begin{table}[hb]
\caption{\label{weakme2}
  $\Delta C=0$ values of the weak transition matrix element
  $I^{\rm pc}=\langle B_f|H^{\rm pc}({\bf 8})|B_i\rangle$. The suffix
  labelling ($a'$) and ($b'$) is explained in the caption of
  Table~\ref{deltaC=0}.}
\begin{center}
\begin{tabular}{lcc}
\hline \hline \\[-11pt]
& matrix element & $I^{\rm pc}$ \\[3pt] \hline \\[-11pt]
SCS & 
$12\langle\Lambda_{c}^+|H^{\rm pc}({\bf 8})|\Xi_{c}^+\rangle$&
  $-16_{a'}+4_{b'}$ \\[3pt]
&$2\sqrt{6}\langle\Sigma_{c}^0|H^{\rm pc}({\bf 8})|\Xi_{c}^0\rangle$&
  $-4_{b'}$ \\[3pt]
&$4\sqrt{3}\langle\Lambda_{c}^+|H^{\rm pc}({\bf 8})|\Xi_{c}^{\prime+}\rangle$&
  $-4_{b'}$ \\[3pt]
&$2\sqrt{6}\langle\Xi_{c}^0|H^{\rm pc}({\bf 8})|\Omega_{c}^0\rangle$&
  $+4_{b'}$ \\[3pt]
&$2\sqrt{2}\langle\Xi_{c}^{\prime\,0}|H^{\rm pc}({\bf 8})|\Omega_{c}^0\rangle$&
  $+4_{b'}$ \\[3pt]
\hline\hline
\end{tabular}
\end{center}
\end{table}

To analyze the $d/f$ structure of the two invariants it is convenient to
switch to the second rank tensor representation of the baryon flavor wave
function and the effective Hamiltonian. For the baryon flavor wave function
this is achieved by the transformation
\begin{equation}
B_{abc}=\tfrac{1}{\sqrt{2}}\,\epsilon_{ibc}B^i_a \quad\Leftrightarrow\quad
B_a^i = \tfrac{1}{\sqrt{2}}\,\epsilon^{ibc}B_{abc} 
\end{equation}
\begin{equation}
\label{boctet}
B=\left(\begin{array}{ccc}
    -\frac{\Lambda^0}{\sqrt{6}}+\frac{\Sigma^0}{\sqrt{2}}& -\Sigma^+ & p \\
    \Sigma^- &-\frac{\Lambda^0}{\sqrt{6}}-\frac{\Sigma^0}{\sqrt{2}}  & n \\
    \Xi^- & -\Xi^0 & \frac{2\Lambda^0}{\sqrt{6}}\\
   \end{array}\right).
\end{equation}
Compared to the conventional $3\times 3$ representation of the baryon octet
we have changed the phases of the $\Sigma^+$, $\Lambda^0$ ad $\Xi^0$
components to be in agreement with our phase convention from
Lichtenberg~\cite{Lichtenberg:1978pc}. In the same vein one can transform the
fourth rank tensor $H^{[ab]}_{[a'b']}$ to a second rank ``spurion'' tensor
$S^i_j$ by writing
\begin{equation}
H^{[ab]}_{[a'b']}=\epsilon_{ia'b'} \,\epsilon^{jab} S^i_j
  \quad\Leftrightarrow \quad
  S_j^i=\tfrac14\epsilon_{jrs}\,\epsilon^{ir's'}H^{rs}_{r's'}.
\end{equation}
The so-called spurion has $S_3^2=1$ as only nonvanishing component. One obtains
\begin{eqnarray}\label{d/f}
B_\ell^{a[bc]}B^{\ell'}_{a[b'c']}H_{[bc]}^{[b'c']}
  &=&2\bar B_i^aB^j_aS_j^i\ =\ 2\tr(\bar BBS)
  \ =\ \tr\big((B\bar B+\bar B B)S-(B\bar B-\bar BB)S\big), \nonumber \\
B_\ell^{j[bc]}B^{\ell'}_{i[bc]}H_{[rj]}^{[ir]}
  &=&B_i^a B^j_aS_j^i\ =\ \tr(B\bar BS)
  \ =\ \tr\big(\tfrac12(B\bar B+\bar BB)S
  +\tfrac12(B\bar B-\bar BB)S\big), \nonumber \\
  \end{eqnarray}
indicating the ratios $d/f=+1$ and $d/f=-1$ for the two couplings
$\tilde C_F=\sqrt2\tr([B,\bar B]S)$ and $\tilde C_D=\sqrt2\tr(\{B,\bar B\}S)$
mentioned by Lichtenberg in Sec.~9.7 of Ref.~\cite{Lichtenberg:1978pc}.

\begin{table}[ht]
\caption{\label{strongme1}
SU(3) properties of the strong transition matrix elements}
\begin{center}
\begin{tabular}{lccc}
\hline \hline \\[-11pt]
matrix element & direct product
& $N_{\rm SU(3)}$ & $\tilde I_1/\tilde I_2$ \\[3pt] \hline \\[-11pt]
%% class 1
$\langle B({\bf 8})M({\bf 8})|B({\bf 8})\rangle$ &
${\bf 8}\otimes{\bf 8}\otimes{\bf 8}$ & 2 & indefinite \\[3pt]
%% class 2
$\langle B_{c}(\overline{\bf 3})M({\bf 8})|B_{c}(\overline{\bf 3})\rangle$ &
$\overline{\bf 3}\otimes{\bf 3}\otimes{\bf 8}$ & 1 & $-5/4$ \\[3pt]
%% class 3
$\langle B_{c}({\bf 6})M({\bf 8})|B_{c}(\overline{\bf 3})\rangle$ &
$\overline{\bf 3}\otimes\overline{\bf 6}\otimes{\bf 8}$ & 1 & $-1/2$ \\[3pt]
%% class 4
$\langle B_{c}(\overline{\bf 3})M({\bf 8})|B_{c}({\bf 6})\rangle$ &
${\bf 6}\otimes{\bf 3}\otimes{\bf 8}$ & 1 & $-1/2$ \\[3pt]
%% class 5
$\langle B_{c}({\bf 6})M({\bf 8})|B_{c}({\bf 6})\rangle$ &
${\bf 6}\otimes\overline{\bf 6}\otimes{\bf 8}$ & 1 & $+1/0$ \\[3pt]
%% class 6
$\langle B({\bf 8})M(\overline{\bf 3})|B_{c}(\overline{\bf 3})\rangle$ &
$\overline{\bf 3}\otimes{\bf 8}\otimes{\bf 3}$ & 1 & $+1$ \\[3pt]
%% class 7
$\langle B_{c}(\overline{\bf 3})M(\overline{\bf 3})|B_{cc}({\bf 3})\rangle$ &
${\bf 3}\otimes{\bf 3}\otimes{\bf 3}$ & 1 & $-2$ \\[3pt]
%% class 8
$\langle B_{c}({\bf 6})M(\overline{\bf 3})|B_{cc}({\bf 3})\rangle$ &
${\bf 3}\otimes\overline{\bf 6}\otimes{\bf 3}$ & 1 & $0/-1$ \\[3pt]
%% class 9
$\langle B_{cc}({\bf 3})M({\bf 8})|B_{cc}({\bf 3})\rangle$ &
${\bf 3}\otimes\overline{\bf 3}\otimes{\bf 8}$ & 1 & $-1$ \\[3pt]
\hline\hline
\end{tabular}
\end{center}
\end{table}

%%%%%%%%%%%%%%%%%%%%%%%%%%%%%%%%%%%%%%%%%%%%%%%%%%%%%%%%%%%%%%%%%%%%%%%%%%%%%%%
\subsection{Strong matrix elements}
%%%%%%%%%%%%%%%%%%%%%%%%%%%%%%%%%%%%%%%%%%%%%%%%%%%%%%%%%%%%%%%%%%%%%%%%%%%%%%%
In Sec.~\ref{current} we have introduced the two basic flavor tensor
invariants $\tilde I_1$ and $\tilde I_2$ to describe the strong matrix element
$\langle B_fM_k|B_i\rangle$. They read
\begin{equation}\label{tilde12}
\tilde I_1=B^{a[bc]} B_{a[bc']}M^{c'}_{c},\qquad
\tilde I_2=B^{a[bc]} B_{b[c'a]}M^{c'}_{c}.
\end{equation}
Depending on the class of the transition, the two basic tensor invariants are
not always independent. This becomes evident when one analyzes the strong
transitions in terms of the three partaking SU(3) multiplets, as shown in
Table~\ref{strongme1}. In the second column we list the direct products of the
participating SU(3) representations and in the third colummn we register the
number $N_{\rm SU(3)}$ of SU(3) reduced matrix elements needed to describe the
transitions. For the cases $N_{SU(3)}=1$ with only one SU(3) reduced matrix
element we list the ratio $\tilde I_1/\tilde I_2$ of the two basic
couplings~(\ref{tilde12}) in column~4. The respective ratios can be obtained
by rewriting the tensor invariant $\tilde I_2$ in terms of $\tilde I_1$
i) following the flavor flow of the transitions, ii) using the Jacobi
identity, and iii) making use of the wave function symmetries of the involved
baryons. By the same reasoning one finds $\tilde I_2=0$ for the class of
decays $\langle B_{c}({\bf 6})M({\bf 8})|B_{c}({\bf 6})\rangle$ and
$\tilde I_1=0$ for the class of decays
$\langle B_{c}({\bf 6})M(\overline{\bf 3})|B_{cc}({\bf 3})\rangle$. This is
specified in column 4 of Table~\ref{strongme1}  by the notation ``$1/0$'' and
``$0/1$''. Note that the class of light baryon transitions
$\langle B({\bf 8})M({\bf 8})|B({\bf 8})\rangle$ is described by two SU(3)
invariant amplitudes, i.e.\ the ratio $\tilde I_1/\tilde I_2$ is not definite
for this class of transitions as annotated in Table~\ref{strongme1}. One can
relate the two invariants $\tilde I_1$ and $\tilde I_2$ to the usual pair of
SU(3) $d$ and $f$ couplings as has been done for the weak transition matrix
elements (see Eq.~(\ref{d/f})). One finds
\begin{eqnarray}\label{trace}
\tilde I_1&=&-\tfrac 12\tr(\bar BB\bar M)
  \ =\ -\tfrac14\tr\left((B\bar B+\bar BB)\bar M\right)
  +\tfrac14\tr\left((B\bar B-\bar BB)\bar M\right)
  \ =\ -\tfrac14I^d+\tfrac14I^f, \nonumber \\
\tilde I_2&=&\tfrac12\tr\left((B\bar B+\bar BB)\bar M\right)\ =\ \tfrac12I^d.
\end{eqnarray}
In the usual way of SU(3) labelling, the invariant $\tilde I_1$ can be seen
to correspond to a coupling ratio $d/f=-1$ while the invariant $\tilde I_2$
corresponds to a pure $d$ coupling. One obtains
\begin{equation}\label{IdIf}
I^d=2\tilde I_2,\qquad I^f=4\tilde I_1+2\tilde I_2.
\end{equation}
It is common knowledge that the generator of SU(3) is proportional to the
antisymmetric coupling of two baryon octets. As Eq.~(\ref{trace}) shows, the
antisymmetric coupling is proportional to the linear combination
$I^f\sim 2\tilde I_1+\tilde I_2$. The normalization of the $f$ coupling must
be chosen such that the expectation value of the charge operator between
proton states is given by the charge of the proton which is $1$ in units of
the elementary charge $e$. Let us verify this for the choice
$I^f=4\tilde I_1+2\tilde I_2$. First we express the $3\times3$ charge operator
$Q$ in terms of its $\pi^0$ and $\eta_8$ components. One has
\begin{equation}\label{charge1}
Q=\begin{pmatrix}2/3&0&0\\ 0&-1/3&0\\ 0&0&-1/3\\\end{pmatrix}
  =\frac{1}{\sqrt2}Q(\pi^0)+\frac{1}{\sqrt6}Q(\eta_8),
\end{equation}
where the charge operator components are given by
\begin{equation}
Q(\pi^0)=\frac1{\sqrt2}\begin{pmatrix}1&0&0\\0&-1&0\\0&0&0\\\end{pmatrix},\qquad
Q(\eta_8)=\frac1{\sqrt6}\begin{pmatrix}1&0&0\\0&1&0\\0&0&-2\\\end{pmatrix}.
\end{equation}
The expectation value of the charge operator between two proton states can
then be calculated to be $1$ according to the decomposition of the charge
operator~(\ref{charge1}). One has
\begin{equation}
\langle p|Q|p\rangle=\frac1{\sqrt2}I^f_{p\pi^0p}
  +\frac1{\sqrt{6}}I^f_{p\eta_8p}=1.
\end{equation}
This is in accordance with Table~\ref{strongme2} where we list the values of
the strong transition matrix elements for the invariants $\tilde I_1$,
$\tilde I_2$, and for the linear combinations of invariants $I^d=2\tilde I_2$
(not shown), $I^f=4\tilde I_1+2\tilde I_2$ and
$I^{\rm CQM}=4\tilde I_1+5\tilde I_2$ which appear in the current algebra
description of the charm baryon nonleptonic decays. The notation
$\langle B_f,M_k|B_i\rangle$ means either $\langle B_fM_k|B_i\rangle$ or
$\langle B_f|M_k|B_i\rangle$, depending on whether one deals with $I^f$ or
$I^{\rm CQM}$, respectively. The same result can be obtained by direct
calculation of the matrix elements $\tilde I_1$ and $\tilde I_2$ for the
charge operator. The result is $\tilde I_1=1/6$ and $\tilde I_2=1/6$ which
confirms again that the invariant $I^f=4\tilde I_1+2\tilde I_2$ has the
correct normalization. Table~\ref{strongme2} shows that the generator matrix
element
$I^f_{\Sigma^-\pi^+\Lambda^0}=\langle\Sigma^-|Q(\pi^-)|\Lambda^0\rangle$
vanishes which follows from the fact that $Q(\pi^-)$ is a generator of the
subgroup SU(2).

\begin{table}[hb]
\caption{\label{strongme2}
  Values of tensor invariants for the strong transitions $B_i\to B_f+M_k$.
  $I^f$ and $I^{\rm CQM}$ are related to $\tilde I_1$ and $\tilde I_2$ by
  $I^f=4\tilde I_1+2\tilde I_2$ and $I^{\rm CQM}= 4\tilde I_1+5\tilde I_2$.}
\begin{center}
\begin{tabular}{lrrrr}
\hline \hline \\[-11pt]
&$\tilde I_1$ & $\tilde I_2$ & $I^f$ & $I^{\rm CQM}$ \\[3pt] \hline \\[-11pt]
%% transition 1
$2\sqrt{2}\langle p,\pi^0|p\rangle$
&$  0 $&$ +1 $&$ +2 $&$ +5 $\\
%% transition 2
$2\sqrt{6}\langle p,\eta_8|p\rangle$
&$ +2 $&$ -1 $&$ +6 $&$ +3 $\\
%% transition 3
$2\sqrt{2}\langle p,K^-|\Sigma^0\rangle$
&$ -1 $&$ +1 $&$ -2 $&$ +1 $\\
%% transition 4
$2\sqrt{6}\langle p,K^-|\Lambda^0\rangle$
&$ +1 $&$ +1 $&$ +6 $&$ +9 $\\ 
%% transition 5
$2\sqrt{2}\langle\Sigma^-,\pi^+|\Sigma^0\rangle$
&$ -1 $&$  0 $&$ -4 $&$ -4 $\\ 
%% transition 6
$2\sqrt{6}\langle\Sigma^-,\pi^+|\Lambda^0\rangle$
&$ -1 $&$ +2 $&$  0 $&$ +6 $\\
%% transition 7
$2\sqrt{6}\langle\Lambda^0,\pi^+|\Sigma^+\rangle$
&$ +1 $&$ -2 $&$  0 $&$ -6 $\\ 
%% transition 8
$2\sqrt{2}\langle\Sigma^0,\pi^+|\Sigma^+\rangle$
&$ -1 $&$  0 $&$ -4 $&$ -4 $\\
%% transition 9
$2\langle\Xi^0,K^+|\Sigma^+\rangle$
&$  0 $&$ +1 $&$ +2 $&$ +5 $\\
%% transition 10
$2\sqrt{2}\langle\Sigma^0,\bar K^0|\Xi^0\rangle$ 
&$  0 $&$ -1 $&$ -2 $&$ -5 $\\
%% transition 11
$2\langle\Xi^0,K^+|\Sigma^+\rangle$
&$  0 $&$ +1 $&$ +2 $&$ +5 $\\
%% transition 12
$2\sqrt{6}\langle\Lambda_{c}^+,\pi^-|\Sigma_{c}^0\rangle$ 
&$ +1 $&$ -2 $&$  0 $&$ -6 $\\
%% transition 13
$2\sqrt{6}\langle\Sigma_{c}^0,\pi^+|\Lambda_{c}^+\rangle$
&$ -1 $&$ +2 $&$  0 $&$ +6 $\\ 
%% transition 14
$12\langle\Xi_{c}^0,K^+|\Lambda_{c}^+\rangle$ 
&$ -5 $&$ +4 $&$-12 $&$  0 $\\ 
%% transition 15
$4\sqrt{3}\langle\Xi_{c}^{\prime\,0},K^+|\Lambda_{c}^+\rangle$ 
&$ +1 $&$ -2 $&$  0 $&$ -6 $\\
%% transition 16
$12\sqrt{2}\langle\Xi_{c}^{+},\pi^0|\Xi_{c}^+\rangle$ 
&$ +5 $&$ -4 $&$+12 $&$  0 $\\
%% transition 17
$2\sqrt{6}\langle\Sigma_{c}^{++},K^-|\Xi_{c}^+\rangle$ 
&$ -1 $&$ +2 $&$  0 $&$ +6 $\\
%% transition 18
$12\langle\Xi_{c}^+,\pi^-|\Xi_{c}^0\rangle$ 
&$ +5 $&$ -4 $&$+12 $&$  0 $\\
%% transition 19
$4\sqrt{3}\langle\Xi_{c}^{\prime+},\pi^-|\Xi_{c}^0\rangle$ 
&$ +1 $&$ -2 $&$  0 $&$ -6 $\\
%% transition 20
$12\langle\Xi_{c}^+,\pi^-|\Xi_{c}^0\rangle$ 
&$ +5 $&$ -4 $&$+12 $&$  0 $\\
%% transition 21
$4\sqrt{3}\langle\Xi_{c}^{\prime+},\pi^-|\Xi_{c}^0\rangle$ 
&$ +1 $&$ -2 $&$  0 $&$ -6 $\\
%% transition 22
$12\langle\Xi_{c}^+,\pi^-|\Xi_{c}^0\rangle$ 
&$ +5 $&$ -4 $&$+12 $&$  0 $\\
%% transition 23
$4\sqrt{3}\langle\Xi_{c}^{\prime+},\pi^-|\Xi_{c}^0\rangle$ 
&$ +1 $&$ -2 $&$  0 $&$ -6 $\\
%% transition 24
$12\langle\Xi_{c}^0,\pi^+|\Xi_{c}^+\rangle$ 
&$ -5 $&$ +4 $&$-12 $&$  0 $\\
%% transition 25
$2\sqrt{12}\langle\Xi_{c}^{\prime\,0},\pi^+|\Xi_{c}^+\rangle$ 
&$ -1 $&$ +2 $&$  0 $&$ +6 $\\
%% transition 26
$4\sqrt{3}\langle\Sigma_{c}^+,K^-|\Xi_{c}^0\rangle$ 
&$ -1 $&$ +2 $&$  0 $&$ +6 $\\
%% transition 27
$12\langle\Lambda_{c}^+,K^-|\Xi_{c}^0\rangle$ 
&$ -5 $&$ +4 $&$-12 $&$  0 $\\
%% transition 28
$2\sqrt{6}\langle\Xi_{c}^{0},\bar K^0|\Omega_{c}^0\rangle$ 
&$ -1 $&$ +2 $&$  0 $&$ +6 $\\
%% transition 29
$2\sqrt{2}\langle\Xi_{c}^{\prime\,0},\bar K^0|\Omega_{c}^0\rangle$ 
&$ -1 $&$  0 $&$ -4 $&$ -4 $\\
%% transition 30
$2\sqrt{2}\langle\Xi_{cc}^{++},\pi^0|\Xi_{cc}^{++}\rangle$ 
&$ +1 $&$ -1 $&$ +2 $&$ -1 $\\
%% transition 31
$2\langle\Xi_{cc}^+,\pi^+|\Xi_{cc}^{++}\rangle$ 
&$ -1 $&$ +1 $&$ -2 $&$ +1 $\\
%% transition 32
$2\sqrt{2}\langle\Xi_{cc}^+,\pi^0|\Xi_{cc}^+\rangle$ 
&$ -1 $&$ +1 $&$ -2 $&$ +1 $\\
%% transition 33
$2\sqrt{6}\langle\Sigma^+,D^0|\Xi_{c}^+\rangle$ 
&$ +1 $&$ +1 $&$ +6 $&$ +9 $\\
%% transition 34
$2\sqrt{6}\langle\Lambda_{c}^+,D^+|\Xi_{cc}^{++}\rangle$ 
&$ -2 $&$ +1 $&$ -6 $&$ -3 $\\
%% transition 35
$2\sqrt{2}\langle\Sigma_{c}^+,D^+|\Xi_{cc}^{++}\rangle$ 
&$  0 $&$ -1 $&$ -2 $&$ -5 $\\
%% transition 36
$2\sqrt{6}\langle\Lambda_{c}^+,D^0|\Xi_{cc}^+\rangle$ 
&$ -2 $&$ +1 $&$ -6 $&$ -3 $\\
%% transition 37
$2\sqrt{2}\langle\Sigma_{c}^+,D^0|\Xi_{cc}^+\rangle$ 
&$  0 $&$ +1 $&$ +2 $&$ +5 $\\[3pt]
\hline\hline
\end{tabular}
\end{center}
\end{table}

We want to ascertain that $I^f=4\tilde I_1 +2\tilde I_2$ is the correctly
normalized generator matrix element also from SU(4). As an example we calculate
the charge of the charm baryon state $|\Xi^+_{c}\rangle$. In SU(4) the charge
operator is given by $Q={\rm diag}\,(2/3,-1/3,-1/3,\,2/3)$. For the matrix
elements of the charge operator one finds $\tilde I_1=1/4$ and $\tilde I_2=0$
which gives $I^f=4\tilde I_1 +2\tilde I_2=1$, as required.
Table~\ref{strongme2} shows that $I^f=0$ for the two classes of transitions 
$\langle B_{c}(\overline{\bf 3})M({\bf 8})|B_{c}({\bf 6})\rangle$ and
$\langle B_{c}({\bf 6})M({\bf 8})|B_{c}(\overline{\bf 3})\rangle$. The
vanishing of the $f$-type matrix elements follows from the fact that the
vector charges $M_k$ associated with the members $M_k$ of light meson octet
$M({\bf 8})$ are generators of SU(3) which do not connect different SU(3)
multiplets.

\begin{figure}\begin{center}
\epsfig{figure=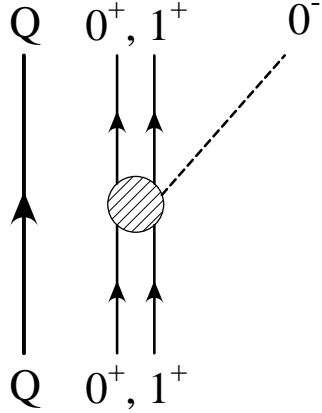, scale=0.30}\end{center}
\caption{\label{HQET} light diquark transitions}
\end{figure}

The second linear combination of $\tilde I_1$ and $\tilde I_2$ that enters the
current algebra calculation is the constituent quark model (CQM) tensor
invariant $I^{\rm CQM}$ which reads
\begin{equation}
I^{\rm CQM}=4\tilde I_1+5\tilde I_2=\tfrac 32 I^d+I^f. 
\end{equation}
The constituent quark model coupling $I^{\rm CQM}$ can be seen to correspond
to a $d/f$ ratio of 3/2. Table~\ref{strongme2} shows that the constituent
quark model coupling $I^{\rm CQM}$ vanishes for the class of decays
$\langle B_{c}(\overline{\bf 3})M({\bf 8})|B_{c}(\overline{\bf 3})\rangle$.
This can be understood from a $LS$ coupling analysis of the light-side
transition  $0^+\to 0^+\,+0^-$. In the constituent quark model (or in the
Heavy Quark Effective Theory (HQET)) the light side transition effectively
decouples from the heavy side charm quark transition as illustrated in
Fig.~\ref{HQET}. The light side diquark has the quantum numbers
$J^P=(0^+,1^+)$ depending on whether the two light quark spins are in the spin
singlet or the spin triplet state. One can then do a $LS$ analysis of the
light side transitions $D(0^+)\to D(0^+)+0^-$, $D(0^+)\to D(1^+)+0^-$,
$D(1^+)\to D(0^+)+0^-$ and $D(1^+)\to D(1^+)+0^-$. From parity the orbital
angular momentum of the final state must be odd. One concludes that the
transition $D(0^+)\to D(0^+)+0^-$ must vanish since even for the lowest
possible angular momentum $L=1$ the final state $D(0^+)+0^-$ cannot couple to
the initial $D(0^+)$ state because the total spin $S$ of the final state is
$S=0$. A similar $LS$ analysis for the other three diquark transitions shows
that these transitions are allowed. The coupling/decoupling arguments
presented here do not hinge on the diquark representation of the light quarks
but follow also in a single quark picture as long as the two light quarks move
independently~\cite{Korner:1994nh,Hussain:1999sp}.

In Sec.~\ref{current} we have also considered the possibility that the strong
transition is described by a general mixture of SU(3) $d$ and $f$ couplings.
We denote the general coupling by $I^{\rm gen}(d/f)$ which we choose to
parametrize by
\begin{equation}
\label{strong2}
I^{\rm gen}(d/f)=d(\tfrac53\,I^{\rm CQM}-\tfrac23\,I^f)+ fI^f  
\end{equation}
with $d+f=1$. We have chosen the parametrization~(\ref{strong2}) such that
i) for $d=0$ and $f=1$ one recovers the $f$-coupling structure
$I^f=4\tilde I_1+2\tilde I_2$ and ii) for $d=3/5$ and $f=2/5$ one recovers
the coupling structure $I^{\rm CQM}=4\tilde I_1+5\tilde I_2$. 

%%%%%%%%%%%%%%%%%%%%%%%%%%%%%%%%%%%%%%%%%%%%%%%%%%%%%%%%%%%%%%%%%%%%%%%%%%%%%%%
\section{\label{scrutiny}Scrutinizing the Cheng et al.\ modification\\
  of the current algebra approach}
%%%%%%%%%%%%%%%%%%%%%%%%%%%%%%%%%%%%%%%%%%%%%%%%%%%%%%%%%%%%%%%%%%%%%%%%%%%%%%%
\setcounter{equation}{0}\def\theequation{D\arabic{equation}}
As the authors of Ref.~\cite{Zou:2019kzq} put it, the CF single charm baryon
decay $\Lambda_{c}^+ \to \Xi^0 K^+$ deserves special attention. First, there is
no factorizing contribution to the decay as evidenced by Table~\ref{TableCF},
i.e.\ the decay is contributed to only by the nonfactorizing $W$-exchange
contributions. Second, as we reconfirm below, the $W$-exchange contribution to
the $S$-wave amplitude $A^{\rm com}=A^{\rm com}_{fki}$ vanishes for this decay
in the standard current algebra approach since $I_3=0$ and $\hat I_3=0$ (see
Table~\ref{TableCF}). Thus the current algebra approach predicts the decay to
be purely $P$-wave. This implies that the asymmetry parameter in this decay is
predicted to be zero. Cheng et al.\ point out that it will be difficult to
reproduce the rather large experimental branching ratio of this decay
(${\cal B}=(5.5\pm0.7)\times10^{-3}$~\cite{Ablikim:2018bir}) with a $P$-wave
contribution alone. They also suggest a way out of what they call a puzzle by
reinstalling an $S$-wave contribution by appealing to the topological
structure of the decay. We believe that the construction of
Ref.~\cite{Zou:2019kzq} is based on an erroneous assumption about the
contribution of the topological diagram IIa to the $S$-wave amplitude
$A^{\rm com}$ as we shall demonstrate in the following.

In order to shed more light on the problem, we list the current algebra
$W$-exchange contributions to the amplitudes $A^{\rm com}$ and
$B^{\rm pole}=B^{\rm pole}_{fki}$ in terms of the intermediate baryon states
for the $s$- and $u$-channel contributions. The intermediate state in the
$s$ channel is given by $\Sigma^+$. In the $u$ channel the intermediate states
are given by the flavor degenerate pair of states $\Xi_{c}^0$ and
$\Xi_{c}^{\prime\,0}$ where $\Xi_{c}^0$ contributes to $A^{\rm com}(u)$ and
$\Xi_{c}^{\prime\,0}$ contributes to $B^{\rm pole}(u)$ since
$\langle\Xi_{c}^{\prime\,0}|K^+|\Lambda_{c}^+\rangle=0$ and
$\langle \Xi_{c}^{0}K^+|\Lambda_{c}^+\rangle=0$
(see Appendix~\ref{weak-strong}). One therefore has 
\begin{eqnarray}\label{ABtest}
A^{\rm com}&=&\frac{1}{f_K}6(4\pi\bar X_2)
  \Big(I^f_{\Xi^0K^+\Sigma^+}\,I^{\rm pc}_{\Sigma^+\Lambda_{c}^+}
  -I^{\rm pc}_{\Xi^0\Xi_{c}^0}I^f_{\Xi_{c}^0K^+\Lambda_{c}^+}\Big),
  \nonumber\\
B^{\rm pole}&=&\frac{1}{f_K} 6 (4\pi)\bar X_2\tfrac 23 (4\pi) \bar Z
  \Big(I^{\rm CQM}_{\Xi^0K^+\Sigma^+}I^{\rm pc}_{\Sigma^+\Lambda_{c}^+}
  R_s(\Sigma^+)+I^{\rm pc}_{\Xi^{0}\,\Xi_{c}^{\prime\,0}}\,
  I^{\rm CQM}_{\Xi_{c}^{\prime\,0}K^+\Lambda_{c}^+}
  R_u(\Xi_{c}^{\prime\,0})\Big).\qquad
\end{eqnarray}
In line with the fact that we are working in the SU(3) limit we define
$\bar X_2$ and $\bar Z$ as SU(3) averages of the bag model integrals listed in
Ref.~\cite{Zou:2019kzq}. Note that the bag model integral $X_1$ does not
appear in Eqs.~(\ref{ABtest}), since $X_1$ vanishes in the SU(3) limit.
Intrinsic SU(3) breaking effects are accounted for in the bag model
calculations but are quite small of ${\cal O}(1-2)\%$, as can be estimated
from the ratio $X_1/X_2$ and the relative size of the bag model integrals
$X_2^d$ and $X_2^d$, and $Z_1$ and $Z_2$ in Ref.~\cite{Zou:2019kzq}. One can
check that Eqs.~(\ref{ABtest}) agree with Eq.~(40) of Ref.~\cite{Zou:2019kzq}
when the SU(3) limit is taken.

The flavor coefficients in Eqs.~(\ref{ABtest}) are given by
\begin{eqnarray}\label{ABtest2}
\mbox{$S$-wave $(s;\Sigma^+)$}:&&
I^f_{\Xi^0K^+\Sigma^+}\,I^{\rm pc}_{\Sigma^+\Lambda_{c}^+}
  =-4/2\sqrt{6}=2I_3+\underline{4I_5}\nonumber\\
\mbox{$S$-wave $(u;\Xi_{c}^0,\underbrace{\Xi_{c}^{\prime\,0}}_0)$}:&&
I^{\rm pc}_{\Xi^0\Xi_{c}^0}I^f_{\Xi_{c}^0K^+\Lambda_{c}^+}
  =-4/2\sqrt{6}=2\hat I_3+\underline{4I_5}\nonumber\\
\mbox{$P$-wave $(s;\Sigma^+)$}:&&
I^{{\rm CQM}(K^+)}_{\Xi^0\Sigma^+}I^{\rm pc}_{\Sigma^+\Lambda_{c}^+}
      = -10/2\sqrt{6}=I_3+\underline{2I_4}+\underline{6I_5}\nonumber\\
\mbox{$P$-wave $(u;\underbrace{\Xi_{c}^0}_0,\Xi_{c}^{\prime\,0})$}:&&
I^{\rm pc}_{\Xi^{0}\,\Xi_{c}^{\prime\,0}}\,
  I^{{\rm CQM}(K^+)}_{\Xi_{c}^{\prime\,0}\,\Lambda_{c}^+}
  =-6/2\sqrt{6}=\hat I_3+2\hat I_4+\underline{6I_5},\qquad
\end{eqnarray}
where we have included the corresponding results in terms of the topological
tensor invariants. As proven in Sec.~\ref{tables}, the two results agree with
each other. We emphasize that the topological tensor invariants depend only on
the initial and final state particles. The summation over intermediate states
is automatically accounted for. It is for this reason that the representation
of the current algebra results in terms of topological tensor invariants is
much compacter than the representation in terms of intermediate states.

The $S$-wave contribution in the first line of Eqs.~(\ref{ABtest}) clearly
vanishes for the decay $\Lambda_{c}^+ \to \Xi^0 K^+$ since
$A^{\rm com}\sim I_3-\hat I_3$, and both $I_3$ and $\hat I_3$ vanish. As
emphasized in Sec.~\ref{current}, the contribution of the nonvanishing
topological invariant $I_5$ cancels out in the total sum of the $s$- and
$u$-channel contributions. Therefore, there is no need to banish the
contribution of the topological diagram III represented by the single
topological invariant $I_5$ as done in Ref.~\cite{Zou:2019kzq} since $I_5$
does not contribute to the $S$-wave amplitude altogether. Instead, Cheng et
al. reinstall an $S$-wave contribution by appealing to the topological
structure of the nonleptonic transitions. Their idea is that diagram IIa (see
also Fig.~1b in Ref.~\cite{Zou:2019kzq}) allows for an $s$-channel $S$-wave
contribution to $\Lambda_{c}^+ \to \Xi^0 K^+$, since $I_4 \neq 0$. However, our
analysis shows that the topological tensor invariant $I_4$ associated with
diagram IIa does not contribute to the $S$-wave $s$-channel amplitude
$A^{\rm com}$. The flaw in the reasoning of Ref.~\cite{Zou:2019kzq} results
from an incomplete knowledge of how the topological diagrams are connected
with the current algebra contributions which are now available from our
analysis.
 
Nevertheless, let us present the results of the modified current algebra
approach of Ref.~\cite{Zou:2019kzq} where we again use the representation in
terms of topological tensor invariants. In the modified current algebra
approach the $u$-channel contributions to $\Lambda_{c}^+ \to \Xi^0 K^+$ are
relinquished by observing that i) there are no contributions from the
topological diagram IIb and from the postulate that ii) contributions from the
topological diagram III must be set to zero. In the modification of the
current algebra approach one therefore has
\begin{eqnarray}
A^{\rm com}&=&\frac{1}{f_K}6(4\pi \bar X_2)\Big(I^f_{\Xi^0K^+\Sigma^+}
  \,I^{\rm pc}_{\Sigma^+\Lambda_{c}^+}-0(u)\Big), \nonumber \\
B^{\rm pole}&=&\frac{1}{f_K} 6(4\pi\bar X_2)\tfrac 32(4\pi \bar Z)
 \Big(I^{\rm CQM}_{\Xi^0K^+\Sigma^+}I^{\rm pc}_{\Sigma^+\Lambda_{c}^+}
 R_s(\Sigma^+)+0(u)\Big),\qquad
\end{eqnarray}
where $0(u)$ stands for the banished $u$-channel contributions.  As a result
of their modifications, in their numerical evaluation Cheng et al.\ obtain a
rather large value for the asymmetry parameter
$\alpha_{\Lambda_{c}^+ \to \Xi^0 K^+}=0.90$ due to the fact that $A^{\rm com}$
is no longer zero. This result is in crass contradiction to the current
algebra result where $\alpha_{\Lambda_{c}^+ \to \Xi^0 K^+}=0$. The issue of a
vanishing or non-vanishing $S$-wave contribution to the decay
$\Lambda_{c}^+ \to \Xi^0 K^+$ can be settled by a measurement of the asymmetry
parameter in this decay. Unfortunately, such a measurement is not available at
present. As concerns the rate, however, Cheng et al.\ succeed in their
original goal to increase the branching ratio to ${\cal B}=7.1\times10^{-3}$
close to the experimental branching ratio
${\cal B}=(5.5\pm0.7)\times10^{-3}$~\cite{Ablikim:2018bir}.

Since the arguments in Ref.~\cite{Zou:2019kzq} are based on an incomplete
knowledge of how the topological invariants contribute to the current algebra
results, it is difficult to follow their reasoning in the treatment of some of
the other charm baryon decays. A case in point is the SCS decay
$\Xi_{c}^+ \to \Sigma^0 \pi^+$. The $W$-exchange contributions proceed via the
$(c \to s;\,s \to u)$ transitions (called ($a$) in Appendix~\ref{tensors})
with the nonvanishing  tensor invariants $I_3=2/4\sqrt{3}$ and
$I_5=1/4\sqrt{3}$ (see Table~\ref{TableSCSa}). The intermediate state in the
$s$-channel is $\Sigma^+$ for both $A^{\rm com}(s)$ and $B^{\rm pole}(s)$
while the intermediate states in $A^{\rm com}(u)$ and $B^{\rm pole}(u)$ are
$\Xi_{c}^{\,0}$ and $\Xi_{c}^{\prime\,0}$, respectively, as follows from the
fact that $\langle\Xi_{c}^{'0}|\pi^+|\Xi_{c}^+\rangle=0$ and
$\langle \Xi_{c}^0 \pi^+|\Xi_{c}^+ \rangle=0$ (see Appendix~\ref{weak-strong}).
One has
\begin{eqnarray}
A^{\rm com}&=&\frac{1}{f_K}6(4\pi\bar X_2)
  \Big(I^f_{\Xi^0\pi^+\Sigma^+}\,I^{\rm pc}_{\Sigma^+\Lambda_{c}^+}
  -I^{\rm pc}_{\Xi^0\Xi_{c}^0}I^f_{\Xi_{c}^0\pi^+\Lambda_{c}^+}\Big),
  \nonumber\\
B^{\rm pole}&=&\frac{1}{f_K} 6 (4\pi)\bar X_2\tfrac 23 (4\pi) \bar Z
 \Big(I^{\rm CQM}_{\Xi^0\pi^+\Sigma^+}I^{\rm pc}_{\Sigma^+\Lambda_{c}^+}
   R_s(\Sigma^+)+I^{\rm pc}_{\Xi^{0}\,\Xi_{c}^{\prime\,0}}\,
   I^{\rm CQM}_{\Xi_{c}^{\prime\,0}\pi^+\Lambda_{c}^+}
   R_u(\Xi_{c}^{\prime\, 0})\Big)\qquad
\end{eqnarray}
and
\begin{eqnarray}
\mbox{$S$-wave $(s;\Sigma^+)$}:&&
I^f_{\Sigma^0\pi^+\Sigma^+}\,I^{\rm pc}_{\Sigma^+\Xi_{c}^+}
  =(-\sqrt{2})(-2\sqrt{6})=2/\sqrt{3}=\underline{2I_3}+\underline{4I_5}
  \nonumber\\
\mbox{$S$-wave $(u;\Xi_{c}^0,\underbrace{\Xi_{c}^{\prime\,0}}_0)$}:&&
I^{\rm pc}_{\Sigma^0\Xi_{c}^0}I^f_{\Xi_{c}^0\pi^+\Xi_{c}^+}
  =(-1/\sqrt{3})(-1)=1/\sqrt{3}=2\hat I_3+\underline{4I_5}\nonumber\\
\mbox{$P$-wave $(s;\Sigma^+)$}:&&
I^{\rm CQM}_{\Sigma^0\pi^+\Sigma^+}I^{\rm pc}_{\Sigma^+\Xi_{c}^+}
  =(-\sqrt{2})(-\sqrt{2/3})= 2/\sqrt{3}=\underline{I_3}+2I_4+\underline{6I_5}
  \nonumber\\
\mbox{$P$-wave $(u;\underbrace{\Xi_{c}^0}_0,\Xi_{c}^{\prime\,0})$}:&&
I^{\rm pc}_{\Sigma^{0}\,\Xi_{c}^{\prime\,0}}\,
I^{\rm CQM}_{\Xi_{c}^{\prime\,0}\pi^+\Xi_{c}^+}
  =(1)(\sqrt{3}/2)=\sqrt{3}/2=\hat I_3+2\hat I_4+\underline{6I_5},\qquad
\end{eqnarray}
leading to
\begin{eqnarray}
\hat A^{\rm com}&=&(\underline{2I_3}+\underline{4I_5})
  -(2\hat I_3+\underline{4I_5})\nonumber\\
\hat B^{\rm pole}&=&(\underline{I_3}+2I_4+\underline{6I_5})\,R_s(\Sigma^+)
  +(\hat I_3+2\hat I_4+\underline{6I_5})\,R_u(\Xi^{\prime\,0}).\qquad
\end{eqnarray}
The results can be checked to be in agreement with the results in
Ref.~\cite{Zou:2019kzq} up to sign differences due to different sign
conventions for the flavor wave functions. The above results for the decay
$\Xi_{c}^+ \to \Sigma^0 \pi^+$ in Ref.~\cite{Zou:2019kzq} were calculated in
the modified current algebra approach as defined earlier on in the paper.
An obvious question is why the $u$-channel contributions were dropped in
Eqs.~(\ref{ABtest}) and not in the decay $\Xi_{c}^+ \to \Sigma^0 \pi^+$. In
both cases the $u$-channel contributions are proportional to the tensor
invariant $I_5$ which are banished in one case but not in the other case.

Another case of puzzlement are the SCS decays $\Omega_{c}^0\to\Sigma^+\bar K^-$
and $\Omega_{c}^0\to\Sigma^0\bar K^0$ treated in a follow-up
paper~\cite{Hu:2020nkg}, again in the framework of the modified current
algebra approach. That the authors of Ref.~\cite{Hu:2020nkg} use the modified
current algebra approach is evident from the fact that they do not even list
the decays $\Omega_{c}^0 \to pK^-$ and $\Omega_{c}^0 \to n \bar K^0$ in their
list of $\Omega_{c}^0$ decays since these decays are proportional to $I_5$
alone (see Table~\ref{TableOmegac}). Up to a normalization factor, the above
two decays $\Omega_{c}^0\to\Sigma^+\bar K^-$ and
$\Omega_{c}^0\to\Sigma^0\bar K^0$ can be seen to have a topological invariant
structure which is identical to the one in the decay
$\Lambda_{c}^+ \to \Xi^0 K^+$ treated above (see Tables~\ref{TableCF}
and~\ref{TableOmegac}). Therefore, one would expect substantial $S$-wave
contributions and large values of the asymmetry parameter in these decays in
the modified current algebra approach, contrary to the numerical results
listed in Ref.~\cite{Hu:2020nkg}.

Apart from the misidentification in the $S$-wave contribution mentioned above,
a second serious objection against the prescription of Cheng et al.\ is the
following. As written down in Eq.~(\ref{expansion}), the topological tensor
invariant $I_5$ projects onto the reduced topological matrix element
$\mathbfcal{T}_5$.  The reduced matrix element  $\mathbfcal{T}_5$ in turn is
calculated in terms of bag model integrals which certainly do not vanish as
shown in Refs.~\cite{Cheng:2020wmk,Zou:2019kzq,Meng:2020euv,Hu:2020nkg}.
However, one cannot arbitrarily set $\mathbfcal{T}_5$ to zero and keep the
other reduced topological matrix elements at their nonzero bag model values.
As an additional justification of their prescription to banish the
contributions proportional to $I_5$, the authors of Ref.~\cite{Zou:2019kzq}
cite the constituent quark model calculation of Ref.~\cite{Korner:1992wi}
where the contributions proportional to $I_5$ were found to be numerically
small. However, in Ref.~\cite{Korner:1992wi} the reduced matrix element
$\mathbfcal{T}_5$  given in terms of an overlap integral $H_3$ was found to be
numerically small which had no implications for the other two topological
contributions IIa and IIb since they are proportional to an unrelated overlap
integral $H_2$.

%%%%%%%%%%%%%%%%%%%%%%%%%%%%%%%%%%%%%%%%%%%%%%%%%%%%%%%%%%%%%%%%%%%%%%%%%%%%%%%
\section{\label{spinkinematics}Amplitudes, rates and asymmetry parameter}
%%%%%%%%%%%%%%%%%%%%%%%%%%%%%%%%%%%%%%%%%%%%%%%%%%%%%%%%%%%%%%%%%%%%%%%%%%%%%%%
\setcounter{equation}{0}\def\theequation{E\arabic{equation}}
As before we shall use the abbreviations $Q_\pm=(m_i\pm m_f)-m_k^2$ such that
the magnitude of the rest frame momenta of the daughter particles read
$p=\sqrt{Q_+Q_-}/2m_i$. We follow the conventions of Ref.~\cite{Korner:1992wi}
except that we change the sign of the p.c.\ amplitude $B$.
\begin{eqnarray}
\mbox{Invariant amplitudes}&:&\langle B_fM|{\cal H}|B_1\rangle
  =\bar u_f(A-B\gamma_5)u_i\nonumber\\
\mbox{Helicity amplitudes}&:& H^{\rm pv}_{1/2\,0}=2\sqrt{Q_+}A,\quad
  H^{\rm pc}_{1/2\,0}=2\sqrt{Q_-}B\nonumber\\
\mbox{Rate}&:&\Gamma=\frac{p}{32\pi n_i^2}\left(|H^{\rm pv}_{1/2\,0}|^2
  +|H^{\rm pc}_{1/2\,0}|^2\right)=\nonumber\\&&
  \quad=\frac{p}{8\pi m_i^2}\left(Q_+|A|^2+Q_-|B|^2\right)\nonumber\\
\mbox{Asymmetry parameter}&:&
\frac{|H_{1/2\,0}|^2-|H_{-1/2\,0}|^2}{|H_{1/2\,0}|^2+|H_{-1/2\,0}|^2}
  =\frac{2\real(H^{\rm pv}H^{\rm pc\,\ast})}{|H^{\rm pv}|^2+|H^{\rm pc}|^2}
  =\nonumber\\&&\qquad\qquad\qquad\qquad\quad\,
  =\frac{2\sqrt{Q_+Q_-}\real(AB^\ast)}{\big(Q_+|A|^2+Q_-|B|^2\big)}
\end{eqnarray}

\end{appendix}

\end{document}